\documentclass[english,twocolumn,showpacs,aps,pre,secnumarabic]{revtex4}
\usepackage[T1]{fontenc}
\usepackage[latin1]{inputenc}
\usepackage{textcomp}
\usepackage{amsmath}
\usepackage{graphicx}
\usepackage{amssymb}
\usepackage{float}
\usepackage{amsfonts}
\usepackage{hyperref}

\usepackage{babel}

\providecommand{\tabularnewline}{\\}



\begin{document}

\title{Master crossover functions for the one-component fluid {}``subclass''}

\author{Yves Garrabos, Carole Lecoutre, Fabien Palencia}

\affiliation{Equipe du Supercritique pour l'Environnement, les Matériaux et l'Espace
- Institut de Chimie de la Matière Condensée de Bordeaux - UPR 9048,
Centre National de la Recherche Scientifique - Université Bordeaux
I, 87 avenue du Docteur Schweitzer, F 33608 PESSAC Cedex, France}

\author{Bernard Le Neindre }

\affiliation{Laboratoire des Interactions Moléculaires et des Hautes Pressions
- UPR 1311, Centre National de la Recherche Scientifique - Université
Paris XIII - avenue Jean Baptiste Clément, F 93430 Villetaneuse, France.}

\author{Can Erkey}

\affiliation{Department of Chemical Engineering, University of Connecticut, Storrs,
Connecticut USA.}

\date{15 August 2006}

\begin{abstract}
Introducing three well-defined dimensionless numbers, we establish
the link between the scale dilatation method able to estimate master
(i.e. unique) singular behaviors of the one-component fluid {}``subclass''
and the universal crossover functions recently estimated {[}Garrabos
and Bervillier, Phys. Rev. E \textbf{74}, 021113 (2006)] from the
bounded results of the massive renormalization scheme applied to the
$\Phi_{d}^{4}\left(n\right)$-model of scalar order parameter ($n=1$)
and three dimensions ($d=3$), representative of the Ising-like universality
class. The master (i.e. rescaled) crossover functions are then able
to fit the singular behaviors of any one-component fluid without adjustable
parameter, only using one critical energy scale factor, one critical
length scale factor, and two dimensionless asymptotical scale factors,
which characterize the fluid critical interaction cell at its liquid-gas
critical point. An additional adjustable parameter accounts for quantum
effects in light fluids at the critical temperature. The effective
extension of the thermal field range along the critical isochore where
the master crossover functions seems to be valid corresponds to a
correlation length greater than three times the effective range of
the microscopic short-range molecular interaction.
\end{abstract}


\maketitle

\section{Introduction}

The universal features of three-dimensional (3D) Ising-like systems
are now well-established by the renormalization group approach \citet{ZinnJustin2002}
of the classical-to-critical crossover behavior \citet{Pelissetto2002}.
In this theoretical context, it was possible to estimate the complete
functions which interpolate between the critical behavior (controlled
by the non-trivial (Wilson-Fisher) fixed point \citet{Wilson1972,Wilson1974})
and a classical behavior (controlled by the Gaussian fixed point).
Such interpolating theoretical expressions were customarily named
classical-to-critical crossover functions. The corresponding crossover
\emph{within the critical domain} was referred as the critical crossover
\citet{Pelissetto1998,Pelissetto2002}, or the asymptotic crossover
\citet{Anisimov2000}.

Our present interest is restricted to the dimensionless expressions
derived from a massive renormalization (MR) scheme applied to the
$\Phi_{d}^{4}\left(n\right)$ model, for three-dimensional systems
($d=3$) and scalar order parameter ($n=1$) \citet{Bagnuls1984a,Bagnuls1985,Bagnuls1987,Bagnuls2002}
($d$ and $n$ are the dimensions of the space and order parameter
density, respectively, which characterize each universality class
\citet{ZinnJustin2002}). In that specific renormalization group approach
\citet{MSRscheme}, the Ising-like universality is linked to the existence
of a unique non-trivial fixed point. For convenient simplification
in the following presentation, the complete Ising-like universality
class is labeled $\left\{ \Phi_{3}\left(1\right)\right\} $-class
(with reference to the $\Phi_{d=3}^{4}(n=1)$-model), while the one-component
fluid {}``subclass'' made of all one-component fluids is labeled
$\left\{ 1f\right\} $-subclass, with the obvious relation $\left\{ 1f\right\} $-subclass
$\subset\left\{ \Phi_{3}\left(1\right)\right\} $-class.

The introduction of the system-dependent parameters for the practical
use of the theoretical functions was discussed in a detailed manner
in Refs. \citet{Bagnuls1984b,Bagnuls1987,Bagnuls2002,Garrabos2006gb}.
More generally, the dimensionless forms of the theoretical expressions
must be used to fit the experimental results in order to preserve
both the number and the critical scaling nature of the fluid-dependent
factors which are free in the massive renormalization scheme. Indeed,
it was precisely shown in Ref. \citet{Garrabos2006gb}, hereafter
labeled I, that the Ising-like universal features \citet{Guida1998},
estimated in the Ising-like preasymptotic domain close to the non-trivial
fixed point, require to characterize each system along its critical
isochore (the thermodynamic line equivalent to $h=0$) using four
parameters. 

Two of them are dimensional parameters, namely the critical temperature
$T_{c}$ which acts as energy unit (introducing the universal Boltzmann
constant $k_{B}$) to express the Hamiltonian in dimensionless form,
and the unknown inverse coupling constant $\left(g_{0}\right)^{-1}$
of the fourth-order term of the dimensionless Hamiltonian. As a matter
of fact, any Hamiltonian representative of a physical system at near
criticality, such as a one-component fluid near its liquid-vapor critical
point, is driven to the non-trivial fixed point under the action of
the renormalization transformations \citet{Wilson1974}. Due to the
fact that renormalizable field theories are short-distance insensitive,
universality emerges in a regime $\xi\Lambda_{0}\gg1$ in which the
correlation length $\xi$ is much larger than the microscopic scale,
which plays the role of the inverse wavenumber cut-off $\left(\Lambda_{0}\right)^{-1}$
in the renormalization scheme. This universality is non-mean field
like in nature (at least for the three-dimensional systems which are
of present interest), because the actual molecular interaction range
at the microscopic scale of the physical system cannot be completely
eliminated. $\left(\Lambda_{0}\right)^{-1}$ remains the single natural
length unit in the theoretical scheme. Indeed, $\left(g_{0}\right)^{-1}$
takes the convenient length dimension at $d=3$ to act as adjustable
length unit, and to express the correlation length in dimensionless
form.

The other two parameters are dimensionless coefficients, namely the
scale factors $\vartheta$ and $\psi$, which provide the analytical
(linear) proportionality between the two dimensionless physical fields
$\Delta\tau^{*}$ and $\Delta h^{*}$ of the Ising-like fluid and
the two renormalized relevant fields $t$ and $h$ of the $\Phi_{d=3}^{4}(n=1)$-model,
respectively \citet{Wilson1972,Wilson1974} (using customarily field
notations, see below). In such a situation, the universal features
close to the non-trivial fixed point are estimated in conformity with
the so-called two-scale-factor universality, where only two asymptotic
critical exponents and one confluent exponent are independent. Then,
the lowest value $\Delta\simeq0.51$ \citet{Guida1998} of the confluent
exponent characterizes the corrections to scaling due to one possible
irrelevant field \citet{Wegner1972}). In order to maintain the coherence
with the previous presentation of these universal features given in
Refs. \citet{Bagnuls2002,Garrabos2006gb}, we have here also selected
$\nu\simeq0.630$ and $\gamma\simeq1.240$ \citet{Guida1998} as independent
leading exponents attached to the correlation length $\ell_{\text{th}}\left(t\right)$)
and the susceptibily $\chi_{\text{th}}\left(t\right)$ along the critical
isochore ($h=0$), respectively.

The finite scale $\left(\Lambda_{0}\right)^{-1}$ is generally unknown
for a real microscopic interaction at short range distance in pure
fluids. Simultaneously the macroscopic size $L$ of the fluid sample
should be larger than $\xi$, i.e. $L\gg\xi\rightarrow\infty$, then
a special attention to account for extensive nature of the thermodynamic
properties of the physical system is needed. Moreover, thanks to the
general point of view of the thermodynamics for 3-D systems, the dimensionless
forms of any physical density variable $\frac{X}{V}$ (where $X$
is the total extensive variable and $V\propto L^{d}$ is the total
volume of the system) can be obtained without reference to the unknown
wavelength number $g_{0}$ defined at the critical point. Indeed,
introducing the total amount of matter $N$ {[}or the total mass $M=Nm_{\bar{p}}$,
where $m_{\bar{p}}$ is the mass of the particle, while the subcript
$\bar{p}$ refers to a particle property] of a system filling a total
volume $V$, the dimensionless order parameter conjugated to the dimensionless
ordering field can always be defined if the amount of matter in a
\emph{reference} volume is known (here the reference volume can be
chosen for example as the volume of a mole, a particle, a cell lattice,
a mass unit of matter, etc.). Therefore, any reference length $a_{0}$,
defined such as $n_{0}$ is the amount of matter in the volume $\left(a_{0}\right)^{d}$,
can be used as explicit length unit for the thermodynamic and correlation
functions. Thus the massive renormalization scheme generates a third
adjustable dimensionless scale factor - namely $u_{0}^{*}=g_{0}a_{0}$
- which relates the dimensionless correlation length $\frac{\xi}{a_{0}}$
of the physical system to the corresponding theoretical function derived
from the massive renormalization scheme. As a correlative result,
when $a_{0}$ takes its physical sense to represent the effective
range of the microscopic molecular interaction between the $n_{0}$
particles, i.e. $a_{0}\propto\left(\Lambda_{0}\right)^{-1}$ while
$n_{0}\propto$ coordination number, the Ising-like singular nature
of the physical system can be characterized by a set of three dimensionless
scale factors $\left\{ u_{0}^{*},\vartheta,\psi\right\} $. However,
in such a three-parameter characterization of the physical system,
it is then essential to recall that the theoretical estimations of
the universal features are only valid within the Ising-like preasymptotic
domain. In this preasymtotic domain, each dimensionless theoretical
function can be approximated by its restricted asymptotic form as
a two-term Wegner-like expansion, leading to three independent critical
exponents (i. e. our selected set $\left\{ \nu,\gamma,\Delta\right\} $
in present work).

Indeed, in the seventies, it was clearly shown by experimentalists
that the singular properties of pure fluids close to their liquid-gas
critical point were satisfied by power laws with universal features
comparable to the ones estimated for the uniaxial three-dimensional
Ising system used as a predictive model (for a review see for example
\citet{Levelt1978}). It was then revealed that two independent leading
amplitudes, attached to the universal values of two independent critical
exponents, are the only two fluid-dependent parameters necessary for
characterizing the asymptotic singular behavior of each one-component
fluid. Therefore, selecting the dimensionless correlation length $\xi^{*}\left(\Delta\tau^{*}\right)$
and the dimensionless isothermal compressibility $\kappa_{T}^{*}\left(\Delta\tau^{*}\right)$
in the homogeneous domain ($\Delta\tau^{*}>0$) along the critical
isochore ($\Delta\rho^{*}=0$), each Ising-like critical fluid can
then be characterized by the related leading amplitudes $\xi^{+}$
and $\Gamma^{+}$ (using standard notations for critical fluids \citet{Privman1991}).
This asymptotic situation characterized by two dimensionless leading
amplitudes $\xi^{+}$ and $\Gamma^{+}$ was in conformity with the
two-scale-factor universality expected for all systems, with short-ranged
interaction, and which have an isolated transition point.

Correlatively, it was demonstrated \citet{Levelt1981} that the two-scale-factor
universality related to the Ising-like nature of the critical phenomena
in pure fluids are {}``observed'' in a very limited range of temperature
and densities around their liquid-gas critical point. Obviously, since
the asymptotical critical domain associated to this limit is so narrow
that experiments are difficult to achieve, it was fundamental to account
for the possible nonuniversal character of the system through the
confluent singularities in the corrections to scaling \citet{Wegner1972}
(ignoring here the background contributions which are significative
only in the case of specific heat \citet{Bagnuls1984b}). That leads
to express the singular properties as truncated forms of the Wegner-like
series. That precisely corresponds to the Ising-like limit of the
asymptotic crossover mentionned just above and investigated in details
in the renormalization theory where the resummation of the Wegner-like
expansions should yield complete crossover functions from asymptotic
(Ising-like) critical behavior to the classical (mean-field) critical
behavior. Different theoretical approaches have been adopted by many
investigators to obtain explicit solutions resumming the complete
Wegner series (see for exemple a review in Ref. \citet{Anisimov2000}
for their application to the fluid case). The practical results essentially
depend on the approximations used in the renormalization scheme and
the way to account for the cutoff effects. Despite these technical
differences to treat the asymptotic crossover, the Ising-like universal
feature was related to the lowest confluent exponent $\Delta$ where
only one (fluid-dependent) confluent amplitude is needed to characterize
the first order term of the confluent correction to scaling \citet{Wegner1972}.
In the characterization of each critical fluid where the leading amplitudes
$\xi^{+}$ and $\Gamma^{+}$ are selected in conformity with the asymptotic
two-scale-factor universality, then it was necessary to add the confluent
amplitude $a_{\chi}^{+}$ of the first order correction term for the
susceptibility (related to the confluent amplitude $a_{\xi}^{+}$
of the correlation length by the universal value of the ratio $\frac{a_{\xi}^{+}}{a_{\chi}^{+}}\approx0.68$
\citet{Guida1998}). The resulting amplitude set $\left\{ \xi^{+},\Gamma^{+},a_{\chi}^{+}\right\} $
defines the complete asymptotic crossover of each one-component fluid.
This amplitude set is Ising-like equivalent (in quantity and nature)
to the previous scale factor set $\left\{ u_{0}^{*},\vartheta,\psi\right\} $
used to characterize the Ising-like critical behavior whitin the preasymptotic
domain of each one-component fluid.

As recalled in I, this three-parameter description of the asymptotic
singular behavior of the correlation length and the susceptibility
of xenon was studied \citet{Bagnuls1984b} using the crossover functions
initially derived by Bagnuls and Bervillier \citet{Bagnuls1984a,Bagnuls1985}
from the massive renormalization scheme. In this pioneering study
of the crossover, for the first time the minimal quantity of Ising-like
non-universal parameters of xenon was introduced as a set of a single
wavelength (defined at the critical point), and two dimensionless
scale factors expressing the analytical approximation between the
two relevant scaling (thermal-like $t$ and magnetic-like $h$) fields
and the corresponding physical ($\Delta\tau^{*}$ and $\Delta h^{*}$)
fields \citet{Wilson1972,Wilson1974}. Subsequently, theoretical and
numerical approaches applied to the asymptotic crossover description
of the singular behavior observed in pure fluids, have confirmed this
characterization with three parameters (see for example Refs. \citet{Luijten1999,Luijten2000,Muser2002,Hahn2001,Zhong2003,Zhong2004}
and a review in Ref. \citet{Anisimov2000}).

Today, with the appropriate introduction of two fluid-dependent factors
in conformity with the two-scale-factor universality, any theoretical
function which fits the temperature dependence of the effective critical
exponent along the critical isochore may be made universal by simply
rescaling the temperature distance to the critical temperature, as
initially proposed by Kouvel and Fisher \citet{Kouvel1964} who introduced
a single crossover temperature scale $\Delta\tau_{X}^{*}$. Unfortunately,
the unsolved problems in these theoretical approaches remain the validity
of the linear approximations of two relevant fields (which correctly
introduce the two system-dependent scale factors), the importance
of the neglected analytical and non-analytical corrections, and, more
generally, the estimation of the extension range in temperature and
densities around the liquid gas critical point where the Ising-like
universal features should be observed.

To complete the above introduction of the non-universal character
of the asymptotic crossover in pure fluids, we also recall that, at
the beginning of the eighties, the description of the behavior of
the singular thermodynamic properties at \emph{finite} distance from
the liquid gas critical point was also made using the theoretical
formulation of the nonasymptotic crossover from a regime of Ising-like
scaled behavior to another regime in which the critical anomalies
due to large fluctuations are ignored \citet{Chen1990b,Chen1990a,Anisimov1992,Anisimov1995,Anisimov2000,Agayan2001}.
The common attempt to address this problem was based on the classical-to-critical
crossover description of the free energy density. Indeed, this approach
is useful for better understanding of crossover critical phenomena
in \emph{{}``complex}'' fluids where the character of the crossover
reflects an interplay between Ising-like universality caused by long-range
fluctuations and a specific supramolecular structure characterized
by an additional nanoscopic or mesoscopic length scale (which can
then differ significantly from $\left(\Lambda_{0}\right)^{-1}$).
Therefore, while the Ising-like two-scale-factor universality was
similarly accounted for introducing the two dimensionless parameters
of proportionality between the respective relevant (physical and renormalized)
fields (for example $c_{t}\left(\sim\vartheta\right)$ and $c_{\rho}\left(\sim\psi\right)$
in the notations of Refs. \citet{Chen1990a,Chen1990b}), the fundamental
difference with the asymptotic crossover description comes from the
introduction of two independent dimensionless parameters (for example
$\bar{u}$ and $\Lambda$ in the notations of Ref. \citet{Anisimov1995}),
in order to control this nonasymptotic crossover character in complex
fluids. However, in an application related to the pure fluid case,
it is not necessary to introduce an additional mesoscopic length scale
to account for the realistic microscopic situation in one-component
fluids \citet{Hirschfelder1954}. Such pure fluids can then be assimilated
to Lennard-Jones-like fluids when they are made of atoms or highly
centro-symetrical molecules, or to short-distance associating fluids
when they include more sophisticated short-range molecular interactions
between unsymetrical molecules, polar molecules, bonding-like molecules,
etc. Moreover, the representation of the experimental phase surface
of any pure fluid by a van der Waals-like equation of state is not
accurate \emph{either close or far away} from the critical point (since
the van der Waals equation of state is theoretically justified only
for infinite range of the molecular interaction). As a final result,
the fluid-dependent parameters needed to describe the classical behavior
of the free energy density have no quantitative signification. The
nonuniversal complexity of the pure fluid was then accounted for by
introducing a significant number of adjustable parameters whose coupling
with the two dimensionless crossover parameters $\bar{u}$ and $\Lambda$
can not be completely defined. Therefore, in spite of the correct
introduction of a crossover function in the definition of variables
and thermodynamic potentials, the only founded theoretical challenge
of the nonasymptotic crossover applied to the one-component fluids
remains to account for the correct Ising-like universal features with
a single crossover scale. The uniqueness of the crossover scale can
thus be defined introducing an arbitrary fixed value of the product
$\bar{u}\Lambda$ \citet{Agayan2001}. The set $\left\{ \bar{u}\left(or\,\Lambda\right),c_{t},c_{\rho}\right\} $
appears {}``Ising-like'' equivalent to the set $\left\{ u_{0}^{*},\vartheta,\psi\right\} $.
This nonasymptotic crossover (which thus must match the asymptotic
critical crossover close to the Wilson-Fisher fixed point \citet{Bagnuls1996})
has to be not completely solved in regards to the most recent theoretical
predictions of universal exponents \citet{Guida1998,Campostrini2002}
and universal amplitude ratios \citet{Guida1998,Bagnuls2002}. Moreover,
the introduction of the single crossover parameter, which is then
related to the mean-field concept of the Ginzburg number \citet{Anisimov1992},
add conceptual difficulties to understand the role of real microscopic
parameters controlling a true rescaled universal behavior in the whole
crossover region \citet{Luijten1996,Pelissetto1998,Pelissetto2002}.

Finally, since the van der Waals dissertation, the real difficulty
for scientists interested in liquid-gas critical phenomena in pure
fluids, comes from the nonclassical (i.e. renormalizable) theories
which are not able to predict the location of the critical point,
while the classical theories provide its uncorrect location. Such
a difficulty has generated a crucial experimental challenge where
the determination of the two characteristic leading amplitudes and
the characteristic crossover parameter of each pure fluid, and alternatively
but equivalently, the localization of its liquid-gas critical point
on the $p,V,T$ phase surface, remain mandatory.

Based on this recurrent situation, an alternative phenomenological
way to characterize the asymptotic singular behavior of the one-component
fluids was also formulated by Garrabos \citet{Garrabos1982} as follows:
{}``If you are able to locate a single liquid gas critical point
on the experimental $p,v_{\bar{p}},T$ phase surface of a fluid particle
of mass $m_{\bar{p}}$, then you are also able to describe the asymptotic
crossover around this isolated point''. $p$ is the pressure, $T$
is the temperature, $v_{\bar{p}}=\frac{V}{N}=\frac{m_{\bar{p}}}{\rho}$
is the volume of the particle, and $\rho$ is the (mass) density.
Accordingly, a minimal set $Q_{c,a_{\bar{p}}}^{\text{min}}$ made
of four critical coordinates \citet{Garrabos1982} {[}see below Eq.
(\ref{qcmin (1)})], provides unequivocal determination of four (two
dimensional and two dimensionless) scale factors {[}see below Eqs.
(\ref{qcmin scale factors (3)}) to (\ref{pT preferred direction (7)})].
Then a scale dilatation method of the physical fields can be used
to observe and quantify the master (i.e., unique) asymptotic crossover
behavior of the $\left\{ 1f\right\} $-subclass \citet{Garrabos1985,Garrabos1986}.
The two dimensional critical parameters, noted $\left(\beta_{c}\right)^{-1}$
and $\alpha_{c}$, take appropriate energy and length dimensions,
respectively to reduce the physical variables, the thermodynamic functions,
and the correlation functions. The two dimensionless critical numbers,
noted $Y_{c}$ and $Z_{c}$, are well-defined characteristic parameters
of the critical interaction cell of volume $\left(\alpha_{c}\right)^{d}$.
An additional adjustable parameter, noted $\Lambda_{qe}^{*}$, accounts
for quantum effects in light fluids at the critical temperature \citet{Garrabos2006qe}.
Conversely, when $Q_{c,a_{\bar{p}}}^{\text{min}}$ and $\Lambda_{qe}^{*}$
were known for the selected fluid, the asymptotic master behavior
characterized by three master (i.e. constant) amplitudes was used
to calculate the amplitude set $\left\{ \xi^{+},\Gamma^{+},a_{\chi}^{+}\right\} $
which characterizes the asymptotic singular behavior of this fluid.
In addition to this intrinsic predictive power, another important
characteristic attached to the scale dilatation method was the Ising-like
analogy in its formal introduction of the two dimensionless scale
factors $Y_{c}$ and $Z_{c}$ and the corresponding ones $\vartheta$
and $\psi$ introduced by linear approximations in the massive renormalization
scheme.

As a matter of fact, for each selected fluid belonging to the $\left\{ 1f\right\} $-subclass,
this analogy can be useful to provide explicit estimation of the unknown
scale factor set $\left\{ \left(g_{0}\right)^{-1},\vartheta,\psi\right\} $
{[}or $\left\{ u_{0}^{*},\vartheta,\psi\right\} $] of the theoretical
crossover functions (using then, the thermodynamic length scale unit
$\alpha_{c}$ of the selected one-component fluid as a reference length
$a_{0}$). Especially in the case of the unique form of the \emph{mean}
theoretical functions estimated in I (which incorporates the error-bar
propagation of the \emph{min} and \emph{max} crossover functions revisited
in \citet{Bagnuls2002}), we can formulate the unambiguous modifications
of the theoretical crossover functions for the $\left\{ \Phi_{3}\left(1\right)\right\} $-class
to exactly match the master two-term Wegner-like expansions valid
within the Ising-like preasymptotic domain of the $\left\{ 1f\right\} $
-subclass.

These formulations were used to study the correlation length in the
homogeneous domain of seven one-component fluids \citet{Garrabos2006corlength}
and the squared capillary length in the non-homogeneous domain of
twenty one-component fluids \citet{Garrabos2006sugden}. Similarly,
a recent application to the practical parachor correlations (i.e.,
equations expressing surface tension as a power law of the density
difference between coexisting gas and liquid phases), have shown that
the corresponding master form acts as a universal equation of state
for the interfacial properties \citet{Garrabos2007pa}. Now, our present
objective is to achieve the complete uniquevocal link between these
updated results of I and the scale dilatation method to predict the
master singular behavior of the $\left\{ 1f\right\} $-subclass. For
these studies, the analytical relations between the relevant scaling
fields of both descriptions must be defined.

The paper is organized as follows. In Section 2 the master description
of the universal features within the Ising-like preasymptotic domain
is recalled. First, starting from the four critical coordinates of
the critical point, we define four scale factors which are needed
to unambiguously determine three dimensionless amplitudes which characterize
the Ising-like preasymptotic domain of each one-component fluid. Second,
we show the master singular behavior of the isothermal compressibility,
applying the scale dilatation method to the related physical quantities.
That complete the master sigular behavior of the correlation length
in conformity with the two-scale-factor universality of the $\left\{ \Phi_{3}\left(1\right)\right\} $-universality
class. In Section 3, a brief presentation of the theoretical crossover
functions for the correlation length and the susceptibility in the
homogeneous phase is given to demonstrate the analytical matching
with the master singular behavior provided by the scale dilatation
method. Introducing three well-defined dimensionless numbers characterizing
the $\left\{ 1f\right\} $-subclass, the unequivocal link between
three theoretical amplitudes, which characterize the $\left\{ \Phi_{3}\left(1\right)\right\} $-universality
class, and three master amplitudes, which characterize the $\left\{ 1f\right\} $-subclass,
is given before concluding in Section 4. Two appendices deal with
first, the equivalence between different one-parameter crossover models,
and second, the determination of the crossover parameter beyond the
preasymptotic domain using the well-known linear model of the parametric
equation of state with effective exponents.

\section{Master singular description of the one-component fluid subclass}

\subsection{The minimal set of critical parameters}

For the $\left\{ 1f\right\} $-subclass, it was hypothesized \citet{Garrabos1982}
\citet{Garrabos1985} that all the information needed to characterize
non-quantum fluid critical phenomena is contained within the four
critical parameters needed to localize the single critical point and
its tangent plane on the experimental phase surface of normalized
equation of state $\Phi_{a_{\bar{p}}}^{p}\left(p,v_{\bar{p}},T\right)=0$
(the needed supplementary information to characterize quantum fluids
is given in Ref. \citet{Garrabos2006qe}; see also below Eqs. (\ref{Lambdaqestar (9)})
and (\ref{lambdac (10)})). This minimal set of four coordinates reads
as follows\begin{equation}
Q_{c,a_{\bar{p}}}^{\text{min}}=\left\{ T_{c},p_{c},v_{\overline{p},c},\gamma_{c}^{'}\right\} \label{qcmin (1)}\end{equation}
 where $v_{\bar{p},c}=\frac{V}{N_{c}}=\frac{m_{\bar{p}}}{\rho_{c}}$
is the critical volume per particle ($V$ is the total volume, $N_{c}$
is the total critical number of particles, and $\rho_{c}$ is the
critical density), and \begin{equation}
\gamma_{c}^{'}=\left(\frac{\partial p}{\partial T}\right)_{v_{\bar{p}}=v_{\bar{p},c};T=T_{c}}=\left(\frac{dp_{\text{sat}}}{dT}\right)_{T=T_{c}}\label{gamaprime slope (2)}\end{equation}
 is the common critical direction of the critical isochore and the
saturation pressure curve at the critical point, in the $p;T$ diagram.
$\gamma_{c}^{'}$ is related to the Riedel factor \citet{Riedel1954},
$\alpha_{R,c}=\left(\frac{d\log p_{\text{sat}}}{d\log T}\right)_{T=T_{c}}$,
through the relation $\alpha_{R,c}=\frac{T_{c}}{p_{c}}\gamma_{c}^{'}$.
The subscript $c$ refers to a critical quantity. From Eq. (\ref{qcmin (1)}),
we can construct a more convenient set,\begin{equation}
Q_{c}^{\text{min}}=\left\{ \left(\beta_{c}\right)^{-1},\alpha_{c},Y_{c},Z_{c}\right\} \label{qcmin scale factors (3)}\end{equation}
 making use of the following four scale factors

\begin{equation}
\left(\beta_{c}\right)^{-1}=k_{B}T_{c}\sim\left[\text{energy}\right]\text{,}\label{energy scale (4)}\end{equation}

\begin{equation}
\alpha_{c}=\left(\frac{k_{B}T_{c}}{p_{c}}\right)^{\frac{1}{d}}\sim\left[\text{length}\right],\label{length scale (5)}\end{equation}

\begin{equation}
Z_{c}=\frac{p_{c}v_{\overline{p},c}}{k_{B}T_{c}}\sim\left[\text{dimensionless}\right]\text{,}\label{critical compression factor (6)}\end{equation}
 \begin{equation}
Y_{c}=\left(\gamma_{c}^{'}\frac{T_{c}}{P_{c}}\right)-1\sim\left[\text{dimensionless}\right]\label{pT preferred direction (7)}\end{equation}
$\left(\beta_{c}\right)^{-1}$ and $\alpha_{c}$ are used to express
dimensionless quantities. $\alpha_{c}$ is a measure of the effective
range of the microscopic short-range molecular interaction (Lennard-Jones
like in nature) \citet{Hirschfelder1954}. $Z_{c}$ is the critical
compression factor, while $Y_{c}=\alpha_{R,c}-1$. In the above dimensionless
form of the thermodynamic functions normalized per particle, $\frac{1}{Z_{c}}$
is the number of particles in the volume\begin{equation}
v_{c,I}=\left(\alpha_{c}\right)^{d}\label{CIC volume (8)}\end{equation}
 which corresponds to the volume of the \emph{critical interaction
cell} \citet{Garrabos1982}.

This actual set $Q_{c}^{\text{min}}$ (made from measured critical
parameters), refers to the characteristic range of the microscopic
molecular interaction in {}``classical'' (i.e. non-quantum) fluids
{[}here the molecular interaction range is measured by $\alpha_{c}$
of Eq. (\ref{length scale (5)})]. To include quantum fluids in the
one-component fluid subclass \citet{Garrabos2006qe}, we need the
phenomenological introduction of a supplementary adjustable parameter,
noted $\Lambda_{qe}^{\ast}$, which accounts for the quantum effects
at this microscopic length scale of the effective molecular interaction.
The (dimensionless) parameter $\Lambda_{qe}^{\ast}$ \citet{Garrabos2006qe}
is given by

\begin{equation}
\Lambda_{qe}^{\ast}=1+\lambda_{c}\label{Lambdaqestar (9)}\end{equation}
 with

\begin{equation}
\lambda_{c}=\lambda_{q,f}\frac{\Lambda_{T,c}}{\alpha_{c}}\label{lambdac (10)}\end{equation}
 $\lambda_{q,f}$ (with $\lambda_{q,f}>0$), is thus a non universal
adjustable number which accounts for statistical contribution due
to the nature (boson, fermion, etc.) of the quantum particle. $\Lambda_{T,c}=\frac{h_{P}}{\left(2\pi m_{\bar{p}}k_{B}T_{c}\right)^{\frac{1}{2}}}$
is the de Broglie thermal wave-vector at $T=T_{c}$, $h_{P}$ is the
Planck constant (the subscript $P$ is here added to make a distinction
with the theoretical ordering field noted $h$).

\subsection{Thermodynamic characterization of the critical interaction cell}

We introduce the (mass) density variable $\rho=\frac{Nm_{\overline{p}}}{V}=\frac{m_{\overline{p}}}{v_{\bar{p}}}$
and we consider the usual compression factor \begin{equation}
Z=\frac{pV}{Nk_{B}T}=\frac{pm_{\overline{p}}}{\rho k_{B}T}\label{compression factor (11)}\end{equation}
generally expressed in thermodynamic textbooks \citet{ZrhoT} as a
function $Z\left(T^{*},\tilde{\rho}\right)$ of the two dimensionless
variables $T^{*}=\frac{T}{T_{c}}$ and $\tilde{\rho}=\frac{\rho}{\rho_{c}}$.
Here we note the distinction underlined using superscript asterisk
for a dimensionless quantity obtained only from$\left(\beta_{c}\right)^{-1}$
and $\alpha_{c}$ units, and decorated tilde for a dimensionless quantity
which can refer to a specific amount of matter, then introducing also
the critical density $\rho_{c}$. Practically, the two dimensionless
critical parameters \begin{equation}
y_{\bar{p},c}^{*}=\left[\left(\frac{\partial Z}{\partial T^{*}}\right)_{\tilde{\rho}=\tilde{\rho}_{c}}\right]_{\text{CP}}=Y_{c}Z_{c}\label{ZT preferred direction (12)}\end{equation}
 \begin{equation}
z_{\bar{p},c}^{*}=\left[\left(\frac{\partial Z}{\partial\tilde{\rho}}\right)_{T^{*}=T_{c}^{*}}\right]_{\text{CP}}=-Z_{c}\label{Zrho preferred direction (13)}\end{equation}
are the two preferred directions \citet{Griffiths1970} of the characteristic
surface related to the total Grand potential $J\left(T,V,\mu_{\bar{p}}\right)$,
expressed per particle. $\mu_{\bar{p}}$ is the chemical potential
per particle related to the specific (i.e., per mass unit) chemical
potential $\mu_{\rho}$ by $\mu_{\rho}=\frac{\mu_{\bar{p}}}{m_{\bar{p}}}$
(where the subscript $\rho$ refers to a specific property). Therefore,
it is essential to note that $y_{\bar{p},c}^{*}=Y_{c}Z_{c}$ and $z_{\bar{p},c}^{*}=-Z_{c}$
are the dimensionless forms of two characteristic molecular (i.e.,
per particle) quantities.

As a matter of fact, when we consider the thermodynamic description
of a one-component fluid at constant volume of matter, the total Grand
potential $J\left(T,V,\mu_{\bar{p}}\right)=-p\left(T,\mu_{\bar{p}}\right)V$
takes, alternatively but equivalently, the role of the total Gibbs
free energy $G\left(T,p,N\right)=\mu_{\bar{p}}\left(T,p\right)N$
usually considered in the thermodynamic description of a one-component
fluid of constant amount of matter. The external pressure $p\left(T,\mu_{\bar{p}}\right)=\frac{-J}{V}$
of the container maintained at constant volume, in contact with a
particle reservoir, is then the thermodynamic potential equivalent
to the molecular chemical potential $\mu_{\bar{p}}\left(T,p\right)=\frac{G}{N}$
of the fluid maintained at constant amount of matter, in contact with
a volume reservoir. Therefore, considering the normalization per particle
of the thermodynamic description of a one component fluid at constant
volume, the molecular (i.e., per particle) Grand potential reads,
$j_{\bar{p},v_{\bar{p}}=\text{cte.}}\left(T\right)=-p\left(T,\mu_{\bar{p}}\right)v_{\bar{p}}$.
Using the associated opposite Massieu form, $z_{\bar{p},v_{\bar{p}}=\text{cte.}}=-\left(\frac{j_{\bar{p},v_{\bar{p}}=\text{cte.}}}{T}\right)$,
and the {}``universal'' Boltzmann constant $k_{B}$ as unique unit,
we obtain the following dimensionless form \begin{equation}
z_{\bar{p},v_{\bar{p}}=\text{cte.}}^{*}=\frac{z_{\bar{p},v_{\bar{p}}=\text{cte.}}}{k_{B}}=\frac{p\left(T,\mu_{\bar{p}}\right)v_{\bar{p}}}{k_{B}T}\equiv Z\label{dimensionless J or Z (14)}\end{equation}
 which demonstrates that the compression factor $Z$ of a constant
amount of fluid matter maintained at constant volume (i.e. a one-component
fluid monitored by the temperature along an isochore) is indeed a
dimensionless molecular potential \citet{Zminimum}. For the critical
filling $N=N_{c}$ of this isochoric container, we obtain $-\left(\frac{j_{\bar{p},v_{\bar{p}}=\text{cte.}}}{T}\right)_{N=N_{c}}^{*}=Z_{\tilde{\rho}=1}\equiv\frac{p^{*}}{T^{*}}v_{\bar{p},c}^{*}=\frac{T_{c}}{P_{c}}\left[\frac{p\left(T\right)}{T}\right]_{\rho=\rho_{c}}Z_{c}$.
Here, $\left[\frac{p\left(T\right)}{T}\right]_{\rho=\rho_{c}}$ acts
as first characteristic (i.e., independent) equation of state for
a critical \emph{isochoric} fluid, where the two extensive variables
$V$ and $N_{c}$ are fixed {[}i.e., a critical fluid at $\tilde{\rho}=1$
in contact with a thermostat (i.e. an energy reservoir) of constant
energy $k_{B}T$]. Multiplying the particle property $y_{\bar{p},c}^{*}$
by the number of particle $\frac{1}{Z_{c}}$ in the critical interaction
cell, it appears that the critical quantity $Y_{c}=\left[\left(\frac{\partial\left(\frac{p^{*}}{T^{*}}\right)}{\partial T^{*}}\right)_{v_{\bar{p}}=v_{\bar{p},c}}\right]_{\text{CP}}$
is readily a characteristic parameter of the critical interaction
cell.

Now considering a critical \emph{isothermal} fluid where the two variables
$V$ and $T_{c}$ are fixed (i.e., a critical fluid at $T^{*}=1$,
filling a constant total volume thermostated at constant critical
energy $k_{B}T_{c}$, in contact with a particle-reservoir), we obtain
$-\left(\frac{j_{\bar{p},V=const}}{T}\right)_{T=T_{c}}^{*}=Z_{T^{*}=1}\equiv\frac{p^{*}}{1}v_{\bar{p}}^{*}=\frac{1}{k_{B}}\frac{\left[p\left(\mu_{\bar{p}}\right)\right]_{T=T_{c}}}{T_{c}}v_{\bar{p}}$.
Here, $\left[\frac{p\left(\mu_{\bar{p}}\right)}{T}\right]_{T=T_{c}}$
acts as second characteristic (i.e., independent) equation of state
for a critical isothermal one component fluid. In such a thermostated
container at fixed total volume, we underline the fact that the only
independent extensive variable to monitor the thermodynamic fluid
state is the number of particles $N$ which fixes the equilibrium
mean value of the molecular chemical potential $\mu_{\bar{p}}$. For
$N=N_{c}$, at $T^{*}=1$ (i.e. the critical point condition), the
critical chemical potential per particle takes the value $\mu_{\bar{p},c}$,
such that $\left(z_{\bar{p},v_{\bar{p}}=v_{\bar{p},c}}^{*}\right)_{T=T_{c}}=\frac{p_{c}\left(\mu_{\bar{p},c}\right)v_{\bar{p},c}}{k_{B}T_{c}}=Z_{c}$.
Within the critical interaction cell filled with $\frac{1}{Z_{c}}$
particles, the normalized Grand potential takes the master critical
value $\frac{1}{Z_{c}}\left(z_{\bar{p},v_{\bar{p}}=v_{\bar{p},c}}^{*}\right)_{T=T_{c}}=1$.

Therefore, as an essential microscopic meaning related to Eq. (\ref{CIC volume (8)}),
we note that the critical set $Q_{c}^{min}$ of Eq. (\ref{qcmin scale factors (3)}),
characterizes the master thermodynamic information contained in the
critical interaction cell volume of each one-component fluid at the
critical point.

Finally, we summarize the two main constraints for the thermodynamic
description of a one-component fluid near its gas-liquid critical
point:

i) The dimensionless reduction of the variables is mandatorily made
by using the two dimensional factors $\left(\beta_{c}\right)^{-1}$
and $\alpha_{c}$ of Eqs. (\ref{energy scale (4)}) and (\ref{length scale (5)}),
respectively (see also Ref. \citet{Privman1991});

ii) The thermodynamic properties expressed per particle are better
suited to understand the microscopic nature of the two dimensionless
numbers $Y_{c}$ and $Z_{c}$. That leads to express dimensionless
properties from reference to the ones estimated for the volume of
the critical interaction cell. Then the thermodynamic origin of the
dimensionless master (i.e., unique) constants is well-identified.

\subsection{The relevant physical fields crossing the liquid-gas critical point}

Such a constrained dimensionless thermodynamic description is appropriately
obtained from the Grand canonical statistical distribution, considering
a one-component fluid in contact with a {}``particle-energy'' reservoir
maintained at constant total volume \emph{$V$}. Selecting the thermodynamic
nature (fixing, either the energy level $k_{B}T$, or the particle
amount $N$) of the reservoir to reach the critical point (either
at constant critical density, or constant critical temperature), the
normalized thermodynamic potential is then related to the intensive
quantities $\left[\frac{p\left(T\right)}{T}\right]_{\rho=\rho_{c}}$
or $\left[\frac{p\left(\mu_{\bar{p}}\right)}{T}\right]_{T=T_{c}}$.
In addition to the temperature variable conjugated to the total entropy,
the other natural (intensive) variable is the chemical potential per
particle $\mu_{\bar{p}}$, conjugated to the natural fluctuating total
number of particles $N$ (leading to the fluctuating number density
$n=\frac{N}{V}$). Therefore, the two relevant physical fields, either
to express the finite distance to the critical point, or to cross
it, along the critical isochore and along the critical isotherm, are\begin{equation}
\Delta\tau^{*}=k_{B}\beta_{c}\left(T-T_{c}\right)\label{CIC thermal field (15)}\end{equation}
 {\small and\begin{equation}
\Delta h^{*}=\beta_{c}\left(\mu_{\bar{p}}-\mu_{\bar{p},c}\right)\label{CIC ordering field (16)}\end{equation}
}respectively. Using the thermodynamic description per particle, the
order parameter density is then proportional to the critical number
density difference $n-n_{c}$ ($n_{c}=\frac{N_{c}}{V}$ is the number
density), and the associated dimensionless order parameter density
is given by \citet{Garrabos1985,Garrabos2002}:\begin{equation}
\Delta m^{*}=\left(n-n_{c}\right)\left(\alpha_{c}\right)^{d}\label{CIC OP (17)}\end{equation}
 We retrieve the distinction (using superscript asterisk or decorated
tilde), either between $\Delta h^{*}$ {[}see Eq. (\ref{CIC ordering field (16)})],
and\begin{equation}
\Delta\tilde{\mu}=\left(\mu_{\rho}-\mu_{\rho,c}\right)\frac{\rho_{c}}{p_{c}}\label{practical ordering field (18)}\end{equation}
 or between $\Delta m^{*}$ {[}see Eq. (\ref{CIC OP (17)})], and\begin{equation}
\Delta\tilde{\rho}=\frac{\rho-\rho_{c}}{\rho_{c}}\label{practical OP (19)}\end{equation}
 where $\Delta\tilde{\mu}$ and $\Delta\tilde{\rho}$ were customarily
defined in a critical fluid description using specific properties
and practical dimensionless variables $\tilde{x}=\frac{x}{x_{c}}$
(see, for example, Refs. \citet{Levelt1978,Anisimov2000}). The corresponding
relations can be expressed as follows,\begin{eqnarray}
\Delta h^{*} & = & Z_{c}\Delta\tilde{\mu}\label{deltahstar vs deltamutilde (20)}\\
\Delta m^{*} & = & \frac{1}{Z_{c}}\Delta\tilde{\rho}\label{deltamstar vs deltarhotilde (21)}\end{eqnarray}
which show that the dimensionless isothermal susceptibilities $\chi_{T}^{*}=\left[\frac{\partial\left(\Delta m^{*}\right)}{\partial\left(\Delta h^{*}\right)}\right]_{T^{*}}$
and $\tilde{\chi}_{T}=\left[\frac{\partial\left(\Delta\tilde{\rho}\right)}{\partial\left(\Delta\tilde{\mu}\right)}\right]_{\tilde{T}}$
differ by a factor $\left(\frac{1}{Z_{c}}\right)^{2}$. Equations
(\ref{deltahstar vs deltamutilde (20)}) and (\ref{deltamstar vs deltarhotilde (21)})
illustrate the primary role of $Z_{c}$ in the dimensionless form
of thermodynamics, due to the fact that $\left(Z_{c}\right)^{-1}$,
i.e., \emph{the particle number within the critical interaction cell
volume}, accounts for extensivity of the critical fluid.

\begin{table*}
\begin{tabular}{|c||c|c|c|c||c||c|c|}
\hline 
{\footnotesize $\mathcal{P}_{\text{qf}}^{\ast}$}  & {\footnotesize $\left\{ \mathcal{Z}_{\chi}^{+};\mathcal{Z}_{\xi}^{+}\right\} $}  & {\footnotesize $\mathcal{Z}_{P}^{\pm}$}  & {\footnotesize $\left\{ \mathcal{Z}_{\chi}^{1+}\right\} $}  & {\footnotesize $\mathcal{Z}_{P}^{1\pm}$}  & {\footnotesize $P^{\pm}$}  & {\footnotesize $P^{0,\pm}$}  & {\footnotesize $P^{1,\pm}$}\tabularnewline
\hline 
{\footnotesize $\ell_{\text{qf}}^{\ast}$}  & {\footnotesize $\mathcal{Z}_{\xi}^{+}=0.570481$}  &  &  & {\footnotesize $\mathcal{Z}_{\xi}^{1,+}=0.37695$}  & $\xi$  & {\footnotesize $\xi_{0}^{\pm}=\alpha_{c}\xi^{\pm}=\alpha_{c}\Lambda_{qe}^{*}(Y_{c})^{-\nu}\mathcal{Z}_{\xi}^{\pm}$}  & {\footnotesize $a_{\xi}^{\pm}=\mathcal{Z}_{\xi}^{1,\pm}(Y_{c})^{\Delta}$}\tabularnewline
 &  & {\footnotesize $\mathcal{Z}_{\xi}^{-}=0.291062$}  &  & {\footnotesize $\mathcal{Z}_{\xi}^{1,-}=0.37695$}  &  &  & \tabularnewline
{\footnotesize $\mathcal{X}_{\text{qf}}^{\ast}$}  & {\footnotesize $\mathcal{Z}_{\chi}^{+}=0.119^{*}$}  &  & {\footnotesize $\mathcal{Z}_{\chi}^{1,+}=0.555^{*}$}  &  & $\kappa_{T}^{*}$  & {\footnotesize $\Gamma^{\pm}=\left(\Lambda_{qe}^{*}\right)^{d-2}(Z_{c})^{-1}(Y_{c})^{-\gamma}\mathcal{Z}_{\chi}^{\pm}$}  & {\footnotesize $a_{\chi}^{\pm}=\mathcal{Z}_{\chi}^{1,\pm}(Y_{c})^{\Delta}$}\tabularnewline
 &  & {\footnotesize $\mathcal{Z}_{\chi}^{-}=0.0248465$}  &  & {\footnotesize $\mathcal{Z}_{\chi}^{1,-}=2.58741$}  &  &  & \tabularnewline
{\footnotesize $\mathbb{\mathcal{C}}_{\text{qf}}^{\ast}$}  &  & {\footnotesize $\mathcal{Z}_{C}^{+}=0.105658$}  &  & {\footnotesize $\mathcal{Z}_{C}^{1,+}=0.522743$}  & $c_{V}^{*}$  & {\footnotesize $\frac{A^{\pm}}{\alpha}=\left(\Lambda_{qe}^{*}\right)^{-d}(Y_{c})^{2-\alpha}\mathcal{Z}_{C}^{\pm}$}  & {\footnotesize $a_{C}^{\pm}=\mathcal{Z}_{C}^{1,\pm}(Y_{c})^{\Delta}$}\tabularnewline
 &  & {\footnotesize $\mathcal{Z}_{C}^{-}=0.196829$}  &  & {\footnotesize $\mathcal{Z}_{C}^{1,-}=0.384936$}  &  &  & \tabularnewline
{\footnotesize $\mathcal{M}_{\text{qf}}^{*}$}  &  & {\footnotesize $\mathcal{Z}_{M}=0.468^{*}$}  &  & {\footnotesize $\mathcal{Z}_{M}^{1}=0.4995$}  & $\Delta\rho_{LV}^{*}$  & {\footnotesize $B=\left(\Lambda_{qe}^{*}\right)^{-1}(Z_{c})^{-\frac{1}{2}}(Y_{c})^{\beta}\mathcal{Z}_{M}$}  & {\footnotesize $a_{M}=\mathcal{Z}_{M}^{1}(Y_{c})^{\Delta}$}\tabularnewline
\hline
\end{tabular}

\caption{Master ($\mathcal{P}_{\text{qf}}^{\ast}$) and physical ($P^{\pm}$)
amplitude values of the singular behavior of the correlation length
(lines 2 and 3), the susceptibility (lines 4 and 5), the specific
heat (lines 6 and 7) and the order parameter density (line 8), along
the critical isochore of any one-component fluid; columns 2 and 4:
independent master amplitudes {[}see Eq. (\ref{Scal1fA (40)}); columns
3 and 5: amplitude values in conformity with the theoretical universal
features estimated within the Ising-like preasymptotic domain \citet{Guida1998,Bagnuls2002};
asterisk indicate the {}``experimental'' master values estimated
using xenon as a standard critical fluid (see Refs. \citet{Garrabos1985,Garrabos1986,Garrabos2006khiT});columns
7 and 8; corresponding physical amplitudes when $Q_{c}^{\text{min}}=\left\{ \left(\beta_{c}\right)^{-1},\alpha_{c},Y_{c},Z_{c}\right\} $
and $\Lambda_{qe}^{*}$ are known for the selected one-component fluid.\label{Table I}}

\end{table*}

\subsection{The scale dilatation method for the $\left\{ 1f\right\} $-subclass}

\begin{figure*}
\includegraphics[width=140mm,height=140mm,keepaspectratio]{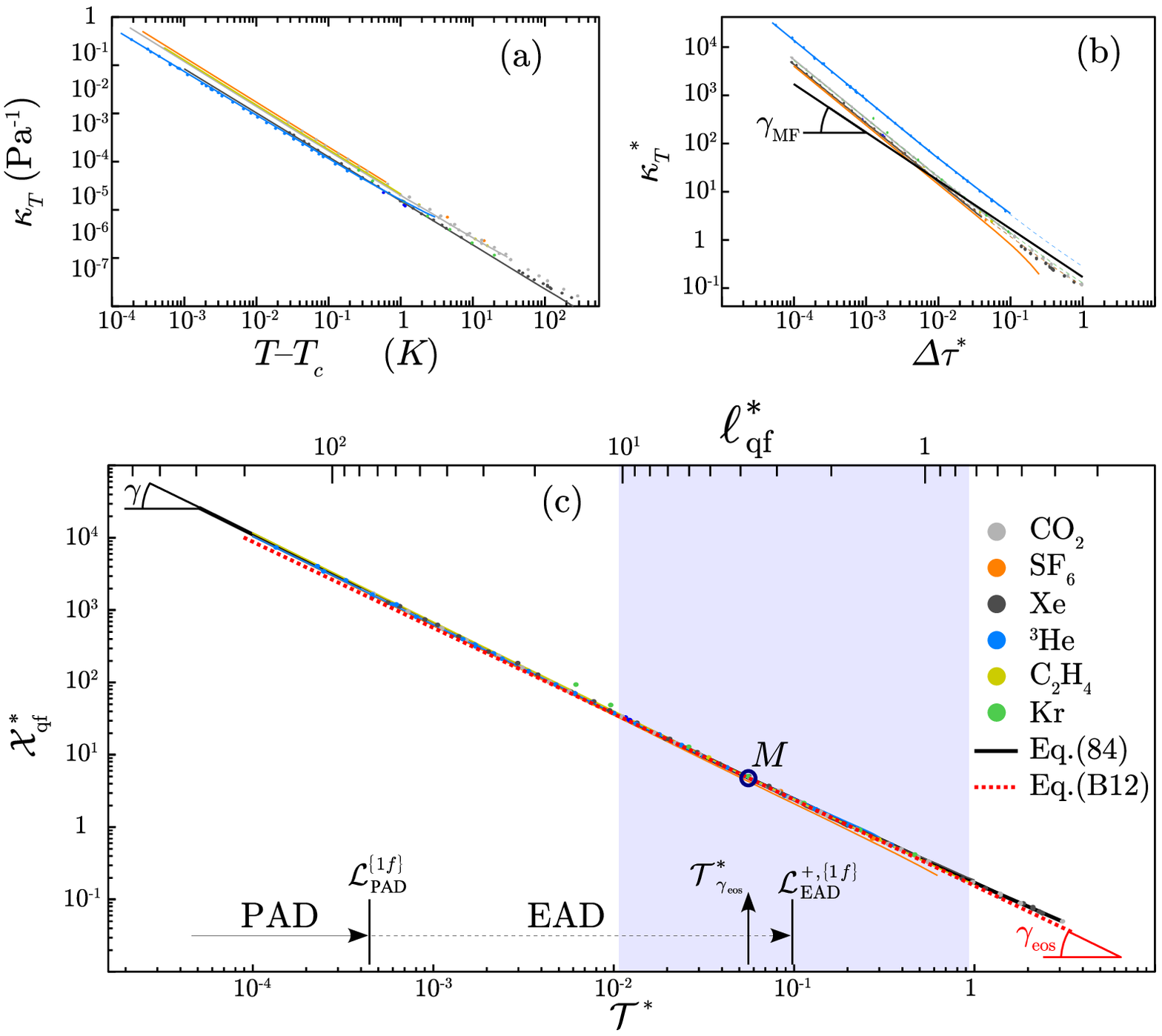}

\caption{(Color online) Singular behavior of the isothermal compressibility
of the one-component fluids. (a) $\kappa_{T}$ as a function of $T-T_{c}>0$
(log-log scale), along the critical isochore for Xe, Kr, $^{3}$He,
SF$_{6}$, CO$_{2}$, and C$_{2}$H$_{4}$ (see inserted Table for
fluid color indexation); (b) Log-Log plot of $\kappa_{T}^{*}\left(\Delta\tau^{*}\right)$;
black full curve: mean-field behavior of equation $\kappa_{T,\text{vdW}}^{*}=\frac{1}{6}\left(\Delta\tau^{*}\right)^{-\gamma_{\text{MF}}}$
with $\gamma_{\text{MF}}=1$. (c) Matched master behavior (log-log
scale) of the renormalized susceptibility $\mathcal{X}_{\text{qf}}^{*}=\left(\Lambda_{qe}^{*}\right)^{2-d}Z_{c}\kappa_{T}^{*}$
{[}see Eqs. (\ref{khiqe master (31)})], as a function of the renormalized
thermal field $\mathcal{T}^{*}$ {[}see Eqs. (\ref{temperature scale (22)})];
black full curve: Eq. (\ref{1f RG c3 fitting eq (84)}); red dashed
curve: tangent of Eq. (\ref{eos tangent line (B12)}) at the point
M (see text and Appendix B); (full) arrow (label PAD): master extension
of the Ising like preasymptotic domain of Eq. (\ref{LcalPAD1f (100)});
(dashed) arrow (label EAD): effective extension of the extended asymptotic
domain of Eq. (\ref{LcalEAD1f (111)}) corresponding to $\ell_{\text{qf}}^{\ast}=\left(\Lambda_{qe}^{*}\right)^{-1}\frac{\xi}{\alpha_{c}}\gtrsim3$
(see Ref. \citet{Garrabos2006corlength}); grey area: master correlation
length range $10.5\lesssim\ell_{\text{qf}}^{\ast}\lesssim0.73$ (thermal
field range $1.9\times10^{-2}\lesssim\mathcal{T}^{*}\lesssim1$) discussed
in Appendix B .\label{Figure 1} }

\end{figure*}

A detailed presentation of the scale dilatation method can be found
in references \citet{Garrabos1982,Garrabos1985,Garrabos1986,Garrabos2002,Garrabos2006qe}.
Hereafter we only recall the main features which close the master
description of the singular behaviors of the $\left\{ 1f\right\} $-subclass
within the preasymptotic domain (with $\gamma$, $\nu$, and $\Delta$
selected as independent critical exponents). The scale dilatation
method uses explicit analytical transformations of each physical field
$\Delta\tau^{*}$ and $\Delta h^{*}$ given by the equations\begin{equation}
\mathcal{T}_{\text{qf}}^{*}\equiv\mathcal{T}^{*}=Y_{c}\left|\Delta\tau^{*}\right|\label{temperature scale (22)}\end{equation}

\begin{equation}
\mathcal{H}_{\text{qf}}^{*}=\left(\Lambda_{qe}^{*}\right)^{2}\mathcal{H}^{*}=\left(\Lambda_{qe}^{*}\right)^{2}\left(Z_{c}\right)^{-\frac{d}{2}}\left|\Delta h^{*}\right|\label{external field scale (23)}\end{equation}
 where $\mathcal{T}_{\text{qf}}^{*}\equiv\mathcal{T}^{*}$ is the
renormalized thermal field, and $\mathcal{H}_{\text{qf}}^{*}$ is
the renormalized ordering field. The subscript $\text{qf}$ distinguishes
between a quantity which refers to a quantum fluid (i.e., $\Lambda_{qe}^{*}\neq1$)
from the one which refers to a non-quantum fluid (i.e., $\Lambda_{qe}^{*}=1$)
\citet{Garrabos2006qe}. Accordingly, the analytic transformation
between the physical order parameter density $\Delta m^{*}$ and the
renormalized order parameter density $\mathcal{M}_{\mbox{qf}}^{*}$,
reads as follows \citet{Garrabos1985,Garrabos2002,Garrabos2006qe}

\begin{equation}
\mathcal{M}_{\text{qf}}^{*}=\Lambda_{qe}^{*}\mathcal{M}^{*}=\Lambda_{qe}^{*}\left(Z_{c}\right)^{\frac{d}{2}}\left|\Delta m^{\ast}\right|\label{order parameter scale (24)}\end{equation}
 Introducing then the dimensionless correlation length $\xi^{\ast}=\frac{\xi}{\alpha_{c}}$,
the renormalized correlation length $\ell_{\text{qf}}^{\ast}$ is
given by the equation

\begin{equation}
\ell_{\text{qf}}^{\ast}=\left(\Lambda_{qe}^{\ast}\right)^{-1}\ell^{\ast}=\left(\Lambda_{qe}^{\ast}\right)^{-1}\xi^{\ast}\label{ksi dilatation (25)}\end{equation}
 which preserves the same length unit for thermodynamic and correlations
functions (with $\ell^{\ast}\equiv\xi^{\ast}$ for the non-quantum
fluid case).

The master asymptotic singular behavior of $\ell_{\text{qf}}^{\ast}\left(\mathcal{T}^{\ast}\right)$
was studied in \citet{Garrabos2006corlength}. Specifically, the observed
asymptotic divergence of $\ell_{\text{qf}}^{\ast}$ was represented
by the following (two-term) Wegner expansion

\begin{equation}
\ell_{\text{qf}}^{\ast}=\mathcal{Z}_{\xi}^{+}\left(\mathcal{T}^{\ast}\right)^{-\nu}\left[1+\mathcal{Z}_{\xi}^{1,+}\left(\mathcal{T}^{\ast}\right)^{\Delta}\right]\label{two-term master equation eleqf (26)}\end{equation}
 where $\nu=0.6303875$ and $\Delta=0.50189$ \citet{Guida1998}.
The leading amplitude $\mathcal{Z}_{\xi}^{+}=0.570481$ and the first
confluent amplitude $\mathcal{Z}_{\xi}^{1,+}=0.37695$ have master
(i.e. unique) values for the $\left\{ 1f\right\} $-subclass. The
associated asymptotic singular behavior of the physical correlation
length was given by \begin{equation}
\xi_{\text{exp}}\left(\Delta\tau^{*}\right)=\xi_{0}^{+}\left(\Delta\tau^{*}\right)^{-\nu}\left[1+a_{\xi}^{+}\left(\Delta\tau^{*}\right)^{\Delta}\right]\label{2term ksiexp (27)}\end{equation}
 Therefore, the term to term comparison of (master) Eq. (\ref{two-term master equation eleqf (26)})
and (physical) Eq. (\ref{2term ksiexp (27)}), results in the following
amplitude combinations\begin{equation}
\frac{\xi_{0}^{+}}{\alpha_{c}}=\xi^{+}=\Lambda_{qe}^{*}\left(Y_{c}\right)^{-\nu}\mathcal{Z}_{\mathcal{\xi}}^{+}\label{ksizeroplus vs Qcmin (28)}\end{equation}
 \begin{equation}
a_{\xi}^{+}=\mathcal{Z}_{\xi}^{1,+}\left(Y_{c}\right)^{\Delta}\label{aksi vs Qcmin (29)}\end{equation}

Applying now the scale dilatation method to any physical (thermodynamic)
property $P\left(\Delta\tau^{*}\right)$, the master singular behavior
for the renormalized (thermodynamic) property $\mathcal{P}_{\text{qf}}^{\ast}\left(\mathcal{T}^{*}\right)$
can be also observed and represented within the preasymptotic domain
by the restricted expansion

\begin{equation}
\mathcal{P}_{\text{qf}}^{\ast}=\mathcal{Z}_{P}^{\pm}\left(\mathcal{T}^{*}\right)^{-e_{P}}\left[1+\mathcal{Z}_{P}^{1,\pm}\left(\mathcal{T}^{*}\right)^{\Delta}\right]\label{master two term power law (30)}\end{equation}
 where $\mathcal{Z}_{P}^{\pm}$ and $\mathcal{Z}_{P}^{1\pm}$ are
two master constants for any one-component fluid (see Table \ref{Table I}).
To close the master description in conformity with the universal features
estimated within this Ising-like preasymptotic domain, we complete
the representation of the master correlation length with the one of
the master susceptibility $\mathcal{X}_{\text{qf}}^{*}$ obtained
from master order parameter density $\mathcal{M}_{\text{qf}}^{*}$,
and master ordering field $\mathcal{H}_{\text{qf}}^{*}$, using the
thermodynamic definition, $\mathcal{X}_{\text{qf}}^{*}=\left(\frac{\partial\mathcal{M}_{\text{qf}}^{*}}{\partial\mathcal{H}_{\text{qf}}^{*}}\right)_{\mathcal{T}^{*}}$.
$\mathcal{X}_{\text{qf}}^{*}$ is related to the dimensionless isothermal
susceptibility $\chi_{T}^{*}=\left(\frac{\partial\left(\Delta m^{*}\right)}{\partial\left(\Delta h^{*}\right)}\right)_{\Delta\tau^{*}}$
by the following equations,\begin{equation}
\begin{array}{rl}
\mathcal{X}_{\text{qf}}^{*} & =\left(\Lambda_{qe}^{*}\right)^{2-d}\varkappa_{\mathcal{T}}^{*}\\
 & =\left(\Lambda_{qe}^{*}\right)^{2-d}\left(Z_{c}\right)^{d}\chi_{T}^{*}\end{array}\label{khiqe master (31)}\end{equation}
 As previously mentioned for the critical isochore case, $\chi_{T}^{*}\left(n_{c}^{*}\right)=\frac{\tilde{\chi}_{T}\left(\tilde{\rho}=1\right)}{\left(Z_{c}\right)^{2}}$,
while $\tilde{\chi}_{T}\left(\tilde{\rho}=1\right)\equiv\kappa_{T}^{*}(\tilde{\rho}=1)$
{[}with $\tilde{\chi}_{T}=\left(\frac{\partial\left(\Delta\tilde{\rho}\right)}{\partial\left(\Delta\tilde{\mu}\right)}\right)_{\Delta\tau^{*}}=\left(\tilde{\rho}\right)^{2}\kappa_{T}^{*}$],
where $\kappa_{T}^{*}$ is the dimensionless isothermal compressibility
$\kappa_{T}^{*}=\frac{1}{\beta_{c}\left(\alpha_{c}\right)^{d}}\left[\frac{1}{\rho}\left(\frac{\partial\rho}{\partial p}\right)_{T}\right]=p_{c}\kappa_{T}$
(with $\kappa_{T}=\frac{1}{\rho}\left(\frac{\partial\rho}{\partial p}\right)_{T}$).
Therefore, the master susceptibility can be also related to the dimensionless
isothermal compressibility by,\begin{equation}
\mathcal{X}_{\text{qf}}^{*}=\left(\Lambda_{qe}^{*}\right)^{2-d}Z_{c}\kappa_{T}^{*}\label{khiTstarmaster vs kappaTstar (32)}\end{equation}
 The master asymptotic singular behavior of $\mathcal{X}_{\text{qf}}^{*}$
reads as follows\begin{equation}
\mathcal{X}_{\text{qf}}^{*}=\mathcal{Z}_{\chi}^{+}\left(\mathcal{T}^{*}\right)^{-\gamma}\left[1+\mathcal{Z}_{\chi}^{1,+}\left(\mathcal{T}^{*}\right)^{\Delta}\right]\label{khi master pplaw (33)}\end{equation}
 where $\gamma=1.2396935$ \citet{Guida1998}. The master values of
the leading and confluent amplitudes are $\mathcal{Z}_{\chi}^{+}=0.119$
and $\mathcal{Z}_{\chi}^{1,+}=0.555$, respectively, where the universal
value of the confluent amplitude ratio $\frac{\mathbb{\mathcal{Z}}_{\xi}^{1,+}}{\mathbb{\mathcal{Z}}_{\chi}^{1,+}}=0.67919$
is given in Ref. \citet{Bagnuls2002}. The associated asymptotic singular
behavior of the isothermal compressibility reads as follows

\begin{equation}
\kappa_{T,\text{exp}}^{*}\left(\Delta\tau^{*}\right)=\Gamma^{+}\left(\Delta\tau^{*}\right)^{-\gamma}\left[1+a_{\chi}^{+}\left(\Delta\tau^{*}\right)^{\Delta}\right]\label{2term kappaTexp (34)}\end{equation}
 The term to term comparison of (master) Eq. (\ref{khi master pplaw (33)})
and (physical) Eq. (\ref{2term kappaTexp (34)}), leads to the following
amplitude estimations\begin{equation}
\Gamma^{+}=\left(\Lambda_{qe}^{*}\right)^{d-2}\left(Z_{c}\right)^{-1}\left(Y_{c}\right)^{-\gamma}\mathcal{Z}_{\chi}^{+}\label{gamma+ vs Qmin (35)}\end{equation}
 \begin{equation}
a_{\chi}^{+}=\mathcal{Z}_{\chi}^{1+}\left(Y_{c}\right)^{\Delta}\label{akhi vs Qmin (36)}\end{equation}
 with\begin{equation}
\frac{\mathbb{\mathcal{Z}}_{\xi}^{1,+}}{\mathbb{\mathcal{Z}}_{\chi}^{1,+}}=\frac{a_{\xi}^{+}}{a_{\chi}^{+}}=0.67919\label{aksiplus vs akhiplus (37)}\end{equation}

The expected asymptotic collapse of the fluid properties on a single
curve due to the scale dilatation method is illustrated in Fig. \ref{Figure 1}
\emph{(log-log scale).} The raw data are reported in Fig. \ref{Figure 1}a
to easily distinguish between singular behavior of $\kappa_{T}$ (expressed
in $\text{Pa}^{-1}$) as a function of $T-T_{c}$ (expressed in $\text{K}$),
for each one-component fluid (see the fluid color indexation inserted
in Fig. \ref{Figure 1}c). Figure \ref{Figure 1}b illustrates the
differences between the corresponding dimensionless behaviors $\kappa_{T}^{*}\left(\Delta\tau^{*}\right)$
which confirm the failure of results provided by the two-parameter
corresponding state principle. This figure also shows the failure
of mean-field like behavior predicted from the van der Waals (vdW)
equation of state which is here represented by the black full curve
of equation $\kappa_{T,\text{vdW}}^{*}\left(\Delta\tau^{*}\right)^{\gamma_{\text{vdW}}}=\Gamma_{\text{vdW}}^{+}=\frac{1}{6}$,
with $\gamma_{\text{vdW}}=\gamma_{\text{MF}}=1$. On the other hand,
Fig. \ref{Figure 1}c demonstrates the collapse of  $\mathcal{X}_{\text{qf}}^{*}\left(\mathcal{T}^{*}\right)$
on a master curve where the scatter corresponds to the estimated $\kappa_{T}$-precision
(5-10\%) for each fluid. We underline the combination of the {}``scaling''
and {}``extensive'' roles of the characteristic factor $Z_{c}$
in the renormalization {[}see Eqs. (\ref{khiqe master (31)}) and
(\ref{khiTstarmaster vs kappaTstar (32)})] of the ordinate axis of
Fig. \ref{Figure 1}c (compare for example with Fig. 3 of Ref. \citet{Luijten2000}
or with Fig. 2 of Ref. \citet{Hahn2001}). The complementary materials
for complete analysis of this Fig. \ref{Figure 1}c will be given
below and in Appendix B.

Therefore, adding the correlation length results given in Fig. 1c
of Ref. \citet{Garrabos2006corlength} to the present isothermal susceptibility
results, we close the asymptotic master behavior generated by the
scale dilatation method, in conformity with the two-scale-factor universality
of the Ising-like systems.

To summarize the main interest of this Ising-like master description
of the $\left\{ 1f\right\} $-subclass (associated to the selected
set $\left\{ \Delta;\nu;\gamma\right\} $ of three independent universal
exponents \citet{Guida1998}), we introduce

i) the physical amplitude set \begin{equation}
S_{A}=\left\{ a_{\chi}^{+};\xi^{+};\Gamma^{+}\right\} \label{SAfluid (38)}\end{equation}
 which characterizes the physical Ising-like universal features of
each selected pure fluid having the critical set $\left\{ Q_{c}^{\text{min}};\Lambda_{qe}^{*}\right\} $;

ii) the corresponding scale factor set\begin{equation}
S_{SF}=\left\{ Y_{c};Z_{c};\Lambda_{qe}^{*}\right\} \label{SFfluid (39)}\end{equation}
which characterizes the dimensional universal features of the critical
interaction cell of each selected pure fluid having $\left(\beta_{c}\right)^{-1}$
and $\alpha_{c}$ as energy and length units, respectively, and

iii) the master amplitude set, \begin{equation}
\mathcal{S}_{A}^{\left\{ 1f\right\} }=\left\{ \begin{array}{cl}
\mathcal{Z}_{\chi}^{1,+}= & 0.555\\
\mathcal{Z}_{\mathcal{\xi}}^{+}= & 0.570481\\
\mathcal{Z}_{\chi}^{+}= & 0.119\end{array}\right\} \label{Scal1fA (40)}\end{equation}
which characterizes the master Ising-like universal features of the
$\left\{ 1f\right\} $-subclass. Three independent relations, i.e.,
{[}Eqs. (\ref{ksizeroplus vs Qcmin (28)}), (\ref{gamma+ vs Qmin (35)}),
and (\ref{akhi vs Qmin (36)})], connecting these three previous sets,
can be written in the following condensed functional form\begin{equation}
\mathcal{S}_{A}^{\left\{ 1f\right\} }=\left\{ S_{A}\,\mathcal{F}\left(S_{SF}\right)\right\} _{\left(\beta_{c}\right)^{-1},\alpha_{c},\Lambda_{qe}^{*}}\label{Scal1fA vs SAfluid (41)}\end{equation}
where the function $\mathcal{F}\left(S_{SF}\right)$ takes an universal
scaling form of the two (fluid-dependent) scale factors $Y_{c}$ and
$Z_{c}$. Accordingly, any physical amplitude of any one-component
fluid can be estimated from the equations given in Table \ref{Table I}
satisfying to the two-scale-factor universality of the $\left\{ \Phi_{3}\left(1\right)\right\} $-class
(where xenon acts as a standard critical fluid to estimate three characteristic
master amplitudes labeled with an asterisk, see Refs. \citet{Garrabos1985,Garrabos1986,Garrabos2006khiT}).
However, the effective extension range where the master behavior is
observed, as an explicit criteria which defines the preasymptotic
range where the two-term Wegner-like expansion is valid, remain not
easy to estimate precisely only using the scale dilatation method.
These two problems can be solved using a master modification of the
mean crossover functions \citet{Garrabos2006gb} obtained from the
massive renormalization scheme, as shown in the next section.

\section{Master modifications of the mean crossover functions}

\subsection{System-dependent parameters of the mean crossover functions.}

For the $\left\{ \Phi_{3}\left(1\right)\right\} $-class, the mean
crossover functions $F_{P}\left(t,h=0\right)$ describing the crossover
behavior of the theoretical properties $P_{\text{th}}\left(t\right)$
as a function of the renormalized temperature-like field $t$, for
zero value of the external ordering (magnetic-like) field $h$, are
given in detail in I. All the theoretical functions $F_{P}\left(t\right)$
have the same functional form whatever $P_{\text{th}}$, and, as noted
in I, a closed presentation of their universal features, only needs
to use for example the mean crossover functions $F_{\ell}\left(t\right)=\frac{1}{\ell_{\text{th}}\left(t\right)}$
for the inverse correlation length, and $F_{\chi}\left(t\right)=\frac{1}{\chi_{\text{th}}\left(t\right)}$
for the inverse susceptibility, at $h=0$, in the homogeneous phase
$T>T_{c}$ {[}$T$ ($T_{c}$) is the temperature (critical temperature)].
These two theoretical functions read as follows:\begin{equation}
\left[\ell_{\text{th}}\left(t\right)\right]^{-1}=\mathbb{Z}_{\xi}^{+}t^{\nu}{\displaystyle \prod_{i=1}^{3}\left(1+X_{\xi,i}^{+}t^{D\left(t\right)}\right)^{Y_{\xi,i}^{+}}}\label{MR ele vs t (42)}\end{equation}
 \begin{equation}
\left[\mathcal{X}_{\text{th}}\left(t\right)\right]^{-1}=\mathbb{Z}_{\chi}^{+}t^{\gamma}{\displaystyle \prod_{i=1}^{3}\left(1+X_{\chi,i}^{+}t^{D\left(t\right)}\right)^{Y_{\chi,i}^{+}}}\label{MR khi vs t (43)}\end{equation}
 $D\left(t\right)$ is a universal mean crossover function for the
confluent exponents $\Delta$ and $\Delta_{\text{MF}}$ which reads
\begin{equation}
D\left(t^{*}\right)=\frac{\Delta_{\text{MF}}S_{2}\sqrt{t}+\Delta}{S_{2}\sqrt{t}+1}\label{Deff exponent MR universal (44)}\end{equation}
 such that $D\left(t=\frac{1}{\left(S_{2}\right)^{2}}\right)=\frac{\Delta_{\text{MF}}+\Delta}{2}$.
All the universal exponents $\nu$, $\gamma$, $\Delta$, $\Delta_{\text{MF}}$,
and the parameters $\mathbb{Z}_{\xi}^{+}$, $X_{\xi,i}^{+}$, $Y_{\xi,i}^{+}$,
$\mathbb{Z}_{\chi}^{+}$, $X_{\chi,i}^{+}$, $Y_{\chi,i}^{+}$, and
$S_{2}$, are given in I.

The temperature-like field $t$ is analytically related to the physical
dimensionless temperature distance\begin{equation}
\Delta\tau^{*}=\frac{T-T_{c}}{T_{c}}\label{dimensionless temperature distance (45)}\end{equation}
 by the following linear approximation\begin{equation}
t=\vartheta\Delta\tau^{*}\label{theta from t vs deltataustar (46)}\end{equation}
 which introduces $\vartheta$ as an adjustable (system-dependent)
parameter. Here $\vartheta$ is a scale factor for the temperature
field. Correlatively, it is important to note that the definition
of $\Delta\tau^{*}$ {[}see Eq. (\ref{dimensionless temperature distance (45)})],
introduces the critical temperature $T_{c}$ as a system-dependent
parameter. Then the relation between the dimensionless thermodynamic
free energies of the $\Phi^{4}$-model and the physical (one-component
fluid) system, only involves the energy unit $\left(\beta_{c}\right)^{-1}=k_{B}T_{c}$.

Similarly, the ordering-like field $h$ is analytically related to
the corresponding physical dimensionless variables $\Delta\tilde{\mu}$
(or $\Delta h^{*}$) by the following linear approximations (including
quantum effects)\begin{equation}
\begin{array}{rl}
h & =\psi_{\rho}\left[\left(\Lambda_{qe}^{*}\right)^{2}\Delta\tilde{\mu}\right]\\
\text{or}\: h & =\psi\left[\left(\Lambda_{qe}^{*}\right)^{2}\Delta h^{*}\right]\end{array}\label{psi from h vs deltahstar (47)}\end{equation}
 which introduce $\psi_{\rho}$ (or $\psi$) as an adjustable (system-dependent)
parameter. $\psi_{\rho}$, respectively $\psi=\left(Z_{c}\right)^{-1}\psi_{\rho}$,
is a scale factor for the ordering field $\Delta\tilde{\mu}$, respectively
$\Delta h^{*}=Z_{c}\Delta\tilde{\mu}$.

Accordingly, the dimensional analysis of each term of the dimensionless
hamiltonian of the $\Phi^{4}$-model leads to the introduction of
a finite arbitrary wave-vector $\Lambda_{0}$, so-called the cutoff
parameter, which is related to the finite short range of the microscopic
interaction (see for example, Ref. \citet{Bagnuls1984a}). Since the
value of the cutoff parameter of a selected physical system is generally
unknown, a convenient method at $d=3$ consists in replacing $\Lambda_{0}$
by $g_{0}$ \citet{Bagnuls1984b,Garrabos2006gb}, which is the critical
coupling constant having the correct wavenumber dimension (see our
introductive part). This system-dependent wavenumber $g_{0}$ provides
the practical {}``adjustable'' link between the theoretical dimensionless
correlation length ($\ell_{\text{th}}$) and the measured physical
correlation length ($\xi_{\text{exp}}$) of the system at $d=3$,
through the fitting equation :\begin{equation}
\left(\Lambda_{qe}^{*}\right)^{-1}\xi_{\text{exp}}\left(\Delta\tau^{*}\right)=\left(g_{0}\right)^{-1}\ell_{\text{th}}\left(t\right)\label{gzero vs elestar (48)}\end{equation}
 In Eq. (\ref{gzero vs elestar (48)}), $\left(g_{0}\right)^{-1}$
appears as a metric prefactor for the theoretical correlation length
function. From Eqs. (\ref{theta from t vs deltataustar (46)}), (\ref{psi from h vs deltahstar (47)}),
and (\ref{gzero vs elestar (48)}), the asymptotical non-universal
nature of each physical system is then characterized by the scale
factor set $\left\{ \vartheta;\left(g_{0}\right)^{-1};\psi_{\rho}\,\left(or\,\psi\right)\right\} $
(with implicit knowledge of $T_{c}$ and $\Lambda_{qe}^{*}$). However,
for the present fluid study where the thermodynamic length unit is
already fixed by Eq. (\ref{length scale (5)}), the above fitting
Eq. (\ref{gzero vs elestar (48)}) introduces one supplementary dimensionless
number defined such as:\begin{equation}
\mathbb{L}^{\left\{ 1f\right\} }=g_{0}\alpha_{c}\label{Ltabfstar (49)}\end{equation}
where the notation $\mathbb{L}^{\left\{ 1f\right\} }$ anticipates
a master nature of this product which we will demonstrate below {[}see
Eq. (\ref{1f ufstar (93)})]. More generally, in order to maintain
unicity of the length unit in the dimensionless description of the
singular behavior, any theroretical density property (which implicitely
refers to the length scale unit $\left(g_{0}\right)^{-1}$) needs
to introduce the proportionality factor $\left(\mathbb{L}^{\left\{ 1f\right\} }\right)^{-d}$
to the corresponding dimensionless physical density which refers to
the length scale unit $\alpha_{c}$. As a direct consequence of the
fitting Eq. (\ref{gzero vs elestar (48)}) for the correlation length,
the order parameter density $m$ must be analytically related to the
corresponding physical dimensionless variables $\Delta\tilde{\rho}$
(or $\Delta m^{*}$) by the following linear approximation (including
quantum effects)\begin{equation}
\begin{array}{rl}
m & =\left(\mathbb{L}^{\left\{ 1f\right\} }\right)^{-d}\left(\psi_{\rho}\right)^{-1}\left[\Lambda_{qe}^{*}\Delta\tilde{\rho}\right]\\
or\, m & =\left(\mathbb{L}^{\left\{ 1f\right\} }\right)^{-d}\psi^{-1}\left[\Lambda_{qe}^{*}\Delta m^{*}\right]\end{array}\label{inverse psi from m vs deltamstar (50)}\end{equation}
 For simplification of the following presentation, we only use $\psi_{\rho}$
related to the practical dimensionless form of the variables (see
above § 2.3).

Finally, adding the knowledge of the energy unit and the length unit
for each pure fluid to the theoretical results obtained from the massive
renormalization scheme, the dimensionless singular behaviors of the
fluid properties are now characterized by the set\begin{equation}
\mathbb{S}_{SF}=\left\{ \vartheta;\mathbb{L}^{\left\{ 1f\right\} };\psi_{\rho}\right\} \label{StabSF (51)}\end{equation}
 made of three dimensionless scale factors (admitting that $\left(\beta_{c}\right)^{-1}$,
$\alpha_{c}$, and $\Lambda_{qe}^{*}$ are known). Therefore, it is
easy to analytically define these three dimensionless parameters which
characterize each Ising-like fluid, thanks to the exact values of
the mean crossover functions within this preasymptotic domain.

\subsection{Three scale-factor characterization \emph{within} the Ising-like
preasymptotic domain.}

As already mentionned in the introduction and discussed in a detailed
manner in I, this asymptotic characterization is valid within the
Ising-like preasymptotic domain where the complete crossover functions
of Eqs. (\ref{MR ele vs t (42)}) and (\ref{MR khi vs t (43)}) can
be approximated by the following restricted (two-term) Wegner-like
expansions \citet{Wegner1972}:\begin{equation}
\ell_{\text{PAD,th}}\left(t\right)=\left(\mathbb{Z}_{\xi}^{+}\right)^{-1}t^{-\nu}{\displaystyle \left[1+\mathbb{Z}_{\xi}^{1,+}t^{\Delta}\right]}\label{MR 2term elestar (52)}\end{equation}
 \begin{equation}
\mathcal{X}_{\text{\text{PAD},th}}\left(t\right)=\left(\mathbb{Z}_{\chi}^{+}\right)^{-1}t^{-\gamma}{\displaystyle \left[1+\mathbb{Z}_{\chi}^{1,+}t^{\Delta}\right]}\label{MR 2term khistar (53)}\end{equation}
 In Eqs. (\ref{MR 2term elestar (52)}) and (\ref{MR 2term khistar (53)}),
$\mathbb{Z}_{\xi}^{1,+}$ {[}see below Eq. (\ref{MR Z1ele amplitude (54)})],
is the amplitude of the first confluent correction to scaling for
the correlation length, which is related to the one for the susceptibility
$\mathbb{Z}_{\chi}^{1,+}$ {[}see below Eq. (\ref{MR Z1khi amplitude (55)})],
by the universal ratio $\frac{\mathbb{Z}_{\xi}^{1,+}}{\mathbb{Z}_{\chi}^{1,+}}=0.67919$
\citet{Bagnuls2002}, with: \begin{equation}
\mathbb{Z}_{\xi}^{1,+}=-{\displaystyle \sum_{i=1}^{3}}X_{\xi,i}^{+}Y_{\xi,i}^{+}\label{MR Z1ele amplitude (54)}\end{equation}
 \begin{equation}
\mathbb{Z}_{\mathcal{\chi}}^{1,+}=-{\displaystyle \sum_{i=1}^{3}}X_{\chi,i}^{+}Y_{\chi,i}^{+}\label{MR Z1khi amplitude (55)}\end{equation}
 The theoretical field extension $t\lesssim\mathcal{L}_{\text{PAD}}^{\text{Ising}}$
of the Ising-like preasymptotic domain where the restricted Eqs. (\ref{MR 2term elestar (52)})
and (\ref{MR 2term khistar (53)}) are valid is defined in I, such
as\begin{equation}
\mathcal{L}_{\text{PAD}}^{\text{Ising}}=\frac{10^{-3}}{\left(S_{2}\right)^{2}}\approx1.9\,10^{-6}\label{LcalPADMR (56)}\end{equation}
Now considering all the theoretical functions estimated for all the
singular properties of the Ising-like systems (see I), we can note
that the universal features in the Ising-like preasymptotic domain
are characterized by the set\begin{equation}
\mathcal{\mathbb{S}}_{A}^{\left\{ MR\right\} }=\left\{ \begin{array}{rl}
\mathbb{Z}_{\chi}^{1,+}= & 8.56347\\
\left(\mathbb{Z}_{\xi}^{+}\right)^{-1}= & 0.471474\\
\left(\mathbb{Z}_{\chi}^{+}\right)^{-1}= & 0.269571\end{array}\right\} \label{StabMRA (57)}\end{equation}
of three theoretical amplitudes associated to the set $\left\{ \Delta;\nu;\gamma\right\} $
of three universal exponents selected as independent. Accordingly,
the restricted forms of two independent fitting equations are\begin{equation}
\begin{array}{cc}
\left(\Lambda_{qe}^{*}\right)^{-1}\frac{\xi_{\text{exp}}}{\alpha_{c}}= & \left(\mathbb{L}^{\left\{ 1f\right\} }\right)^{-1}\left(\mathbb{Z}_{\xi}^{+}\right)^{-1}\left(\vartheta\Delta\tau^{*}\right)^{-\nu}\\
 & {\displaystyle \left[1+\mathbb{Z}_{\xi}^{1,+}\left(\vartheta\Delta\tau^{*}\right)^{\Delta}\right]}\end{array}\label{ksi two term fit PAD (58)}\end{equation}
 \begin{equation}
\begin{array}{cc}
\left(\Lambda_{qe}^{*}\right)^{2}\kappa_{T,\text{exp}}^{*}= & \left(\mathbb{L}^{\left\{ 1f\right\} }\right)^{d}\left(\psi_{\rho}\right)^{2}\left(\mathbb{Z}_{\chi}^{+}\right)^{-1}\left(\vartheta\Delta\tau^{*}\right)^{-\gamma}\\
 & \left[1+\mathbb{Z}_{\mathcal{\chi}}^{1,+}\left(\vartheta\Delta\tau^{*}\right)^{\Delta}\right]\end{array}\label{khi two term fit PAD (59)}\end{equation}
where $\xi_{\text{exp}}$ and $\kappa_{T,\text{exp}}^{*}$ are given
by the restricted Wegner-like expansions of Eqs. (\ref{2term ksiexp (27)})
and (\ref{2term kappaTexp (34)}), respectively. That provides the
following \emph{hierarchical} relations\begin{equation}
a_{\chi}^{+}=\mathbb{Z}_{\mathcal{\chi}}^{1,+}\vartheta^{\Delta}\label{aksiplus vs PADSF (60)}\end{equation}
 \begin{equation}
\frac{\xi_{0}^{+}}{\alpha_{c}}=\xi^{+}=\left(\mathbb{Z}_{\xi}^{+}\right)^{-1}\left(\mathbb{L}^{\left\{ 1f\right\} }\right)^{-1}\Lambda_{qe}^{*}\vartheta^{-\nu}\label{ksiplus vs PADSF (61)}\end{equation}
 \begin{equation}
\Gamma^{+}=\left(\mathbb{Z}_{\chi}^{+}\right)^{-1}\left(\mathbb{L}^{\left\{ 1f\right\} }\right)^{d}\left(\Lambda_{qe}^{*}\right)^{d-2}\left(\psi_{\rho}\right)^{2}\vartheta^{-\gamma}\label{gammaplus vs PADSF (62)}\end{equation}
 with $\frac{a_{\xi}^{+}}{a_{\chi}^{+}}=\frac{\mathbb{Z}_{\xi}^{1,+}}{\mathbb{Z}_{\mathcal{\chi}}^{1,+}}=0.67919$
\citet{Guida1998,Bagnuls2002}. We underline the fact that Eq. (\ref{aksiplus vs PADSF (60)})
(or equivalently equation $a_{\xi}^{+}=\mathbb{Z}_{\mathcal{\xi}}^{1,+}\vartheta^{\Delta}$
in the correlation length case), is to be first validated (to confer
unequivocal Ising-like equivalence between the first (system-dependent)
scale factor $\vartheta$ and $a_{\chi}^{+}$). Then Eq. (\ref{ksiplus vs PADSF (61)})
fixes the asymptotic amplitude of the dimensionless correlation length
and generates a single scale factor attached to the selected (physical)
length unit, which is then mandatory common to the thermodynamic and
correlations functions. Finally, the validation of Eq. (\ref{gammaplus vs PADSF (62)})
provides unequivocal Ising-like equivalence between the second (system-dependent)
scale factor $\psi_{\rho}$ and $\Gamma^{+}$ (accounting for {}``critical''
and {}``extensive'' nature of the susceptibility). 

Equations (\ref{aksiplus vs PADSF (60)}) to (\ref{gammaplus vs PADSF (62)})
satisfy the following condensed functional form\begin{equation}
\mathcal{\mathbb{S}}_{A}^{\left\{ MR\right\} }=\left\{ S_{A}\,\mathbb{F}\left(\mathbb{S}_{SF}\right)\right\} _{\left(\beta_{c}\right)^{-1},\alpha_{c},\Lambda_{qe}^{*}}\label{StabMRA vs SAfluid (63)}\end{equation}
 where the function $\mathbb{F}$ takes an universal scaling form
of the dimensionless asymptotic scale factors $\vartheta$ and $\psi_{\rho}$.
The universal character of Eq. (\ref{StabMRA vs SAfluid (63)}) occurs
for any one-parameter crossover modeling. That infers Ising-like equivalence
between all estimated crossover functions only using three model-dependent
characteristic numbers. This result is shown in Appendix A, considering
the asymptotic crossover infered by the minimal-subtraction renormalization
scheme \citet{MSRscheme,Zhong2003} and the phenomenological approach
given by a parametric model of the equation of state \citet{Agayan2001}.

Obviously, from Eqs. (\ref{theta from t vs deltataustar (46)}) and
(\ref{LcalPADMR (56)}), it is easy to define the extension range\begin{equation}
\Delta\tau^{*}<\mathcal{L}_{\text{PAD}}^{f}=\frac{\mathcal{L}_{\text{PAD}}^{\text{Ising}}}{\vartheta}\simeq\frac{1.9\times10^{-6}}{\vartheta}\label{PAD extend (64)}\end{equation}
of the Ising-like preasymptotic domain of the selected fluid (labeled
with superscript $f$). Therefore, for each one-component fluid for
which $\vartheta$ (or equivalently one confluent amplitude among
$a_{\chi}^{+}$ or $a_{\xi}^{+}$) is an unknown parameter, the remaining
question of concern is: How to define the validity range $\Delta\tau^{*}<\mathcal{L}_{\text{PAD}}^{f}$
where the theoretical Ising-like characterization by three scale factors
can replace the experimental characterization by three asymptotic
amplitudes?

\subsection{Three free-parameter characterization \emph{beyond} the Ising-like
preasymptotic domain }

As noted in Ref. \citet{Bagnuls2002}, in the absence of information
concerning the true extension of the Ising-like behavior for a real
system belonging to the 3D Ising-like universality class, the introduction
of the scale factors $\vartheta$, $\psi_{\rho}$, and the wavelength
unit $g_{0}$ throughout Eqs. (\ref{theta from t vs deltataustar (46)})
to (\ref{gzero vs elestar (48)}) cannot be easily controlled. Alternatively,
it was proposed to introduce three adjustable dimensionless parameters
$\mathbb{L}_{0,\mathcal{L}}^{*}$, $\mathbb{X}_{0,\mathcal{L}}^{*}$,
and $\vartheta_{\mathcal{L}}$, using the following fitting equations:\begin{equation}
\begin{array}{cl}
\frac{\alpha_{c}}{\xi_{\text{exp}}^{*}\left(\Delta\tau^{*}\right)}= & \left(\mathbb{L}_{0,\mathcal{L}}^{*}\right)^{-1}\mathbb{Z}_{\xi}^{+}\left(\Delta\tau^{*}\right)^{\nu}\\
 & \prod_{i=1}^{K}\left(1+X_{\xi,i}^{+}\left(t\right)^{D\left(t\right)}\right)^{Y_{\xi,i}^{+}}\end{array}\label{ksi vs elezero fitting eq (65)}\end{equation}
 \begin{equation}
\begin{array}{cl}
\frac{1}{\kappa_{T,\text{exp}}^{*}\left(\Delta\tau^{*}\right)}= & \left(\mathbb{X}_{0,\mathcal{L}}^{*}\right)^{-1}\mathbb{Z}_{\mathcal{\chi}}^{+}\left(\Delta\tau^{*}\right)^{\gamma}\\
 & \prod_{i=1}^{K}\left(1+X_{\mathcal{\chi},i}^{+}\left(t\right)^{D\left(t\right)}\right)^{Y_{X,i}^{+}}\end{array}\label{khiT vs khizero fitting eq (66)}\end{equation}
 with\begin{equation}
t=\vartheta_{\mathcal{L}}\Delta\tau^{*}\label{thetaLcalli vs tstar (67)}\end{equation}
 $\mathbb{L}_{0,\mathcal{L}}^{*}$ and $\mathbb{X}_{0,\mathcal{L}}^{*}$
are two adjustable metric prefactors (with same value above and below
$T_{c}$). $\vartheta_{\mathcal{L}}$ is a global crossover parameter
in a sense where it is attached to an unknown effective parameter
$\mathcal{L}^{f}$ which measures the extent of fitting agreement
involving an undefined number of terms in the Wegner-like expansion
(see I for details). The determination of $\vartheta_{\mathcal{L}}$
is then equivalent to the determination of $\mathcal{L}^{f}$. However,
within the Ising-like preasymptotic domain, the restricted forms of
the fitting Eqs. (\ref{ksi vs elezero fitting eq (65)}) and (\ref{khiT vs khizero fitting eq (66)})
are\begin{equation}
\frac{\xi_{\text{exp}}}{\alpha_{c}}=\mathbb{L}_{0,\mathcal{L}}^{*}\left(\mathbb{Z}_{\xi}^{+}\right)^{-1}\left(\Delta\tau^{*}\right)^{-\nu}{\displaystyle \left[1+\mathbb{Z}_{\xi}^{1,+}\left(\vartheta_{\mathcal{L}}\Delta\tau^{*}\right)^{\Delta}\right]}\label{ksi two term fitting (68)}\end{equation}
 \begin{equation}
\kappa_{T,\text{exp}}^{*}=\mathbb{X}_{0,\mathcal{L}}^{*}\left(\mathbb{Z}_{\chi}^{+}\right)^{-1}\left(\Delta\tau^{*}\right)^{-\gamma}{\displaystyle \left[1+\mathbb{Z}_{\mathcal{\chi}}^{1,+}\left(\vartheta_{\mathcal{L}}\Delta\tau^{*}\right)^{\Delta}\right]}\label{kapaT two term fitting (69)}\end{equation}
 Therefore, the physical leading amplitudes can be calculated using
the (independent) equations:\begin{equation}
\frac{\xi_{0}^{+}}{\alpha_{c}}=\mathbb{L}_{0,\mathcal{L}}^{*}\left(\mathbb{Z}_{\xi}^{+}\right)^{-1}\label{ksieroplus vs elezero (70)}\end{equation}
 \begin{equation}
\Gamma^{+}=\mathbb{X}_{0,\mathcal{L}}^{*}\left(\mathbb{Z}_{\chi}^{+}\right)^{-1}\label{gammaplus vs khizero (71)}\end{equation}
i.e., without explicit reference to $\vartheta_{\mathcal{L}}$ (however
the subscript $\mathcal{L}$ recalls for the implicit $\vartheta_{\mathcal{L}}$
dependence due to the fitting in the temperature range $\Delta\tau^{*}\leq\mathcal{L}^{f}$,
with $\mathcal{L}^{f}>\mathcal{L}_{\text{PAD}}^{f}$). Noticeable
distinction occurs for the confluent corrections to scaling since
the first confluent amplitudes are only $\vartheta_{\mathcal{L}}$-dependent
and can be calculated using the equations:\begin{equation}
a_{\xi}^{+}=\left(\vartheta_{\mathcal{L}}\right)^{\Delta}\mathbb{Z}_{\xi}^{1,+}\label{aksiplus vs tethalcali (72)}\end{equation}
 \begin{equation}
a_{\chi}^{+}=\left(\vartheta_{\mathcal{L}}\right)^{\Delta}\mathbb{Z}_{\mathcal{\chi}}^{1,+}\label{akhiplus vs tethalcali (73)}\end{equation}
 interrelated by the universal ratio $\frac{\mathbb{Z}_{\xi}^{1,+}}{\mathbb{Z}_{\chi}^{1,+}}=0.67919$
\citet{Bagnuls2002}.

For better understanding of the \emph{scaling} nature of the analytical
transformations of the physical variables {[}such as Eqs. (\ref{theta from t vs deltataustar (46)})
or (\ref{thetaLcalli vs tstar (67)})], we select Eq. (\ref{akhiplus vs tethalcali (73)})
as the independent equation for the critical crossover characterization.
We must then rewrite the above Eqs. (\ref{ksieroplus vs elezero (70)})
to (\ref{akhiplus vs tethalcali (73)}) in the following hierarchical
forms\begin{equation}
\left(\vartheta_{\mathcal{L}}\right)^{-\Delta}a_{\chi}^{+}=\mathbb{Z}_{\mathcal{\chi}}^{1,+}=\text{universal}\,\text{cst.}\label{MR renormalized akhi1+ (74)}\end{equation}
 \begin{equation}
\left(\mathbb{L}_{0,\mathcal{L}}^{*}\right)^{-1}\frac{\xi_{0}^{+}}{\alpha_{c}}=\left(\mathbb{Z}_{\xi}^{+}\right)^{-1}=\text{universal}\,\text{cst.}\label{MR renormalized ksizero+ (75)}\end{equation}
 \begin{equation}
\left(\mathbb{X}_{0,\mathcal{L}}^{*}\right)^{-1}\Gamma^{+}=\left(\mathbb{Z}_{\mathcal{\chi}}^{+}\right)^{-1}=\text{universal}\,\text{cst.}\label{MR renormaized gamma+ (76)}\end{equation}
where the l.h.s. of the above equations contain all the system-dependent
information, first for Ising-like critical crossover, then for asymptotic
behavior of correlation functions, and finally for asymptotic behavior
of thermodynamic functions. Moreover, this information is given in
a \emph{dual} form, i.e., as a product between a {}``physical''
amplitude ($a_{\chi}^{+}$, $\xi^{+}$, or $\Gamma^{+}$) and either
a {}``crossover'' factor ($\vartheta_{\mathcal{L}}$), which acts
as a scale factor for the confluent correction contribution, or a
{}``pre''-factor ($\mathbb{L}_{0,\mathcal{L}}^{*}$ or $\mathbb{X}_{0,\mathcal{L}}^{*}$)
which acts as a simple factor of proportionality for the corresponding
leading amplitude ($\xi^{+}$ or $\Gamma^{+}$). The following set\begin{equation}
\mathbb{S}_{1C,\mathcal{L}}=\left\{ \vartheta_{\mathcal{L}};\mathbb{L}_{0,\mathcal{L}}^{*};\mathbb{X}_{0,\mathcal{L}}^{*}\right\} \label{Stab1C2M (77)}\end{equation}
 is equivalent to the previous set $\mathbb{S}_{SF}$ of Eq. (\ref{StabSF (51)}),
except that the subscript $1C,\mathcal{L}$ recalls for a single crossover
parameter obtained over an extended temperature range $\mathcal{L}^{f}>\mathcal{L}_{\text{PAD}}^{f}$,
\emph{beyond} the Ising-like preasymptotic domain. The following condensed
functional form \begin{equation}
\mathcal{\mathbb{S}}_{A}^{\left\{ MR\right\} }=\left\{ S_{A}\,\mathbb{F}_{\mathcal{L}}\left(\mathbb{S}_{1C,\mathcal{L}}\right)\right\} _{\left(\beta_{c}\right)^{-1},\alpha_{c},\Lambda_{qe}^{*}}\label{StabMRA vs SAfluidStab1C2M (78)}\end{equation}
 can be used in a equivalent scaling manner to Eq. (\ref{StabMRA vs SAfluid (63)})
when the crossover parameter $\vartheta_{\mathcal{L}}$ is unique
within the $\mathcal{L}^{f}$ range.

To our knowledge, the unicity of the crossover parameter along the
critical isochore of a one-component fluid has never been directly
evidenced from the singular behavior of the correlation length or
any other thermodynamic property. However, from simultaneous fitting
analysis of several singular properties of xenon and helium 3, an
indirect probe of a single value for one adjustable parameter related
to the scale factor $\vartheta$ was obtained, using the crossover
functions estimated in the massive renormalization scheme \citet{Bagnuls1984b,Garrabos2006qe}
and the minimal-subtraction renormalization scheme \citet{Zhong2003,Zhong2004}.
But these results were never used to accurately analyze the expected
equivalence between Eqs. (\ref{StabMRA vs SAfluid (63)}) and (\ref{StabMRA vs SAfluidStab1C2M (78)}),
and then to estimate the other two scale factors $\mathbb{L}^{\left\{ 1f\right\} }$
and $\psi_{\rho}$, which is the only correct way to verify the asymptotic
condition $\vartheta=\vartheta_{\mathcal{L}}$ within the Ising-like
preasymptotic domain \citet{thetaunicity}.

An analytic determination of $\vartheta_{\mathcal{L}}$, made beyond
the Ising-like preasymptotic domain without use of any adjustable
parameter, is under investigation for the case of the isothermal compressibility
of xenon \citet{Garrabos2006khiT}. The main objective is to carefully
correlate the local value of this crossover parameter with the local
value of the correlation length before to validate its uniqueness
by identification with the asymptotic scale factor $\vartheta$, calculated
by using Eq. (\ref{theta from t vs deltataustar (46)}). However,
such a challenging demonstration of $\vartheta\equiv\vartheta_{\mathcal{L}}$
in the temperature range $\Delta\tau^{*}\leq\mathcal{L}_{\text{EAD}}^{f}$,
i.e., within the so-called Ising-like \emph{extended} asymptotic domain
(EAD) in the following, as a formulation of the three-parameter characterization
of xenon selected as a standard one-component fluid, remain two preliminary
attempts to test the equivalence between Eqs. (\ref{StabMRA vs SAfluid (63)})
and (\ref{StabMRA vs SAfluidStab1C2M (78)}). That needs to be examinated
using a more general approach, as the one proposed below, where we
will introduce three master constants which relate unequivocally dimensionless
lengths and relevant fields of both (theoretical and master) descriptions,
to identify the theoretical crossover of the $\left\{ \Phi_{3}\left(1\right)\right\} $-class
with the master crossover of the $\left\{ 1f\right\} $-subclass.

\subsection{Identification of the theoretical and master asymptotic scaling \emph{within}
the Ising-like preasymptotic domain}

Now, while reconsidering our previous analysis of the relations between
physical and master properties, we must rewrite Eqs. (\ref{ksizeroplus vs Qcmin (28)}),
(\ref{gamma+ vs Qmin (35)}), and (\ref{akhi vs Qmin (36)}), in the
following hierarchical forms\begin{equation}
Y_{c}\left(a_{\chi}^{+}\right)^{-\frac{1}{\Delta}}=\left(\mathcal{Z}_{\chi}^{1,+}\right)^{\frac{1}{\Delta}}=\text{master}\,\text{cst.}\label{1f renormalized akhi1+ (79)}\end{equation}
 \begin{equation}
\frac{1}{\alpha_{c}}\left(Y_{c}\right)^{\nu}\left[\left(\Lambda_{qe}^{*}\right)^{-1}\xi_{0}^{+}\right]=\mathcal{Z}_{\mathcal{\xi}}^{+}=\text{master}\,\text{cst.}\label{1f renormalized ksizero+ (80)}\end{equation}
 \begin{equation}
Z_{c}\left(Y_{c}\right)^{\gamma}\left[\left(\Lambda_{qe}^{*}\right)^{2-d}\Gamma^{+}\right]=\mathcal{Z}_{\chi}^{+}=\text{master}\,\text{cst.}\label{1f renormalized gamma+ (81)}\end{equation}
Comparison of Eqs. (\ref{MR renormalized akhi1+ (74)}) to (\ref{MR renormaized gamma+ (76)})
with Eqs. (\ref{1f renormalized akhi1+ (79)}) to (\ref{1f renormalized gamma+ (81)}),
shows that their r.h.s. differences only concern the respective numerical
values of the characteristic \emph{master} set $\mathcal{S}_{A}^{\left\{ 1f\right\} }$
of Eq. (\ref{Scal1fA (40)}), and \emph{universal} set $\mathcal{\mathbb{S}}_{A}^{\left\{ MR\right\} }$
of Eq. (\ref{StabMRA (57)}). For their l.h.s. comparison, neglecting
the quantum corrections in a first approach (i.e. fixing $\Lambda_{qe}^{*}=1$),
the term to term identification between measurable amplitudes underlines
the analogy between the explicit parameter set $\left\{ Y_{c};Z_{c}\right\} $,
related to the master description, and the implicit one $\left\{ \vartheta;\psi_{\rho}\right\} $,
related to the massive renormalization description. We can then note
that the $\left\{ 1f\right\} $-master formulation compares to the
$\Phi_{3}\left(1\right)$-universal formulation, only if we have correctly
accounted for the asymptotic scaling nature of each dimensionless
number needed by the massive renormalization scheme. In order to reveal
such a scaling nature, it is essential to note that the scale dilatation
method replaces the renormalized fields (such as $t$, $h$, $m$,
etc.) needed to observe the {}``universal'' behavior of the $\Phi_{3}^{4}\left(1\right)$-universality
class, by the $\left\{ 1f\right\} $-fields (such as, $\mathcal{T}^{*}$,
$\mathcal{H}_{\text{qf}}^{*}$, $\mathcal{M}_{\text{qf}}^{*}$, etc.)
needed to observe the {}``master'' behavior of $\left\{ 1f\right\} $-subclass.
The common physical variables are $\Delta\tau^{*}$, $\Delta\tilde{\mu}$,
and $\Delta\tilde{\rho}$. Therefore, it remains to give explicit
forms for the following exchanges between the theoretical variables
and the $\left\{ 1f\right\} $-subclass variables

\begin{eqnarray}
 & t\rightarrow\mathcal{T}^{*}\label{tphi4 vs tau1f exchange (82)}\\
h\rightarrow\mathcal{H}_{\text{qf}}^{*} & or & m\rightarrow\mathcal{M}_{\text{qf}}^{*}\label{hphi4 vs h1f exchange (83)}\end{eqnarray}
(see Ref. \citet{Garrabos2006corlength} for the correlation length
case). The next subsection is dedicated to the isothermal susceptibility
case (which then closes the description of the $\left\{ 1f\right\} $-subclass
along the critical isochore in conformity with the universal features
estimated for the Ising-like universality class).

\begin{table*}
\begin{tabular}{|c||c|c|c|}
\hline 
(a)  & {\footnotesize $F_{P}$}  & {\footnotesize $\mathbb{Z}_{P}^{\pm}$}  & $\mathbb{Z}_{P}^{1,\pm}$\tabularnewline
\hline 
$\mathcal{\mathbb{S}}_{A}^{\left\{ MR\right\} }$  & (\ref{StabMRA (57)})  & $\left\{ \left(\mathbb{Z}_{\xi}^{+}\right)^{-1}=0.471474;\,\left(\mathbb{Z}_{\chi}^{+}\right)^{-1}=0.269571\right\} $  & $\mathbb{Z}_{\chi}^{1,+}=8.56347$\tabularnewline
\hline
\hline 
(b)  & {\footnotesize $\mathcal{P}_{\text{qf}}^{\ast}$}  & {\footnotesize $\mathbb{Z}_{P}^{\left\{ 1f\right\} }$}  & $\mathbb{\mathcal{Z}}_{P}^{1,\pm}$\tabularnewline
\hline 
$\mathcal{S}_{2P1S}^{\left\{ 1f\right\} }$  & (\ref{Scal1fPSF (94)})  & $\left\{ \mathbb{Z}_{\xi}^{\left\{ 1f\right\} }=25.6936;\,\mathbb{Z}_{\chi}^{\left\{ 1f\right\} }=1950.7\right\} $  & $\Theta^{\left\{ 1f\right\} }=4.288\times10^{-3}$\tabularnewline
\hline 
 & {\footnotesize $\ell_{\text{qf}}^{\ast}$}  & {\footnotesize $\mathbb{Z}_{\xi}^{\left\{ 1f\right\} }=\left[\mathcal{Z}_{\xi}^{\pm}\mathbb{Z}_{\xi}^{\pm}\left(\Theta^{\left\{ 1f\right\} }\right)^{\nu}\right]^{-1}=\mathbb{L}^{\left\{ 1f\right\} }=25.6936$}  & $\Theta^{\left\{ 1f\right\} }=\left[\frac{\mathcal{Z}_{\xi}^{1,\pm}}{\mathbb{Z}_{\xi}^{1,\pm}}\right]^{\frac{1}{\Delta}}$\tabularnewline
 & {\footnotesize $\varkappa_{\text{qf}}^{\ast}$}  & {\footnotesize $\mathbb{Z}_{\chi}^{\left\{ 1f\right\} }=\left[\mathcal{Z}_{\chi}^{\pm}\mathbb{Z}_{\chi}^{\pm}\left(\Theta^{\left\{ 1f\right\} }\right)^{\gamma}\right]^{-1}=\left(\mathbb{L}^{\left\{ 1f\right\} }\right)^{-d}\left(\Psi^{\left\{ 1f\right\} }\right)^{-2}=1950.7$}  & $\Theta^{\left\{ 1f\right\} }=\left[\frac{\mathcal{Z}_{\chi}^{1,\pm}}{\mathbb{Z}_{\chi}^{1,\pm}}\right]^{\frac{1}{\Delta}}$\tabularnewline
 & {\footnotesize $\mathbb{\mathcal{C}}_{\text{qf}}^{\ast}$}  & {\footnotesize $\mathbb{Z}_{C}^{\left\{ 1f\right\} }=\frac{\mathcal{Z}_{C}^{\pm}}{\alpha\mathbb{Z}_{C}^{\pm}\left(\Theta^{\left\{ 1f\right\} }\right)^{2-\alpha}}=\left(\mathbb{L}^{\left\{ 1f\right\} }\right)^{d}=16961.9$}  & $\Theta^{\left\{ 1f\right\} }=\left[\frac{\mathcal{Z}_{C}^{1,\pm}}{\mathbb{Z}_{C}^{1,\pm}}\right]^{\frac{1}{\Delta}}$\tabularnewline
 & {\footnotesize $\mathcal{M}_{\text{qf}}^{*}$}  & {\footnotesize $\mathbb{Z}_{M}^{\left\{ 1f\right\} }=\frac{\mathcal{Z}_{M}}{\mathbb{Z}_{M}\left(\Theta^{\left\{ 1f\right\} }\right)^{\beta}}=\left(\mathbb{L}^{\left\{ 1f\right\} }\right)^{d}\Psi^{\left\{ 1f\right\} }=2.94878$}  & $\Theta^{\left\{ 1f\right\} }=\left[\frac{\mathcal{Z}_{M}^{1}}{\mathbb{Z}_{M}^{1}}\right]^{\frac{1}{\Delta}}$\tabularnewline
\hline 
$\mathcal{S}_{SF}^{\left\{ 1f\right\} }$  & (\ref{Scal1fSF (108)})  & {\footnotesize $\left\{ \mathbb{L}^{\left\{ 1f\right\} }=25.6936;\,\Psi^{\left\{ 1f\right\} }=1.73847\,10^{-4}\right\} $}  & {\footnotesize $\Theta^{\left\{ 1f\right\} }=4.288\,10^{-3}$}\tabularnewline
$\mathcal{S}_{A}^{\left\{ 1f\right\} }$  & (\ref{Scal1fA (40)})  & $\left\{ \mathbb{\mathcal{Z}}_{\xi}^{+}=0.570481;\, Z_{\chi}^{+}=0.119\right\} $  & $\mathbb{\mathcal{Z}}_{\chi}^{1,+}=0.555$\tabularnewline
\hline
\hline 
(c)  & {\footnotesize $P_{\text{exp}}^{*}$}  & {\footnotesize $\mathbb{P}_{0,\mathcal{L}}^{*}$}  & \tabularnewline
\hline 
$\mathbb{S}_{1C,\mathcal{L}}$  & (\ref{Stab1C2M (77)})  & $\left\{ \mathbb{L}_{0,\mathcal{L}}^{*};\,\mathbb{X}_{0,\mathcal{L}}^{*}\right\} $  & {\footnotesize $\vartheta_{\mathcal{L}}\equiv\vartheta=Y_{c}\Theta^{\left\{ 1f\right\} }$}\tabularnewline
\hline 
 & $\xi^{*}$  & {\footnotesize $\begin{array}{rl}
\mathbb{L}_{0,\mathcal{L}}^{*}=\Lambda_{qe}^{*}(Y_{c})^{-\nu}\mathcal{Z}_{\xi}^{\pm}\mathbb{Z}_{\xi}^{\pm} & =\Lambda_{qe}^{*}(Y_{c})^{-\nu}\left[\mathbb{L}^{\left\{ 1f\right\} }\times\left(\Theta^{\left\{ 1f\right\} }\right)^{\nu}\right]^{-1}\\
 & =1.20999\Lambda_{qe}^{*}(Y_{c})^{-\nu}\end{array}$}  & \tabularnewline
 & $\kappa_{T}^{*}$  & {\footnotesize $\begin{array}{rl}
\mathbb{X}_{0,\mathcal{L}}^{*}=\frac{\left(\Lambda_{qe}^{*}\right)}{Z_{c}\left(Y_{c}\right)^{\gamma}}^{d-2}\mathcal{Z}_{\chi}^{\pm}\mathbb{Z}_{\chi}^{\pm} & =\frac{\left(\Lambda_{qe}^{*}\right)}{Z_{c}\left(Y_{c}\right)^{\gamma}}^{d-2}\left(\mathbb{L}^{\left\{ 1f\right\} }\right)^{d}\left(\Psi^{\left\{ 1f\right\} }\right)^{2}\left(\Theta^{\left\{ 1f\right\} }\right)^{-\gamma}\\
 & =0.44144\frac{\left(\Lambda_{qe}^{*}\right)}{Z_{c}\left(Y_{c}\right)^{\gamma}}^{d-2}\end{array}$}  & \tabularnewline
 & $c_{V}^{*}$  & {\footnotesize $\begin{array}{rl}
\frac{\mathbb{C}_{0,\mathcal{L}}^{*}}{\alpha}=\left(\Lambda_{qe}^{*}\right)^{-d}(Y_{c})^{2-\alpha}\frac{\mathcal{Z}_{C}^{\pm}}{\alpha\mathbb{Z}_{C}^{\pm}} & =\left(\Lambda_{qe}^{*}\right)^{-d}(Y_{c})^{2-\alpha}\left(\mathbb{L}^{\left\{ 1f\right\} }\right)^{d}\left(\Theta^{\left\{ 1f\right\} }\right)^{2-\alpha}\\
 & =0.564481\left(\Lambda_{qe}^{*}\right)^{-d}(Y_{c})^{2-\alpha}\end{array}$}  & \tabularnewline
 & $\Delta\rho_{LV}^{*}$  & {\footnotesize $\begin{array}{rl}
\mathbb{M}_{0,\mathcal{L}}^{*}=\frac{(Y_{c})^{\beta}}{\Lambda_{qe}^{*}\left(Z_{c}\right)^{\frac{1}{2}}}\frac{\mathcal{Z}_{M}}{\mathbb{Z}_{M}} & =\frac{(Y_{c})^{\beta}}{\Lambda_{qe}^{*}\left(Z_{c}\right)^{\frac{1}{2}}}\left(\mathbb{L}^{\left\{ 1f\right\} }\right)^{d}\Psi^{\left\{ 1f\right\} }\left(\Theta^{\left\{ 1f\right\} }\right)^{\beta}\\
 & =0.499185\frac{(Y_{c})^{\beta}}{\Lambda_{qe}^{*}\left(Z_{c}\right)^{\frac{1}{2}}}\end{array}$}  & \tabularnewline
\hline 
$\mathbb{S}_{SF}$  & (\ref{StabSF (51)})  & {\footnotesize $\left\{ \mathbb{L}^{\left\{ 1f\right\} }=25.6936;\,\psi_{\rho}=\left(Z_{c}\right)^{-\frac{1}{2}}\Psi^{\left\{ 1f\right\} }\right\} $}  & {\footnotesize $\vartheta=Y_{c}\Theta^{\left\{ 1f\right\} }$}\tabularnewline
$S_{SF}$  & (\ref{SFfluid (39)})  & $\left\{ Y_{c};\, Z_{c};\,\mathbb{L}^{\left\{ 1f\right\} }=25.6936;\,\Lambda_{qe}^{*}\right\} $  & \tabularnewline
$S_{A}$  & (\ref{SAfluid (38)})  & $\left\{ \xi^{+};\,\Gamma^{+}\right\} $  & $a_{\chi}^{+}$\tabularnewline
\hline
\end{tabular}

\caption{Three parameter characterization (colum 3: leading amplitudes or prefactors;
colum 4: scale factor or crossover parameter; see text): (a) for the
mean crossover functions $F_{P}\left(t\right)$ defined in I; (b)
for the master crossover functions $\mathcal{P}_{\text{qf}}^{*}\left(\mathcal{T}^{*}\right)$
{[}see Eq. (\ref{theoretical master function (109)})] ; lines 4,
9 and 10: independent parameters; lines 5 to 8): related parameters;
The two relations $\left(\mathbb{Z}_{C}^{\left\{ 1f\right\} }\right)^{\frac{1}{d}}=\mathbb{Z}_{\xi}^{\left\{ 1f\right\} }$
and $\frac{\left(\mathbb{Z}_{\xi}^{\left\{ 1f\right\} }\right)^{d}}{\mathbb{Z}_{\chi}^{\left\{ 1f\right\} }}=\left(\mathbb{Z}_{M}^{\left\{ 1f\right\} }\right)^{2}$,
are in conformity with the two-scale-factor universality; (c) similar
to (b) for the physical crossover functions $P_{\text{exp}}^{*}\left(\Delta\tau^{*}\right)$
{[}see Eq. (\ref{theoretical physical function (110)})]; The two
relations $\left(\frac{\mathbb{C}_{0,\mathcal{L}}^{*}}{\alpha}\right)^{\frac{1}{d}}\mathbb{L}_{0,\mathcal{L}}^{*}=1$
and $\left(\mathbb{L}_{0,\mathcal{L}}^{*}\right)^{-d}\frac{\mathbb{X}_{0,\mathcal{L}}^{*}}{\left(\mathbb{M}_{0,\mathcal{L}}^{*}\right)^{2}}=1$,
are in conformity with the two-scale-factor universality; All the
values of $\mathbb{P}_{0,\mathcal{L}}^{*}$, $\vartheta_{\mathcal{L}}\equiv\vartheta$
and $\psi_{\rho}$ can be estimated from $Q_{c}^{\text{min}}=\left\{ \left(\beta_{c}\right)^{-1},\alpha_{c},Y_{c},Z_{c}\right\} $
and $\Lambda_{qe}^{*}$ of the selected one-component fluid.\label{Table II}}

\end{table*}

\subsection{Master modification of the theoretical crossover for the isothermal
susceptibility}

We start with the following modification of Eq. (\ref{khiT vs khizero fitting eq (66)})\begin{equation}
\frac{1}{\mathcal{X}_{\text{qf}}^{*}\left(\left|\mathcal{T}^{*}\right|\right)}=\mathbb{Z}_{\chi}^{\left\{ 1f\right\} }\mathbb{Z}_{\chi}^{\pm}t^{\gamma}{\displaystyle \prod_{i=1}^{N}\left(1+X_{i,\chi}^{\pm}t^{D^{\pm}\left(t\right)}\right)^{Y_{i,\chi}^{\pm}}}\label{1f RG c3 fitting eq (84)}\end{equation}
 and the following modification of Eq. (\ref{thetaLcalli vs tstar (67)})
\begin{equation}
t=\Theta^{\left\{ 1f\right\} }\left|\mathcal{T}^{*}\right|\label{1f lin transf RG thermal field (85)}\end{equation}
 by introducing the prefactor $\mathbb{Z}_{\chi}^{\left\{ 1f\right\} }$
and the scale factor $\Theta^{\left\{ 1f\right\} }$ as \emph{master}
(i.e. unique) parameters for the $\left\{ 1f\right\} $-subclass.
We note that $\Theta^{\left\{ 1f\right\} }$, characteristic of the
(critical) isochoric line (with same value above and below $T_{c}$),
reads as follows\begin{equation}
\Theta^{\left\{ 1f\right\} }=\left[\frac{\mathcal{Z}_{\varkappa}^{1\pm}}{\mathbb{Z}_{\chi}^{1,\pm}}\right]^{\frac{1}{\Delta}}\label{1f MR crossover scale factor (86)}\end{equation}
 whatever the selected one-component fluid is. By virtue of the universal
feature of confluent amplitude ratios (see Table I), the numerical
value\begin{equation}
\Theta^{\left\{ 1f\right\} }=4.288\,10^{-3}\label{master crossover parameter for 1f (87)}\end{equation}
 is the same whatever the property and the phase domain. However,
we also note that $\Theta^{\left\{ 1f\right\} }$ contributes to the
leading term. Thus, in addition to Eq. (\ref{1f RG c3 fitting eq (84)}),
we define $\mathbb{Z}_{\chi}^{\left\{ 1f\right\} }$ such that\begin{equation}
\mathbb{Z}_{\chi}^{\left\{ 1f\right\} }=\left[\mathcal{Z}_{\varkappa}^{\pm}\mathbb{Z}_{\chi}^{\pm}\left(\Theta^{\left\{ 1f\right\} }\right)^{\gamma}\right]^{-1}\label{master constraint RG c3 for khi (88)}\end{equation}
 The numerical value,\begin{equation}
\mathbb{Z}_{\chi}^{\left\{ 1f\right\} }=1950.70\label{1f master prefactor value for khi (89)}\end{equation}
is the same in the homogeneous phase and in the non homogeneous phase.
The curve labeled MR in Figure \ref{Figure 1} was obtained from Eqs.
(\ref{1f RG c3 fitting eq (84)}) and (\ref{1f lin transf RG thermal field (85)})
using the numerical values of $\Theta^{\left\{ 1f\right\} }$ and
$\mathbb{Z}_{\chi}^{\left\{ 1f\right\} }$ given by Eqs. (\ref{master crossover parameter for 1f (87)})
and (\ref{1f master prefactor value for khi (89)}), respectively.

We recall that our previous analysis \citet{Garrabos2006corlength}
of the correlation length has introduced a similar prefactor $\mathbb{Z}_{\xi}^{\left\{ 1f\right\} }$
through the following modification of Eq. (\ref{ksi vs elezero fitting eq (65)})\begin{equation}
\frac{1}{\ell_{\text{qf}}^{*}\left(\left|\mathcal{T}^{*}\right|\right)}=\mathbb{Z}_{\xi}^{\left\{ 1f\right\} }F_{\ell}\left(t\right)\label{eleqf vs ztabksi1f fitting eq (90)}\end{equation}
 with\begin{equation}
\mathbb{Z}_{\xi}^{\left\{ 1f\right\} }=\left[\mathcal{Z}_{\xi}^{\pm}\mathbb{Z}_{\xi}^{\pm}\left(\Theta^{\left\{ 1f\right\} }\right)^{\nu}\right]^{-1}\label{master constraint RG c3 for ksi (91)}\end{equation}
 which has the same numerical value\begin{equation}
\mathbb{Z}_{\xi}^{\left\{ 1f\right\} }=25.6936\label{1f master prefactor value for ksi (92)}\end{equation}
for the homogeneous and non homogeneous domains. Of course, we retrieve
here the previous Eq. (\ref{Ltabfstar (49)}) \begin{equation}
\mathbb{Z}_{\xi}^{\left\{ 1f\right\} }\equiv\mathbb{L}^{\left\{ 1f\right\} }=\left(g_{0}\alpha_{c}\right)_{\forall\text{fluid}}\label{1f ufstar (93)}\end{equation}
 which now is valid \emph{whatever the fluid under} \emph{consideration}.
The set of master (two pre- + one scale) factors\begin{equation}
\mathcal{S}_{2P1S}^{\left\{ 1f\right\} }=\left\{ \begin{array}{rl}
\Theta^{\left\{ 1f\right\} }= & 4.288\times10^{-3}\\
\mathbb{Z}_{\xi}^{\left\{ 1f\right\} }= & 25.6936\\
\mathbb{Z}_{\chi}^{\left\{ 1f\right\} }= & 1950.70\end{array}\right\} \label{Scal1fPSF (94)}\end{equation}
closes the universal behavior of the $\left\{ 1f\right\} $-subclass,
as shown by the results reported in Table \ref{Table II} for all
the properties calculated along the critical isochore (for notations
see below and Refs. \citet{Garrabos2006corlength,Garrabos2006khiT,Garrabos2006qe}).
Equation (\ref{Ltabfstar (49)}) {[}or Eq. (\ref{1f ufstar (93)})]
appears then as the basic hypothesis which defines the critical length
unicity \citet{Privman1991} between correlation functions and thermodynamic
functions of the one component fluid subclass. $\mathbb{L}^{\left\{ 1f\right\} }$
takes an equivalent nature to the \emph{length reference} used in
the renormalization scheme applied to the $\Phi_{3}\left(1\right)$-class,
whatever the selected physical system.

The major interest of Eqs. (\ref{master constraint RG c3 for khi (88)})
and (\ref{master constraint RG c3 for ksi (91)}) is that they introduce
the needed {}``cross-relation'' between pure asymptotic scaling
description and first confluent correction to scaling, in order to
obtain only two independent leading amplitudes within the Ising-like
preasymptotic domain. Such a cross-relation occurs if the non-universal
scale factor associated with the irrelevant-field which induces the
correction-to-scaling term of lowest relative order $\left(\Delta\tau^{*}\right)^{\Delta}$
in a Wegner-like expansion, is the same as the non-universal scale
factor associated with the relevant (thermal) field which gives the
leading scaling term $\left(\Delta\tau^{*}\right)^{-e_{P}}$.

In that universal description of the confluent corrections to scaling,
each crossover function includes the (two-term) master behavior expected
using the scale dilatation method. By comparing the leading terms
on each member of Eqs. (\ref{1f RG c3 fitting eq (84)}), (\ref{khi master pplaw (33)}),
and (\ref{kapaT two term fitting (69)}), we obtain the relations\begin{equation}
\Gamma^{\pm}=\left(\mathbb{Z}_{\chi}^{\left\{ 1f\right\} }\mathbb{Z}_{\chi}^{\pm}\right)^{-1}\vartheta^{-\gamma}=\mathbb{X}_{0,\mathcal{L}}^{*}\left(\mathbb{Z}_{\chi}^{\pm}\right)^{-1}\label{1f gammaplus amplitude (95)}\end{equation}
where the fluid-dependent metric prefactor $\mathbb{X}_{0,\mathcal{L}}^{*}$
of Eq. (\ref{kapaT two term fitting (69)}) now reads as follows \begin{equation}
\mathbb{X}_{0,\mathcal{L}}^{*}=\mathcal{Z}_{\chi}^{\pm}\mathbb{Z}_{\chi}^{\pm}\left(Y_{c}\right)^{-\gamma}\label{1f khitablzero amplitude (96)}\end{equation}
In Eq. (\ref{1f khitablzero amplitude (96)}), the critical contribution
of the scale factor $Y_{c}$ is explicit. The remaining adjustable
crossover parameter $\vartheta_{\mathcal{L}}$ of Eq. (\ref{thetaLcalli vs tstar (67)})
is characteristic of the Ising-like extended asymptotic domain $\Delta\tau^{*}\lesssim\mathcal{L}_{\text{EAD}}^{f}$
where the theoretical crossover functions and experimental data agree.
Within the Ising-like preasymptotic domain {[}see Eq. (\ref{LcalPADMR (56)})]
where the two-term Wegner-like expansions are expected to be valid,
the comparison of the first confluent amplitudes for master and theoretical
descriptions, enables one to write $\vartheta_{\mathcal{L}}$ as follows\begin{equation}
\vartheta_{\mathcal{L}}\equiv\vartheta\;\left(=Y_{c}\Theta^{\left\{ 1f\right\} }\right)\label{crossover scale factor (97)}\end{equation}
 with\begin{equation}
t\equiv\left(\frac{\mathcal{Z}_{\chi}^{1,\pm}}{\mathbb{Z}_{\chi}^{1,\pm}}\right)^{\frac{1}{\Delta}}\mathcal{T}^{*}\;\left(=\Theta^{\left\{ 1f\right\} }\left|\mathcal{T}^{*}\right|\right)\label{scaling versus renorm thermal field (98)}\end{equation}
\begin{figure*}
\includegraphics[width=1.3\columnwidth,keepaspectratio]{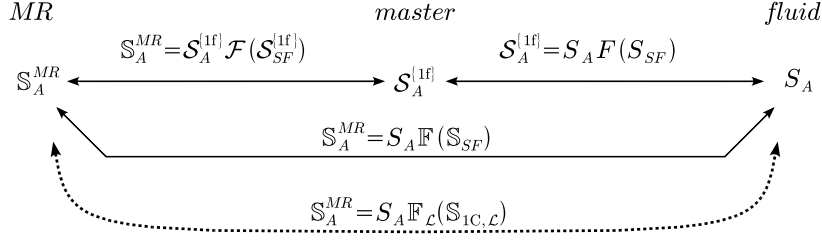}

\caption{Schematic links between three amplitude characterization $\mathbb{S}_{A}^{\left\{ MR\right\} }$
of Eq. (\ref{StabMRA (57)}), $\mathcal{S}_{A}^{\left\{ 1f\right\} }$
of Eq. (\ref{Scal1fA (40)}) and $\mathcal{S}_{A}$ of Eq. (\ref{SAfluid (38)}),
of the theoretical, master and physical singular behaviors, respectively,
for a fluid of critical parameters given by $Q_{c}^{\text{min}}=\left\{ \left(\beta_{c}\right)^{-1},\alpha_{c},Y_{c},Z_{c}\right\} $
of Eq. (\ref{qcmin scale factors (3)}) and $\Lambda_{qe}^{*}$, and
which belongs to the one-component fluid subclass.\label{Figure 2} }

\end{figure*}

As considered from basic input of the scale dilatation method, Eq.
(\ref{scaling versus renorm thermal field (98)}) agrees with the
scale dilatation of the temperature field\begin{equation}
\mathcal{T}^{*}=Y_{c}\left|\Delta\tau^{*}\right|\label{thermal scale dilatation (99)}\end{equation}
 Note that the extension $\mathcal{T}^{*}\lesssim\mathcal{L}_{\text{PAD}}^{\left\{ 1f\right\} }$
of the Ising-like preasymptotic domain of the $\left\{ 1f\right\} $-subclass
can then be immediately obtained from Eq. (\ref{LcalPADMR (56)}),
with\begin{equation}
\mathcal{L}_{\text{PAD}}^{\left\{ 1f\right\} }=\frac{\mathcal{L}_{\text{PAD}}^{\text{Ising}}}{\Theta^{\left\{ 1f\right\} }}\approx4.43\times10^{-4}\label{LcalPAD1f (100)}\end{equation}
 (see for example the full arrow labeled {}``PAD'' in Fig. \ref{Figure 1}c).

\subsection{Closed master modification of the mean crossover functions and master
extension $\mathcal{L}_{\text{EAD}}^{\left\{ 1f\right\} }$ of the
extended asymptotic domain}

Obviously, the equivalent approach at exact criticality and along
the critical isotherm occurs in virtue of the two-scale-factor universality
which implies a second unequivocal relation between $\psi_{\rho}$
and $Z_{c}$. However, we can anticipate such a result only from the
thermodynamic definitions of the susceptibilities $\chi_{\text{th}}=\left(\frac{\partial m}{\partial h}\right)_{t}$
and $\mathcal{X}_{\text{qf}}^{*}=\left(\frac{\partial\mathcal{M}_{\text{qf}}^{*}}{\partial\mathcal{H}_{\text{qf}}^{*}}\right)_{\mathcal{T}^{*}}$,
introducing the scale factor $\Psi^{\left\{ 1f\right\} }$ through
the following linearized equations\begin{equation}
h=\Psi^{\left\{ 1f\right\} }\mathcal{H}_{\text{qf}}^{*}=\Psi^{\left\{ 1f\right\} }\left(\Lambda_{qe}^{*}\right)^{2}\mathcal{H}^{*}\label{1f lin transf RG ordering field (101)}\end{equation}
 \begin{equation}
\begin{array}{cl}
m & =\left(\mathbb{L}^{\left\{ 1f\right\} }\right)^{-d}\left(\Psi^{\left\{ 1f\right\} }\right)^{-1}\left|\mathcal{M}_{\text{qf}}^{*}\right|\\
 & =\left(\mathbb{L}^{\left\{ 1f\right\} }\right)^{-d}\left(\Psi^{\left\{ 1f\right\} }\right)^{-1}\Lambda_{qe}^{*}\left|\mathcal{M}^{*}\right|\end{array}\label{1f lin transf RG order parameter (102)}\end{equation}
where $\Psi^{\left\{ 1f\right\} }$ is a \emph{master} (i.e. unique)
parameter characteristic of the (critical) isothermal line for the
$\left\{ 1f\right\} $-subclass ($\Psi^{\left\{ 1f\right\} }$ has
the same value whatever the sign of the order parameter). From comparison
between either Eqs. (\ref{deltahstar vs deltamutilde (20)}), (\ref{external field scale (23)}),
(\ref{psi from h vs deltahstar (47)}) and (\ref{1f lin transf RG ordering field (101)})
or Eqs. (\ref{deltamstar vs deltarhotilde (21)}), (\ref{order parameter scale (24)}),
(\ref{Ltabfstar (49)}) and (\ref{1f lin transf RG order parameter (102)}),
it is immediate to show that $\chi_{\text{th}}=\left(\mathbb{L}^{\left\{ 1f\right\} }\right)^{-d}\left(\Psi^{\left\{ 1f\right\} }\right)^{-1}\mathcal{X}_{\text{qf}}^{*}$
and, correlatively, to obtain the following expected relation\begin{equation}
\psi_{\rho}=\left(Z_{c}\right)^{-\frac{1}{2}}\Psi^{\left\{ 1f\right\} }\label{ordering scale factor (103)}\end{equation}
The unequivocal link between the scale factors needed, either by the
theoretical description, or by the master description, is given by
Eqs. (\ref{1f ufstar (93)}), (\ref{crossover scale factor (97)})
and (\ref{ordering scale factor (103)}). Therefore, the leading theoretical
and master amplitudes of the susceptibility and the order parameter
are related by the equations :\begin{equation}
\mathbb{\mathcal{Z}}_{\chi}^{\pm}\mathbb{Z}_{\chi}^{\pm}=\left(\mathbb{L}^{\left\{ 1f\right\} }\right)^{d}\left(\Psi^{\left\{ 1f\right\} }\right)^{2}\left(\Theta^{\left\{ 1f\right\} }\right)^{-\gamma}\label{leading khical vs khitab (104)}\end{equation}
 \begin{equation}
\frac{\mathbb{\mathcal{Z}}_{M}}{\mathbb{Z}_{M}}=\left(\mathbb{L}^{\left\{ 1f\right\} }\right)^{d}\Psi^{\left\{ 1f\right\} }\left(\Theta^{\left\{ 1f\right\} }\right)^{\beta}\label{leading Mcal vs Mtab (105)}\end{equation}
 while the leading theoretical and master amplitudes of the correlation
length and the heat capacity are related by the equations :\begin{equation}
\mathbb{\mathcal{Z}}_{\xi}^{\pm}\mathbb{Z}_{\xi}^{\pm}=\left[\mathbb{L}^{\left\{ 1f\right\} }\left(\Theta^{\left\{ 1f\right\} }\right)^{\nu}\right]^{-1}\label{leading ksical vs ksitab (106)}\end{equation}
 \begin{equation}
\frac{\mathbb{\mathcal{Z}}_{C}^{\pm}}{\alpha\mathbb{Z}_{C}^{\pm}}=\left(\mathbb{L}^{\left\{ 1f\right\} }\right)^{d}\left(\Theta^{\left\{ 1f\right\} }\right)^{2-\alpha}\label{leading Ccal vs Ctab (107)}\end{equation}
where the master prefactors $\mathbb{Z}_{C}^{\left\{ 1f\right\} }$
and $\mathbb{Z}_{M}^{\left\{ 1f\right\} }$ are for the heat capacity
case and the order parameter case, respectively {[}see below, Eq.
(\ref{theoretical master function (109)})]. Finally, the characteristic
set\begin{equation}
\mathcal{S}_{SF}^{\left\{ 1f\right\} }=\left\{ \begin{array}{rl}
\Theta^{\left\{ 1f\right\} }= & 4.288\times10^{-3}\\
\mathbb{L}^{\left\{ 1f\right\} }= & 25.6936\\
\Psi^{\left\{ 1f\right\} }= & 1.73847\times10^{-4}\end{array}\right\} \label{Scal1fSF (108)}\end{equation}
 is Ising-like equivalent to the one of Eq. (\ref{Scal1fPSF (94)})
and closes the modifications of the theoretical functions of the $\Phi_{3}\left(1\right)$-class
in order to provide accurate description of the master singular behavior
of the $\left\{ 1f\right\} $-subclass.

Accordingly, each modified function reads as follows\begin{equation}
\mathcal{P}_{\text{qf}}^{\ast}\left(\mathcal{T}^{*}\right)=\mathbb{Z}_{P}^{\left\{ 1f\right\} }F_{P}\left(t\right)\label{theoretical master function (109)}\end{equation}
 with $t=\Theta^{\left\{ 1f\right\} }\mathcal{T}^{*}$ and $F_{P}\left(t\right)$
defined in I. All the master prefactors $\mathbb{Z}_{P}^{\left\{ 1f\right\} }$
can then be calculated using the relations given in part (a), column
3, of Table \ref{Table II}. Within the Ising-like preasymptotic domain,
Eq. (\ref{theoretical master function (109)}) can be approximated
by Eq. (\ref{master two term power law (30)}).

Alternatively but equivalently, each physical property can also be
fitted by the following modified function\begin{equation}
P_{\text{exp}}^{*}\left(\left|\Delta\tau^{*}\right|\right)=\mathbb{P}_{0,\mathcal{L}}^{*}\mathbb{Z}_{P}^{\pm}\left|\Delta\tau^{*}\right|^{-e_{P}}{\displaystyle \prod_{i=1}^{N}\left(1+X_{i,P}^{\pm}t^{D\left(t\right)}\right)^{Y_{i,P}^{\pm}}}\label{theoretical physical function (110)}\end{equation}
 with $t=\vartheta\left|\Delta\tau^{*}\right|=\Theta^{\left\{ 1f\right\} }Y_{c}\left|\Delta\tau^{*}\right|$
and where the function $D\left(t\right)$ {[}see Eq. (\ref{Deff exponent MR universal (44)})]
and the universal quantities $\mathbb{Z}_{P}^{\pm}$, $e_{P}$, $X_{i,P}^{\pm}$,
$Y_{i,P}^{\pm}$, are given in I. All the physical prefactors $\mathbb{P}_{0,\mathcal{L}}^{*}$
can also be calculated using the equations given in part (b), column
3, of Table \ref{Table II}, where the physical prefactors $\mathbb{C}_{0,\mathcal{L}}$
and $\mathbb{M}_{0,\mathcal{L}}$ are for the heat capacity case and
the order parameter case, respectively {[}see Eq. (\ref{theoretical physical function (110)})].

As a summarizing remark related to the schematic Fig. \ref{Figure 2},
the theoretical amplitude set $\mathcal{\mathbb{S}}_{A}^{\left\{ MR\right\} }$
of Eq. (\ref{StabMRA (57)}), the master amplitude set $\mathcal{S}_{A}^{\left\{ 1f\right\} }$
of Eq. (\ref{Scal1fA (40)}), and the physical amplitude set $S_{A}$
of Eq. (\ref{SAfluid (38)}), are unequivocally related only using
$Y_{c}$ and $Z_{c}$ (or $\vartheta$ {[}see Eq. (\ref{crossover scale factor (97)})]
and $\psi_{\rho}$ {[}see Eq. (\ref{ordering scale factor (103)})])
as entry parameters (assuming that $\left(\beta_{c}\right)^{-1}$,
$\alpha_{c}$, and $\Lambda_{qe}^{*}$ are known).

In addition, we can also account for the results of previous analyses
of different singular properties for several one-component fluids
where each master singular behavior is well-fitted by the corresponding
crossover functions in the extended asymptotic domain which corresponds
to $\ell_{\text{qf}}^{\ast}\gtrsim3-4$ (see for example the dashed
arrow labeled {}``EAD'' in Fig. \ref{Figure 1}c, for the susceptibility
case). Indeed, the effective extension $\mathcal{L}_{\text{EAD}}^{+,\left\{ 1f\right\} }$,
where this modified theoretical description seems to be valid, corresponds
to the temperature-like range such as\begin{equation}
\mathcal{T}^{*}\lesssim\mathcal{L}_{\text{EAD}}^{+,\left\{ 1f\right\} }\simeq0.07-0.1\label{LcalEAD1f (111)}\end{equation}
 Equations (\ref{LcalPAD1f (100)}) and (\ref{LcalEAD1f (111)}) are
of crucial importance for experimentalists interested on liquid-gas
critical point phenomena since they are the {}``master'' (experimental)
answer to the unsolved theoretical question: \emph{How large is the
range in which the asymptotic universal features are valid in pure
fluids}? Moreover, when $Q_{c}^{\text{min}}=\left\{ \left(\beta_{c}\right)^{-1},\alpha_{c},Y_{c},Z_{c}\right\} $
and $\Lambda_{qe}^{*}$ are known, we note that each modified crossover
function of Eq. (\ref{theoretical master function (109)}) can act
\emph{beyond} the Ising-like preasymptotic domain, i. e., within the
two-decade range $10^{-2}\lesssim\mathcal{T}^{*}\lesssim1$ corresponding
to the grey areas of Fig. \ref{Figure 1}, to confirm that the critical
Ising-like anomalies characterized by a limited numbers of critical
parameters would dominate in a large range around the liquid-gas critical
point. Such a modified theoretical analysis of the available fluid
data at finite temperature distance appears then similar to the one
initially proposed to provide the first test of the scaling hypothesis
for the one-component fluids by using effective universal equations
of state with only two adjustable dimensionless parameters. As a typical
example, we analyze the isothermal susceptibility for twelve different
fluids in the Appendix B, using the well-known linear model of a parametric
equation of state (eos) \citet{Levelt1978} with $\gamma_{\text{eos}}=1.19$
(and $\beta_{\text{eos}}=0.355$ to close \emph{{}``thermodynamics}''
scaling laws). Furthermore, Eqs. (\ref{LcalPAD1f (100)}) and (\ref{LcalEAD1f (111)})
offer explicit Ising-like criteria to control the development of any
empirical multiparameter equation of state where such a minimal critical
parameter set $Q_{c}^{\text{min}}$ is customarily used (see for example
Ref. \citet{Kiselev2003} and references therein).

\section{Conclusions}

We have shown that the needed information to describe the singular
behavior of one-components fluids within the Ising-like preasymptotic
domain was provided by a minimum set of four scale factors which characterize
the thermodynamics inside the volume of the critical interaction cell.
We have illustrated the Ising-like scaling nature of the scale dilatation
method able to demonstrate the master singular behavior of the one
component fluid subclass. Using the mean crossover function for susceptibility
in the homogeneous phase, which complements a previous study of the
correlation length in the homogeneous phase, we have demonstrated
that the universal features predicted by the massive renormalization
scheme is then accounted for by introducing one common crossover parameter
and appropriate prefactors, only two among the latter being fluid-dependent.
Defining three master constants able to relate the theoretical fields
and the master fields, the corresponding master modifications of the
mean crossover functions were obtained from identification to the
asymptotical master singular behavior of the one-component fluid subclass.
The four critical coordinates which localize the gas-liquid critical
point on the pressure, volume, temperature phase surface provide then
the four scale factors needed to calculate the singular behavior of
any correlation function or thermodynamical property, in a well-controlled
effective extension of the asymptotic critical domain for any one-component
fluid belonging to this subclass, in agreement with the idea first
introduced by one of us. In the case where quantum effects can be
non negligeable, a single supplementary adjustable parameter seems
needed to correctly account for them.

\textbf{Aknowledgements}

The authors are indebted to C. Bervillier for valuable discussion,
constructive comments, and critical reading of the manuscript.

\appendix

\section{Scaling equivalence for a one-parameter crossover modeling within
the preasymptotic domain}

The use of Eq. (\ref{aksiplus vs PADSF (60)}) in the hierachical
Eqs. (\ref{aksiplus vs PADSF (60)}) to (\ref{gammaplus vs PADSF (62)}),
needs that the characteristic scale factor $\vartheta$ is the first
mandatory parameter to be determined, whatever the renormalization
scheme (at $h=0$). For scaling understanding, Eq. (\ref{aksiplus vs PADSF (60)})
must be expressed in the universal form of Eq. (\ref{MR renormalized akhi1+ (74)}),
i.e., such as\begin{equation}
\mathbb{Z}_{\chi}^{1,+}=a_{\chi}^{+}\left[\vartheta^{-\Delta}\right]\label{(A1)}\end{equation}

\begin{table}
\begin{tabular}{|c||c|c|c|c|c|}
\hline 
$M$ & $\Delta$ & $g_{\chi,M}^{1,+}$ & $\left(g_{\chi,M}^{1,+}\right)^{-\frac{1}{\Delta}}$ & $\left(\frac{g_{\chi,M}^{1,+}}{\mathbb{Z}_{\mathcal{X}}^{1,+}}\right)^{\frac{1}{\Delta}}$ & $Ref$\tabularnewline
 &  &  &  & $\left(=\vartheta\Delta\tau_{\chi,M}^{*}\right)$ & \tabularnewline
\hline 
$\text{MSR}$ & $0.504$ & $0.525$ & $3.591$ & $3.9\times10^{-2}$ & \citet{Hahn2001}\tabularnewline
\hline 
$\text{CPM}$ & $0.51$ & $0.590$ & $2.814$ & $4.9\times10^{-3}$ & \citet{Agayan2001}\tabularnewline
\hline
\hline 
 &  & $\mathbb{Z}_{\mathcal{X}}^{1,+}$ & $\left(\mathbb{Z}_{\mathcal{X}}^{1,+}\right)^{-\frac{1}{\Delta}}$ &  & \tabularnewline
\hline 
MR & $0.50189$ & $8.56347$ & $0.013859$ & $1$ & \citet{Garrabos2006gb}\tabularnewline
\hline
\end{tabular}

\caption{Estimated universal values of the confluent exponent (column 2) and
confluent {}``crossover parameter'' (column 3) of the scaling forms
of Eq. (\ref{(A2)}) for the first confluent correction term in the
susceptibility case. Column 1: label $M$ of the different crossover
models (see references given in the last column). Column 5: order
of magnitude for the ratio of the crossover parameters obtained using
MSR or CPM fitting, from reference to the MR fitting (see text).\label{Table III}}

\end{table}

\begin{figure}
\includegraphics[width=80mm,keepaspectratio]{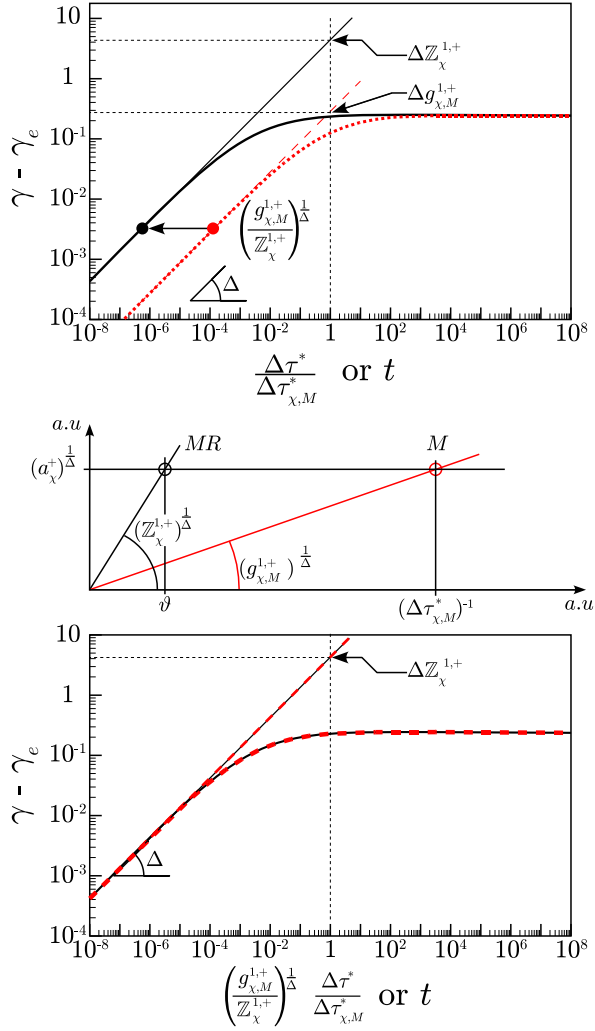}

\caption{(Color online) Schematic illustrations (for $\gamma-\gamma_{e}$ and
$\left(a_{\chi}^{+}\right)^{\frac{1}{\Delta}}$) of the Ising-like
preasymtotic equivalence between the dimensionless crossover temperature
scale $\Delta\tau_{\chi,M}^{*}$ needed by the one-parameter crossover
model $M$ (dashed red line), and the scale factor $\vartheta$ needed
by the massive renormalization (MR) scheme (full black line) {[}see
Eqs. (\ref{MR renormalized akhi1+ (74)}), (\ref{(A2)}), and (\ref{(A3)}),
and Table \ref{Table III}]; The small difference on the respective
$\Delta$ values is neglected.\label{Figure 3}}

\end{figure}

Such a theoretical scaling form of Eq. (\ref{(A1)}) {[}or Eq. (\ref{MR renormalized akhi1+ (74)})]
is then provided from any phenomenological model which use a single
crossover (temperature-like) parameter $\Delta\tau_{\chi,M}^{*}$
related to the (system-dependent) Ginzburg number $G$ (the subscript
$M$ refers to the selected model). Although crossover phenomenon
can be general upon approach of the Ising-like critical point, such
a modeling, in which $G$ is a tunable parameter, is essential to
check carefully its description with the objective to discuss the
shape and the extension of the crossover curves (leading for example
to distinguish a wide variety of Ising-like experimental systems,
including simple fluids, binary liquids, micellar solutions, polymer
mixtures, etc.). However, for the one-component fluid case, our interest
can be restricted to the crossover temperature scale estimated by
three crossover modeling selected in Table \ref{Table III}, i.e.,
i) the massive renormalization scheme (labeled MR) \citet{Bagnuls2002,Garrabos2006gb}
and ii) the minimal subtraction renormalization scheme (labeled MSR)
\citet{MSRscheme,Zhong2003}, both modeling without tunable $G$,
and iii) the parametric model of the equation of state (labeled CPM)
\citet{Agayan2001}, with tunable $G$. The universal form \citet{Agayan2001,Zhong2003,Garrabos2006gb}
of the first confluent amplitude for the susceptibility case, is then
given by the equation\begin{equation}
g_{\chi,M}^{1,+}=a_{\chi}^{+}\left[\left(\Delta\tau_{\chi,M}^{*}\right)^{\Delta}\right]\label{(A2)}\end{equation}
 where $g_{\chi,M}^{1,+}$ is an universal constant given in Table
\ref{Table III}. The differences in the estimates of $g_{\chi,M}^{1,+}$
account for differences in several theoretical aspects: the extension
of the renormalization procedures, the nature of the asymptotic limit
of $\frac{\Delta\tau^{*}}{G}$, the nature of the non universal corrections,
the numerical calculations, etc.. Therefore, we cannot expect practical
understanding from each value given in Table \ref{Table III}. However,
in spite of these numerical differences, the scaling form of Eqs.
(\ref{MR renormalized akhi1+ (74)}), (\ref{(A2)}), and (\ref{(A3)})
provides analytic equivalence between the three models since each
model exactly accounts for the same Ising-like critical crossover
using a single crossover parameter, especially for temperature dependence
of the effective exponent \citet{Kouvel1964}. The crossover temperature
scale $\Delta\tau_{\chi,M}^{*}$ takes a small finite value and can
then be ''comparable'' to $\vartheta$, via the {}``sensor'' $\Delta\tau_{\Delta}^{*}=\frac{t_{0}^{*}}{\vartheta}$
{[}see Eq. (39) in I] of the mean crossover functions (see also Ref.
\citet{Garrabos2006khiT}). As illustrated by the point to point transformations
in Fig. \ref{Figure 3}a and b, and numerical values given in column
5, Table \ref{Table III}, $\Delta\tau_{\chi,M}^{*}$ is then scaled
by $\vartheta$ through the {}``universal'' scaling equation

\begin{equation}
\left(\frac{g_{\chi,M}^{1,+}}{\mathbb{Z}_{\chi}^{1,+}}\right)^{\frac{1}{\Delta}}=\vartheta\Delta\tau_{\chi,M}^{*}=\text{universal}\,\text{cst}\label{(A3)}\end{equation}
where $\Delta\tau_{\chi,\text{MSR}}^{*}=b_{+}^{*}\frac{\mu^{2}}{a}\left(1-\frac{u}{u^{*}}\right)^{\frac{1}{\Delta}}$
for the minimal subtraction renormalization scheme, and $\Delta\tau_{\chi,\text{CPM}}^{*}=\frac{c_{t}^{*}}{\left(\bar{u}\Lambda\right)^{2}}\left(1-\bar{u}\right)^{\frac{1}{\Delta}}$
for crossover parametric model, are the so-called effective Ginzburg
numbers (see the Refs. \citet{Agayan2001,Zhong2003} for the notations
and definitions of the above quantities).

Correlatively but uniquely when Eqs. (\ref{StabMRA vs SAfluidStab1C2M (78)})
or (\ref{(A2)}) are valid (i.e., when the Ising-like critical crossover
is characterized by a single parameter), we must extend the scaling
analysis to the leading amplitudes, expressing again Eqs. (\ref{ksiplus vs PADSF (61)})
and (\ref{gammaplus vs PADSF (62)}) in the {}``universal'' form
of Eq. (\ref{StabMRA vs SAfluid (63)}), i.e. such as :\begin{equation}
\left(\mathbb{Z}_{\xi}^{+}\right)^{-1}=\xi_{0}^{+}\left(g_{0}\vartheta^{\nu}\right)=\xi^{+}\left[\mathbb{L}^{\left\{ 1f\right\} }\vartheta^{\nu}\right]\label{(A4)}\end{equation}
\begin{equation}
\left(\mathbb{Z}_{\chi}^{+}\right)^{-1}=\Gamma^{+}\left[\left(\mathbb{L}^{\left\{ 1f\right\} }\right)^{-d}\left(\psi_{\rho}\right)^{-2}\vartheta^{\gamma}\right]\label{(A5)}\end{equation}
Obviously, as for the confluent amplitude, we can close the asymptotic
identification between the three (MR, MSR, CPM) modeling, introducing
two supplementary universal numbers which relate unequivocally the
scale factors $\mathbb{L}^{\left\{ 1f\right\} }$ and $\psi_{\rho}$
of the massive renormalization scheme, to the equivalent two free
parameters of another crossover approach (see also Refs. \citet{Garrabos2002,Garrabos2006qe}
and the § B3 below).

\section{Effective crossover function beyond the Ising-like preasymptotic
domain}

\subsection{Effective exponent and effective amplitude}

\begin{figure*}
\includegraphics{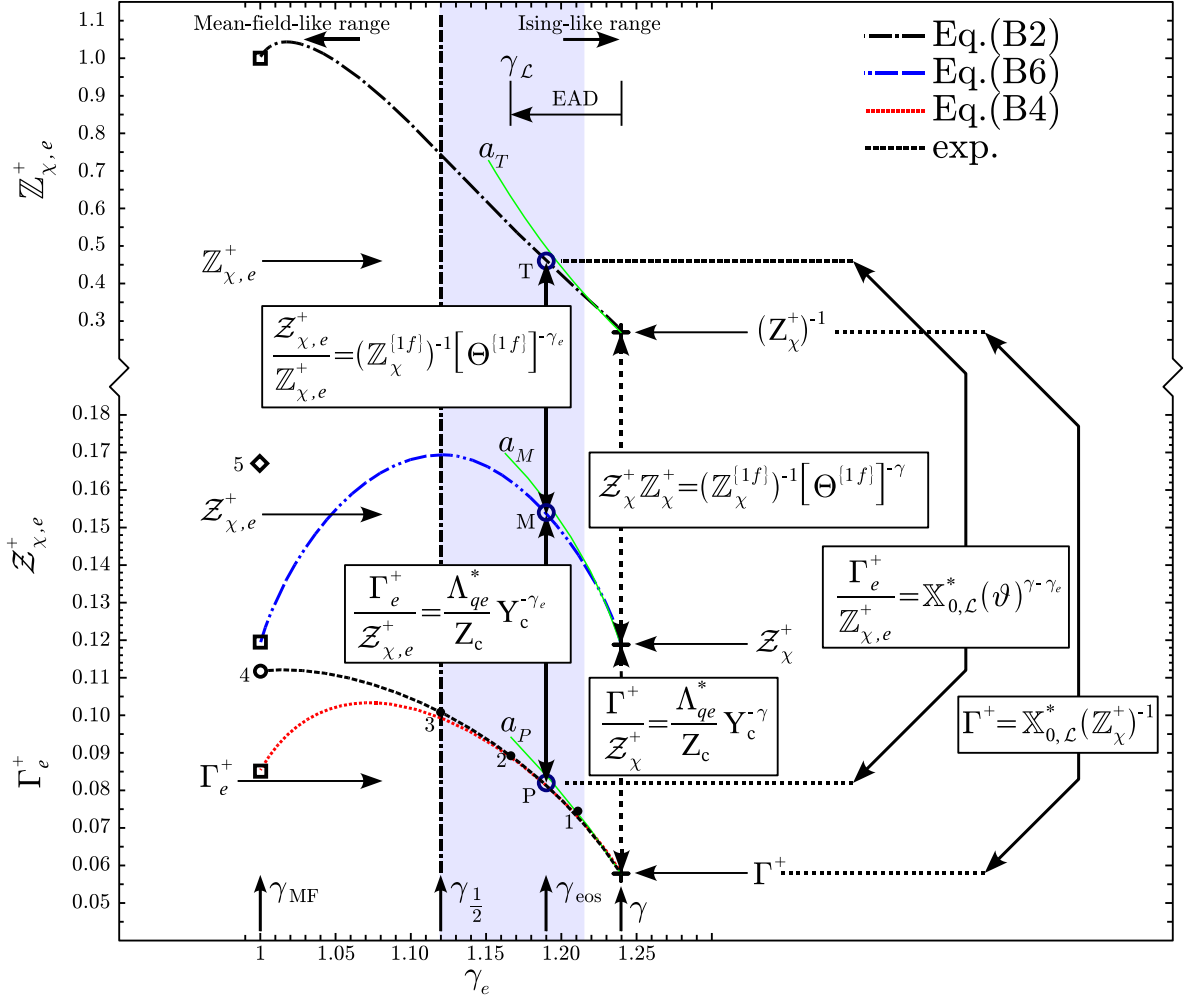}

\caption{(Color online) Theoretical estimations of the effective mean {[}$\mathbb{Z}_{\chi,e}^{+}$,
upper black dot-dashed curve, Eq. (\ref{zkhipluseff MR (B2)})], master
{[}$\mathcal{Z}_{\chi,e}^{+}$, median blue double dot-dashed curve,
Eq. (\ref{Zcalpluseff master (B6)})] and physical (xenon) {[}$\Gamma_{e}^{+}$,
lower red dashed curve, Eq. (\ref{gamapluseff experimental (B4)})]
amplitudes as a function of the effective exponent $\gamma_{e}$ for
the susceptibility case along the critical isochore in the homogeneous
domain; Double (dotted at $\gamma_{e}=\gamma$, full at $\gamma_{e}=\gamma_{\text{eos}}$)
arrays: point-to-point (plusses at $\gamma_{e}=\gamma$, open circles
at $\gamma_{e}=\gamma_{\text{eos}}$, open squares at $\gamma_{e}=\gamma_{\text{MF}}$)
transformations between effective functions using $Y_{c}$ and $Z_{c}$,
or, alternatively but equivalently, $\vartheta$ and $\mathbb{X}_{0,\mathcal{L}}^{*}$
(each relation associated to the transformation at $\gamma$ and $\gamma_{e}$
constant value is illustrated in an attached rectangular box); Lower
black dashed bold curve (labeled exp): $\Gamma_{e}^{+}$from Güttinger
and Cannell's fit for xenon susceptibility \citet{Guttinger1981}
(see also text and Table \ref{Table IV}); M-coordinates $\gamma_{\text{eos}}=1.19$
and $\mathcal{Z}_{\chi,e}^{+}=0.15374$: tangent line at the point
M to the theoretical curve of Eq. (\ref{1f RG c3 fitting eq (84)})
in Fig. \ref{Figure 1}; Others quantities, points, and symbols: see
text. \label{Figure 4}}

\end{figure*}

According to the above asymptotic analysis of the equivalence between
crossover modeling, the scale transformations of the variables which
produce the universal collapse of the Ising-like crossover curves
can be illustrated by using, not only effective exponents \citet{Kouvel1964},
but also effective amplitudes (see also Ref. \citet{Garrabos2006khiT}).
Indeed, from $\chi_{\text{th}}\left(t\right)$ of Eq. (\ref{MR khi vs t (43)}),
the local value of the effective (theoretical) exponent $\gamma_{e,\text{th}}\left(t\right)$
is defined by the equation\begin{equation}
\gamma_{e,\text{th}}\left(t\right)=-\frac{\partial Ln\left[\chi_{\text{th}}\left(t\right)\right]}{\partial Lnt}\label{gammaeth definition (B1)}\end{equation}
 The local value of its attached effective (theoretical) amplitude
$\mathbb{Z}_{e,\chi}^{+}\left(t\right)$ is defined by the equation\begin{equation}
\mathbb{Z}_{\chi,e}^{+}\left(t\right)=\frac{\chi_{\text{th}}\left(t\right)}{t^{-\gamma_{e,th}}}\label{zkhipluseff MR (B2)}\end{equation}
Therefore, $\gamma_{e,\text{th}}\left(t\right)$ and $\mathbb{Z}_{\chi,e}^{+}\left(t\right)$
have equivalent {}``universal'' features as $\chi_{\text{th}}\left(t\right)$.
By eliminating $t$ {[}then simultaneously eliminating the scale factor
$\vartheta_{\mathcal{L}}$ since $t=\vartheta_{\mathcal{L}}\Delta\tau^{*}$],
the classical-to-critical crossover is characterized by a single (i.e.
universal) function $\mathbb{Z}_{\chi,e}^{+}\left(\gamma_{e,\text{th}}\right)$
over the complete range $\gamma_{\text{MF}}\leq\gamma_{e,\text{th}}\left(t\right)\leq\gamma$.
This result is here represented by the top (black dot-dashed) curve
in Fig. \ref{Figure 4}. Its limiting Ising-like critical point takes
{}``universal'' coordinates $\left\{ \gamma;\left(\mathbb{Z}_{\chi}^{+}\right)^{-1}\right\} $
(see the top cross in Fig. \ref{Figure 4}).

In a similar way, from the physical function $\kappa_{T,\text{exp}}^{*}\left(\Delta\tau^{*}\right)$
of Eq. (\ref{khiT vs khizero fitting eq (66)}) which fits the experimental
results using $\vartheta_{\mathcal{L}}$ {[}see Eq. (\ref{thetaLcalli vs tstar (67)})]
and $\mathbb{X}_{0,\mathcal{L}}^{*}$ {[}see Eq. (\ref{1f khitablzero amplitude (96)})],
the local (physical) exponent is defined by\begin{equation}
\gamma_{e,\text{exp}}\left(\Delta\tau^{*}\right)=-\frac{\partial Ln\left[\kappa_{T,\text{exp}}^{*}\left(\Delta\tau^{*}\right)\right]}{\partial Ln\left(\Delta\tau^{*}\right)}\label{gammaeexp definition (B3)}\end{equation}
 and its related local (physical) amplitude by\begin{equation}
\Gamma_{e}^{+}\left(\Delta\tau^{*}\right)=\frac{\kappa_{T,\text{exp}}^{*}\left(\Delta\tau^{*}\right)}{\left(\Delta\tau^{*}\right)^{-\gamma_{e,\text{exp}}}}\label{gamapluseff experimental (B4)}\end{equation}
Eliminating $\Delta\tau^{*}$ from Eqs. (\ref{gammaeexp definition (B3)})
and (\ref{gamapluseff experimental (B4)}), the corresponding physical
function $\Gamma_{e}^{+}\left(\gamma_{e,\text{exp}}\right)$ is represented
in Fig. \ref{Figure 4} by the bottom (red dashed) curve, selecting
xenon as a typical example \citet{Garrabos2006khiT}. Its related
Ising-like critical point takes the physical coordinates $\left\{ \gamma;\Gamma^{+}\right\} $,
as represented by the bottom cross in Fig. \ref{Figure 4} (with $\Gamma^{+}\left(\text{Xe}\right)=0.0578204$).
For quantitative comparison in this {}``physical'' part of Figure
\ref{Figure 4}, we also have represented the experimental lower (black
dashed) curve for $\Gamma_{e}^{+}$ values obtained from the Güttinger
and Cannell's fit of their susceptibility measurements \citet{Guttinger1981}
(bold part of the curve), and from several $pVT$ measurements reported
in Table \ref{Table IV} (full points labeled $1$ to $4$, open circle
labeled P). 

Finally, considering the master singular behavior $\mathcal{X}_{\text{qf}}^{*}\left(\mathcal{T}^{*}\right)$
of Eq. (\ref{1f RG c3 fitting eq (84)}) using $\Theta^{\left\{ 1f\right\} }$
{[}see Eq. (\ref{thetaLcalli vs tstar (67)})] and $\mathbb{Z}_{\chi}^{\left\{ 1f\right\} }$
{[}see Eq. (\ref{1f khitablzero amplitude (96)})], we can define
the local (master) exponent by\begin{equation}
\gamma_{e,1f}\left(\mathcal{T}^{*}\right)=-\frac{\partial Ln\left[\mathcal{X}_{\text{qf}}^{*}\left(\mathcal{T}^{*}\right)\right]}{\partial Ln\left(\mathcal{T}^{*}\right)}\label{gammae1f definition (B5)}\end{equation}
and its related local (master) amplitude by\begin{equation}
\mathbb{\mathcal{Z}}_{\chi,e}^{+}\left(\mathcal{T}^{*}\right)=\frac{\mathcal{X}_{\text{qf}}^{*}\left(\mathcal{T}^{*}\right)}{\left(\mathcal{T}^{*}\right)^{-\gamma_{e,1f}}}\label{Zcalpluseff master (B6)}\end{equation}
After $\mathcal{T}^{*}$ elimination between Eqs. (\ref{gammae1f definition (B5)})
and (\ref{Zcalpluseff master (B6)}), the master function $\mathbb{\mathcal{Z}}_{\chi,e}^{+}\left(\gamma_{e,1f}\right)$
can also be represented by the unique median (blue double dot-dashed)
curve in Fig. \ref{Figure 4}. Its Ising-like critical point takes
the master coordinates $\left\{ \gamma;\mathbb{\mathcal{Z}}_{\chi}^{+}\right\} $,
corresponding to the median cross in Fig. \ref{Figure 4}.

Our main interest can then be focused on the point to point transformation
at constant $\gamma_{e}$ between these three curves, using only two
fluid-dependent parameters, either $\vartheta_{\mathcal{L}}$ and
$\mathbb{X}_{0,\mathcal{L}}^{*}$ for the physical quantities, or
$\Theta^{\left\{ 1f\right\} }$ and $\mathbb{Z}_{\chi}^{\left\{ 1f\right\} }$
for the master quantities. We recall that when $\vartheta_{\mathcal{L}}$
(respectively $\Theta^{\left\{ 1f\right\} }$) and $\mathbb{Z}_{\xi}^{\left\{ 1f\right\} }\equiv\mathbb{L}^{\left\{ 1f\right\} }$
 are known, $\mathbb{X}_{0,\mathcal{L}}^{*}$ gives unequivocal determination
of $\psi_{\rho}$ (respectively $\Psi^{\left\{ 1f\right\} }$). Now,
introducing also $Y_{c}$ and $Z_{c}$, the complete set of the relations
between the - theoretical, master, and physical - amplitudes are summarized
in Fig. \ref{Figure 4}. Consequently, this figure closes the master
description of $\mathcal{X}_{\text{qf}}^{*}\left(\gamma_{e,1f}\right)$
establishing unequivocal link between the three parameter sets $\left\{ \vartheta_{\mathcal{L}};\mathbb{X}_{0,\mathcal{L}}^{*}\right\} $,
$\left\{ \Theta^{\left\{ 1f\right\} };\mathbb{Z}_{\chi}^{\left\{ 1f\right\} }\right\} $,
and $\left\{ Y_{c};Z_{c}\right\} $, and also contains explicit equations
of the schematic links given in Fig. \ref{Figure 2} for the isothermal
susceptibility case {[}with the implicit master condition $\mathbb{Z}_{\xi}^{\left\{ 1f\right\} }\equiv\mathbb{L}^{\left\{ 1f\right\} }=g_{0}\alpha_{c}$
fixing $g_{0}$].

Hereafter we discuss the experimental results obtained at \emph{large}
distance to the critical point, i.e., \emph{beyond} the Ising-like
preasymtotic domain where practical estimations of $\gamma_{e}$ are
significantly different from $\gamma$ (an analysis of the Ising-like
preasymptotic domain very close to the Ising-like limit $\gamma_{e}\rightarrow\gamma$
will be in consideration in Ref. \citet{Garrabos2006khiT}; see also
below § B.3). Especially we focus our present attention on the range
$1.215\gtrsim\gamma_{e}\gtrsim\gamma_{\frac{1}{2}}\approx1.12$ corresponding
to the grey area in Fig. \ref{Figure 4} (obviously equivalent to
the grey area in Fig. \ref{Figure 1}c).

\begin{figure*}
\includegraphics[width=1\textwidth,keepaspectratio]{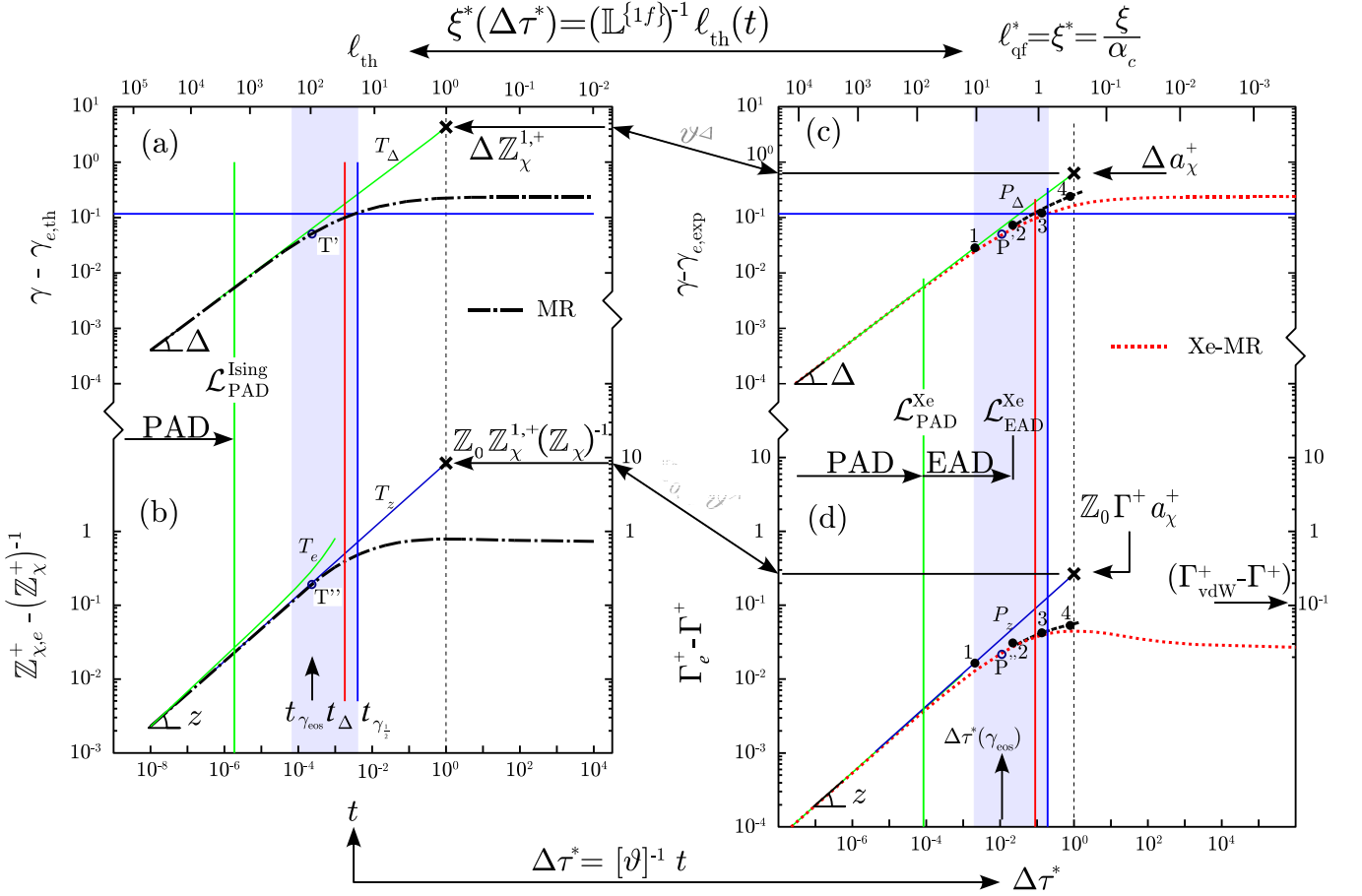}

\caption{(Color online) a) dot-dashed (black) curve (labeled MR): $\gamma-\gamma_{e,\text{th}}$
as a function of $t$, calculated from the theoretical crossover function
of Eq. (\ref{MR khi vs t (43)}) for susceptibility (log-log scale);
(green) line $T_{\Delta}$: limiting singular behavior {[}see Eq.
(\ref{theoretical gamma difference (B18)})] within the Ising-like
preasymptotic domain of extension $t^{*}<\mathcal{L}_{\text{PAD}}^{\text{Ising}}$
{[}vertical green line, see Eq. (\ref{LcalPADMR (56)})]; curve $T_{\Delta}$
of slope $\Delta$ crossing the vertical line $t=1$ (\textbf{x}):
value of the first confluent amplitude of Eq. (\ref{MR Z1khi amplitude (55)});
vertical (blue and pink) lines: $t_{\gamma_{\frac{1}{2}}}$ and $t_{\Delta}$-coordinates
for $\gamma_{e,\text{th}}=\gamma_{\frac{1}{2}}$ and $D\left(t_{\Delta}\right)=\Delta_{\frac{1}{2}}$,
respectively. b) dot-dashed (black) curve (labeled MR): same as a)
for $\left(\mathbb{Z}_{\chi,e}^{+}\right)^{-1}-\left(\mathbb{Z}_{\chi}^{+}\right)^{-1}$
as a function of $t$, calculated from Eq. (\ref{MR khi vs t (43)});
(green and blue) lines $T_{e}$ and $T_{z}$: logarithmic singularity
and power law approximation, {[}see Eqs. (\ref{theoretical Zkhi difference (B14)})
and (\ref{practical Zkhi difference (B15)}), respectively, and text
for detail]; c) and d) dotted (red) curves (labeled Xe-MR): corresponding
xenon quantities $\gamma-\gamma_{e}$ and $\Gamma_{e}^{+}-\Gamma^{+}$
as a function of $\Delta\tau^{*}=\left[\vartheta\left(\text{Xe}\right)\right]^{-1}t$;
points P and 1 to 4 : $pVT$ results (see Ref. \citet{Garrabos2006khiT}
for detail) given in Table \ref{Table IV} (see also Figure \ref{Figure 4});
(green and blue) lines $P_{\Delta}$ and $P_{z}$: xenon counterpart
of the theoretical (green and blue) lines $T_{\Delta}$ and $T_{z}$.
\textbf{x} at $t=\Delta\tau^{*}=1$ : point-to-point transformations
between first confluent amplitudes $\mathbb{Z}_{\chi}^{1,+}$ and
$a_{\chi}^{+}$ (upper arrow) and between leading amplitudes $\left(\mathbb{Z}_{\chi}^{+}\right)^{-1}$
and $\Gamma^{+}$ (lower arrow).\label{Figure 5}}

\end{figure*}

We start with the xenon ($\text{Xe}$) case selected as a standard
one-component fluid. We can then estimate the (theoretical, master
and physical) crossover functions for the correlation length and the
isothermal compressibility of xenon, using $\left\{ Y_{c}=4.91373;\, Z_{c}=0.28601\right\} $
and $\left\{ \vartheta_{\mathcal{L}}=0.021069;\,\mathbb{X}_{0,\mathcal{L}}^{*}=0.214492\right\} $,
(or $\left\{ \vartheta=0.021069;\,\psi_{\rho}=3.2507\times10^{-4};\, g_{0}=29.1473\,\text{nm}^{-1}\right\} $),
with $\alpha_{c}=0.881508\,\text{nm}$ and $\mathbb{L}_{0,\mathcal{L}}^{*}=0.443526$
(for detail, see Ref. \citet{Garrabos2006khiT}). As a basic application,
we can define the correspondence between \emph{theoretical} and \emph{physical}
temperature range and between \emph{theoretical} and \emph{physical}
correlation length range for description of either $\gamma-\gamma_{e,\text{th}}$
and $\mathbb{Z}_{\chi,e}^{+}-\left(\mathbb{Z}_{\chi}^{+}\right)^{-1}$
as a function of $t$ and as a function of $\ell_{\text{th}}$ , or
$\gamma-\gamma_{e,\text{exp}}$ and $\Gamma_{e}^{+}-\Gamma^{+}$ as
a function of $\Delta\tau^{*}$ and as a function of $\xi^{*}$ .
Each respective result is illustrated by a (black dot-dashed or red
doted) curve in each part a to d of Fig. \ref{Figure 5}. Now, the
grey areas in Fig. \ref{Figure 5} correspond, either to the theoretical
ranges $10^{-4}\lesssim t\lesssim2\times10^{-3}$ (bottom axis) and
$180\gtrsim\ell_{\text{th}}\gtrsim18$ (top axis) in parts a and b,
or the physical (xenon) ranges $5\times10^{-3}\lesssim\Delta\tau^{*}\lesssim10^{-1}$
(bottom axis) and $10.5\gtrsim\xi^{*}\gtrsim0.73$ (top axis) in parts
c and d.

\begin{table*}
\begin{tabular}{|c|c|c|c||c|c|}
\hline 
 & $\gamma_{e,pVT}$  & $\Gamma_{e,pVT}^{+}$  & $\left\langle \Delta\tau_{pVT}^{*}\right\rangle $  & $\Delta\tau_{\text{th}}^{*}\left(\gamma_{e,\text{th}}\right)$  & $\Gamma_{e,\text{th}}^{+}$\tabularnewline
\hline 
 &  &  & $\left\langle \Delta\tau_{pVT}^{*}\right\rangle =\sqrt{\Delta\tau_{\text{min}}^{*}\Delta\tau_{\text{max}}^{*}}$  & $\gamma_{e,\text{th}}\left(t\right)=\gamma_{e,pVT}\left(\Delta\tau^{*}\right)$  & \tabularnewline
\hline
\hline 
$1$  & $1.211\pm0.01$  & $0.0743\pm0.015$  & $2.07\times10^{-3}$  & $2.95\times10^{-3}$  & $0.07263$\tabularnewline
\hline 
$2$  & $1.16665$  & $0.089$  & $2.24\times10^{-2}$  & $3.338\times10^{-2}$  & $0.08859$\tabularnewline
\hline 
$3$  & $1.1198\left(=\gamma_{\frac{1}{2}}\right)$  & $0.101$  & $1.21\times10^{-1}$  & $1.928\times10^{-1}$  & $0.09960$\tabularnewline
\hline 
$4$  & $1\left(=\gamma_{\text{MF}}\right)$  & $0.11$  & $7.1\,10^{-1}$  & $\infty$  & $0.08507$\tabularnewline
\hline
\hline 
$5\left(\text{vdW}\right)$  & $1\left(=\gamma_{\text{vdW}}\right)$  & $\frac{1}{6}\left(=\Gamma_{\text{vdW}}^{+}\right)$  &  & $\infty$  & $0.08507$\tabularnewline
\hline 
$P\left(\text{eos}\right)$  & $1.19\left(=\gamma_{\text{eos}}\right)$  & $0.0793\left(=\Gamma_{\text{eos}}^{+}\right)$  & $1.13\times10^{-2}$  & $1.135\times10^{-2}$  & $0.08084$\tabularnewline
\hline
\end{tabular}

\caption{Column 1: index of the points in Figs. \ref{Figure 4} and \ref{Figure 5};
Columns 2 and 3: Effective power law description of $pVT$ measurements
in xenon (see Ref. \citet{Garrabos2006khiT} for detail and data sources);
Columns 4 to 6: calculated values of the (geometrical) mean temperature
$\left\langle \Delta\tau_{pVT}^{*}\right\rangle =\sqrt{\Delta\tau_{\text{min}}^{*}\Delta\tau_{\text{max}}^{*}}$
of $pVT$ measurements (column 4), theoretical local temperature $\Delta\tau_{\text{th}}^{*}\left(\gamma_{e,\text{th}}\right)$
satisfying the condition $\gamma_{e,\text{th}}\left(t\right)=\gamma_{e,pVT}\left(\Delta\tau^{*}\right)$
(column 5), and theoretical local amplitude $\Gamma_{e,\text{th}}^{+}$
for $\gamma_{e,\text{th}}\left(t\right)=\gamma_{e,pVT}\left(\Delta\tau^{*}\right)$
(column 6).\label{Table IV}}

\end{table*}

The following compares these theoretical predictions to the $\gamma_{e,pVT}$
and $\Gamma_{e,pVT}^{+}$ values obtained from $pVT$ measurements
\citet{Beattie1951,Weinberger1952,Habgood1954,Michels1954,Rabinovich1973}
(see also details in Ref. \citet{Garrabos2006khiT}). We recall that
the $pVT$ measurements were performed at \emph{finite} distance to
the critical point, such that the $\kappa_{T,pVT}^{*}$ data obtained
from $pVT$ data can be fitted by an effective power law\begin{equation}
\kappa_{T,pVT}^{*}=\Gamma_{e,pVT}^{+}\left(\Delta\tau^{*}\right)^{-\gamma_{e,pVT}}\label{kapaT pVT power law (B7)}\end{equation}
only valid in a restricted temperature range defined by $\Delta\tau_{\text{min}}^{*}\leq\Delta\tau^{*}\leq\Delta\tau_{\text{max}}^{*}$.
The measured (exponent and amplitude) parameters $\left\{ \gamma_{e,pVT};\Gamma_{e,pVT}^{+}\right\} $
are then associated to the temperature range $\left\{ \Delta\tau_{\text{min}}^{*};\Delta\tau_{\text{max}}^{*}\right\} $
of central value $\left\langle \Delta\tau_{e,pVT}^{*}\right\rangle =\sqrt{\Delta\tau_{\text{min}}^{*}\Delta\tau_{\text{max}}^{*}}$
(in log scale) located \emph{beyond} the Ising like preasymptotic
domain. Therefore, we can represent these results by points of respective
coordinates $\left\{ \gamma_{e,pVT};\Gamma_{e,pVT}^{+}\right\} $,
$\left\{ \gamma_{e,pVT};\left\langle \Delta\tau_{pVT}^{*}\right\rangle \right\} $
and $\left\{ \Gamma_{e,pVT}^{+};\left\langle \Delta\tau_{pVT}^{*}\right\rangle \right\} $
in each appropriate binary diagram.

The four points (labeled $1$ to $4$) illustrated in Figs. \ref{Figure 4},
\ref{Figure 5}c and \ref{Figure 5}d, correspond to the xenon results
reported on lines labeled $1$ to $4$, respectively, of Table \ref{Table IV}.
The points labeled $1$ and $2$ follow the general trend of the theoretical
curves. This result confirms that, in spite of a large correlated
error-bar in the adjustable exponent and amplitude parameters, the
variations of their respective central values agree with a two-parameter
description within the {}``Ising-like'' side of the crossover domain
where $\gamma>\gamma_{e,pVT}>1.17$ . However, the point labeled $3$,
and more significantly the point labeled $4$, show that the $pVT$
experimental results are not in agreement with the mean-field behavior
predicted by the crossover function within the {}``mean-field-like''
side where $\gamma_{\frac{1}{2}}\gtrsim\gamma_{e,pVT}>\gamma_{\text{MF}}$.
The failure of the classical corresponding state theory is also illustrated
by the point labeled $5$ in Fig. \ref{Figure 4}, which corresponds
to the result obtained from the van der Waals equation of state {[}see
the line labeled $5\left(\text{vdW})\right)$ in Table \ref{Table IV}].

To translate the $\mathcal{L}_{\text{EAD}}^{+,\left\{ 1f\right\} }$-
\emph{master} value {[}Eq. (\ref{LcalEAD1f (111)})] in a $\gamma_{\mathcal{L}}$-
\emph{master} value which delimits the effective range of the extended
asymptotic domain in Fig. \ref{Figure 4}, one needs to consider the
upper horizontal axis of Figs. \ref{Figure 5}c and \ref{Figure 5}d
which measures the \emph{master} correlation length $\xi^{*}=\frac{\xi}{\alpha_{c}\left(\text{Xe}\right)}$
{[}i.e. the dimensionless ratio which compares the size of the critical
fluctuation to the actual range of the microscopic interaction, with
$\Lambda_{qe}^{*}\left(\text{Xe}\right)=1$ in xenon case]. As a matter
of fact, the value $\mathcal{L}_{\text{EAD}}^{\text{Xe}}\approx2\times10^{-2}$
corresponds to the value $\frac{\mathcal{L}_{\text{EAD}}^{+,\left\{ 1f\right\} }}{Y_{c}\left(\text{Xe}\right)}$
where $\xi^{*}\approx3$. Therefore, the associated \emph{local} value
is $\gamma_{\mathcal{L}}\approx1.16-1.17$. This value descriminates
the {}``non Ising-like'' range $\gamma_{e}<\gamma_{\mathcal{L}}$
(including the value $\gamma_{\frac{1}{2}}=\frac{\gamma+\gamma_{\text{MF}}}{2}\approx1.12$)
where the effective classical-to-critical crossover for xenon is no
longer accounted for by the theoretical crossover function, as shown
in Fig. \ref{Figure 4} where it is observed an increasing difference
between the curves labeled $\text{Xe}$-MR and the dotted curve labeled
$\text{Xe}$-exp when $\gamma_{e}\rightarrow\gamma_{\text{MF}}=1$.

Accounting for an extended (Ising-like) asymptotic domain defined
by $\gamma_{e}<\gamma_{\mathcal{L}}$, we are also able to revisit
the results previously obtained using an universal scaled form of
the equation of state with {}``universal'' values of the exponents
significantly different to the {}``Ising'' ones. As a typical example,
the xenon results obtained from the restricted linear model of a parametric
equation of state with $\gamma_{\text{eos}}=1.19$ {[}see the line
labeled $P\left(\text{eos}\right)$ in Table \ref{Table IV}] are
in excellent agreement with the theoretical crossover function, as
illustrated by the point labeled P in Fig. \ref{Figure 4}. Moreover,
using as a $\Delta\tau^{*}$-coordinate the theoretical value $\Delta\tau_{\text{th}}^{*}\left(\gamma_{\text{eos}}\right)=1.135\times10^{-2}$
{[}Eq. (\ref{gammaeexp definition (B3)})], we can show that these
results are also well accounted for in the theoretical temperature
dependence (see the corresponding points labeled P' and P'' in Figs.
\ref{Figure 5}c and \ref{Figure 5}d, respectively). Such results
confirm that \emph{two} xenon-parameters involved in an universal
form of the equation of state can be used as Ising-like characteristic
factors to be related to the two scale factors $Y_{c}$ and $Z_{c}$,
as illustrated in the next Section for the case of the linear parametric
equation of state.

\subsection{Master crossover provided by a restricted linear model of a parametric
equation of state}

In the seventies, first analyses of the two-scale factor universality
for one-component fluids used effective scaled forms of the equation
of state (eos) to fit the $pVT$ data measured at \emph{finite} distance
to the critical point (for detail see Refs. \citet{Green1967,Levelt1974,Levelt1975,Levelt1976,Levelt1978,Levelt1981,Linearmodelparameter,Vicentini1969}).
Such a thermodynamic approach of universality was based on a limited
number of characteristic parameters for each pure fluid, using \emph{effective
universal values} for the critical exponents. We limit the present
purpose to the well-known restricted linear model of the parametric
equation of state \citet{Linearmodelparameter}, with application
to several different fluids \citet{Levelt1975}. The two main interests
for such a choice are the following:

i) The effective thermodynamic exponents have been precisely fixed
at (non Ising) values of $\gamma_{\text{eos}}=1.190$, $\beta_{\text{eos}}=0.355$,
$\alpha_{\text{eos}}=0.100$ (the subscript $\text{eos}$ recalls
the origin of these effective values). As shown in Fig. \ref{Figure 4},
the value $\gamma_{\text{eos}}=1.190$ is precisely within the selected
$\gamma_{e}$ range \emph{beyond} the Ising-like preasymptotic domain,
but well \emph{inside} the extended asymptotic domain $\gamma_{e}<\gamma_{\mathcal{L}}$,
which corresponds to $\Delta\tau^{*}\lesssim\mathcal{L}_{\text{EAD}}^{\text{Xe}}$;

ii) The effective values of the thermodynamic amplitude $\Gamma_{\text{eos}}^{+}$
{[}see below Eq. (\ref{kapaT eos power law (B8)})] of the isothermal
compressibility were then obtained only using two adjustable (fluid
dependent) parameters (namely $k$ and $a$), which are the two characteristic
parameters involved in the scaled equation of state. As shown in Fig.
\ref{Figure 4}, {}``equivalent'' values of $\Gamma_{e}^{+}\left(\gamma_{\text{eos}}\right)$
at $\gamma_{\text{eos}}=1.190$ can be simultaneously obtained by
a scale transformation between the point P (on the physical curve)
and the point M (on the master curve) which also involves only two
characteristic parameters (namely $Y_{c}$ and $Z_{c}$) {[}admitting
that the parameter $\Lambda_{qe}^{*}$ which accounts for quantum
effects is known].

Therefore, both in quantity (two), and in nature (Ising like), the
fluid dependent parameters $k$ and $a$ appear {}``equivalent''
to $Y_{c}$ and $Z_{c}$, except the noticeable distinction of their
respective determination, outside the Ising like preasymptotic domain
for the $\left\{ k;a\right\} $ pair, asymptotically close to the
critical point for the $\left\{ Y_{c};Z_{c}\right\} $ pair.

Now we compare the respective values of $\Gamma_{\text{eos}}^{+}$
and $\Gamma_{e}^{+}\left(\gamma_{\text{eos}}\right)$ for twelve selected
fluids. From the linear model of the parametric equation of state,
Eq. (\ref{kapaT pVT power law (B7)}) can be rewritten as\begin{equation}
\kappa_{T}^{*}=\Gamma_{\text{eos}}^{+}\left(\Delta\tau^{*}\right)^{-\gamma_{\text{eos}}}\label{kapaT eos power law (B8)}\end{equation}
 where $\Gamma_{\text{eos}}^{+}$ is related to the characteristic
parameters $k$ and $a$ as follows\begin{equation}
\Gamma_{\text{eos}}^{+}=\frac{k}{a}\label{Gamaplus linear model eos (B9)}\end{equation}
 Considering then the restricted form of the linear model such as
analyzed in Ref. \citet{Levelt1975}, $k$ can be estimated from the
relation\begin{equation}
k=\left(\frac{x_{0}}{b_{\text{SLH}}^{2}-1}\right)^{-\beta_{\text{eos}}}\label{k vs xzero (B10)}\end{equation}
 where $b_{\text{SLH}}^{2}=1.3908$ is an universal quantity while
$x_{0}$ is a fluid-dependent parameter related to the value of the
effective amplitude of the coexistence curve (associated to the value
$\beta_{\text{eos}}=0.355$ of the effective exponent). The values
of $x_{0}$ and $a$ can be found in Ref. \citet{Levelt1975}. They
are reported with the corresponding $k$ values in Table \ref{Table V}
(columns $2$ to $4$, respectively) for the selected twelve fluids
(column 1). The related values of $\Gamma_{\text{eos}}^{+}$ obtained
by using Eq. (\ref{Gamaplus linear model eos (B9)}) are given in
Table \ref{Table V} (column $5$).

\begin{table*}
\begin{tabular}{|c||c|c|c|c|c|c|c|c|c|}
\hline 
Fluid  & $x_{0}$  & $a$  & $k$  & $\Gamma_{\text{eos}}^{+}$  & $Y_{c}$  & $Z_{c}$  & $\Gamma_{e}^{+}\left(\gamma_{\text{eos}}\right)$  & $\Delta\tau^{*}\left(\gamma_{\text{eos}}\right)$  & $r\%(\Gamma_{e}^{+})$\tabularnewline
\hline 
 & \citet{Levelt1975}  & \citet{Levelt1975}  & Eq. (\ref{k vs xzero (B10)})  & Eq. (\ref{Gamaplus linear model eos (B9)})  &  &  & Eq. (\ref{Gamapluseos vs scale transformation (B11)})  & Eq. (\ref{deltataustargamaeos (B13)})  & \tabularnewline
\hline
\hline 
$^{3}$He$^{\left(*\right)}$  & $0.489$  & $4.63$  & $0.9235$  & $0.1995$  & $2.3984$  & $0.30129$  & $0.20003$$^{\left(*\right)}$  & $2.326\,10^{-2}$  & $-0.283$\tabularnewline
\hline 
Ar  & $0.183$  & $16.5$  & $1.309$  & $0.07934$  & $4.3288$  & $0.2896$  & $0.09284$  & $1.289\,10^{-2}$  & $-14.5$\tabularnewline
\hline 
Kr & $0.183$  & $16.5$  & $1.309$  & $0.07934$  & $4.9437$  & $0.2913$  & $0.07887$  & $1.128\,10^{-2}$  & $0.6$\tabularnewline
\hline 
Xe  & $0.183$  & $16.5$  & $1.309$  & $0.07934$  & $4.9137$  & $0.2860$  & $0.08084$  & $1.135\,10^{-2}$  & $-1.86$\tabularnewline
\hline 
O$_{2}$ & $0.183$  & $15.6$  & $1.309$  & $0.08392$  & $4.9864$  & $0.28797$  & $0.07890$  & $1.119\,10^{-2}$  & $6.36$\tabularnewline
\hline 
N$_{2}$ & $0.164$  & $18.2$  & $1.361$  & $0.07478$  & $5.3701$  & $0.28887$  & $0.07201$  & $1.039\,10^{-2}$  & $3.84$\tabularnewline
\hline 
CH$_{4}$ & $0.164$  & $17.0$  & $1.361$  & $0.08006$  & $4.9838$  & $0.28678$  & $0.07928$  & $1.119\,10^{-2}$  & $0.99$\tabularnewline
\hline 
C$_{2}$H$_{4}$ & $0.166$  & $17.5$  & $1.355$  & $0.07744$  & $5.3487$  & $0.2813$  & $0.07431$  & $1.043\,10^{-2}$  & $4.22$\tabularnewline
\hline 
CO$_{2}$ & $0.141$  & $21.8$  & $1.436$  & $0.06587$  & $6.0104$  & $0.27438$  & $0.06631$  & $0.928\,10^{-2}$  & $0.653$\tabularnewline
\hline 
NH$_{3}$ & $0.109$  & $21.4$  & $1.361$  & $0.07353$  & $6.3019$  & $0.24294$  & $0.07079$  & $0.885\,10^{-2}$  & $3.88$\tabularnewline
\hline 
H$_{2}$O & $0.100$  & $22.3$  & $1.622$  & $0.07275$  & $6.8552$  & $0.22912$  & $0.0679$  & $0.814\,10^{-2}$  & $7.14$\tabularnewline
\hline 
D$_{2}$O & $0.100$  & $22.3$  & $1.622$  & $0.07275$  & $7.0728$  & $0.22783$  & $0.0658$  & $0.799\,10^{-2}$  & $10.6$\tabularnewline
\hline
\end{tabular}

\caption{Two-parameter universality of the effective amplitude of the isothermal
compressibility estimated from the linear model of a parametric equation
of state and the master modification of the theoretical function.\label{Table V}}

\end{table*}

The unequivocal scale transformation between the points P and M is
given by the relation (see Fig. \ref{Figure 4})\begin{equation}
\Gamma_{e}^{+}(\gamma_{\text{eos}})=\mathbb{\mathcal{Z}}_{\chi,e}^{+}\left(\gamma_{\text{eos}}\right)\frac{\Lambda_{qe}^{*}\left(Y_{c}\right)^{-\gamma_{eos}}}{Z_{c}}\label{Gamapluseos vs scale transformation (B11)}\end{equation}
 where $\mathbb{\mathcal{Z}}_{\chi,e}^{+}\left(\gamma_{\text{eos}}\right)$
is the effective master amplitude for the $\mathcal{T}^{*}\left(\gamma_{e,1f}\right)$-value
satisfying the condition $\gamma_{e,1f}=\gamma_{\text{eos}}=1.19$.
For practical use of Eq. (\ref{Gamapluseos vs scale transformation (B11)}),
the crucial advantage is given by the uniquivocal scale transformation
between the points T (on the theoretical curve) and M (on the master
curve) illustrated in Fig. \ref{Figure 4}. That provides immediately
$\mathbb{\mathcal{Z}}_{\chi,e}^{+}\left(\gamma_{\text{eos}}\right)=\mathbb{\mathcal{\mathbb{Z}}}_{\chi,e}^{+}\left(\gamma_{\text{eos}}\right)\left[\mathbb{Z}_{\chi}^{\left\{ 1f\right\} }\left(\Theta^{\left\{ 1f\right\} }\right)^{\gamma_{\text{eos}}}\right]^{-1}$
and $\mathcal{T}^{*}\left(\gamma_{\text{eos}}\right)=t\left(\gamma_{\text{eos}}\right)\left(\Theta^{\left\{ 1f\right\} }\right)^{-1}$.
Using Eqs. (\ref{gammaeth definition (B1)}) and (\ref{zkhipluseff MR (B2)}),
the theoretical function $\chi_{\text{th}}\left(t\right)$ leads to
the values $t\left(\gamma_{\text{eos}}\right)=2.392\times10^{-4}$
and $\mathbb{\mathcal{\mathbb{Z}}}_{\chi,e}^{+}\left(\gamma_{\text{eos}}\right)=0.456414$
related to coordinates of the points labeled T' and T'' in Figs.
\ref{Figure 5}a and \ref{Figure 5}b, respectively. Using the numerical
values of $\mathbb{Z}_{\chi}^{\left\{ 1f\right\} }$ and $\Theta^{\left\{ 1f\right\} }$
given in Table \ref{Table II}, we obtain $\mathcal{T}^{*}\left(\gamma_{\text{eos}}\right)=5.579\times10^{-2}$
and $\mathbb{\mathcal{Z}}_{\chi,e}^{+}\left(\gamma_{\text{eos}}\right)=0.15374$.
Subsidiarily, in Fig. \ref{Figure 1}c, we note that the master curve
$\chi_{\text{qf}}^{*}\left(\mathcal{T}^{*}\right)$ of Eq. (\ref{1f RG c3 fitting eq (84)})
has a tangent curve of slope $-\gamma_{\text{eos}}$ at the point
M of $\mathcal{T}^{*}\left(\gamma_{\text{eos}}\right)$-coordinate
which corresponds to the effective power law\begin{equation}
\mathbb{\mathcal{X}}_{\text{eos}}^{+}\left(\mathcal{T}^{*}\right)=0.15374\left(\mathcal{T}^{*}\right)^{-1.19}\label{eos tangent line (B12)}\end{equation}

The values of $Y_{c}$ and $Z_{c}$ for the selected fluids are reported
in Table \ref{Table V} (columns $6$ and $7$, respectively). The
estimated values of $\Gamma_{e}^{+}(\gamma_{\text{eos}})$ using Eq.
(\ref{Gamapluseos vs scale transformation (B11)}) are given in Table
\ref{Table V} (column $8$). Each physical curve $\kappa_{T}^{*}\left(\Delta\tau^{*}\right)$
of Eq. (\ref{khiT vs khizero fitting eq (66)}) have a tangent curve
of slope $-\gamma_{\text{eos}}$ at the point of $\Delta\tau^{*}$-coordinate:\begin{equation}
\Delta\tau^{*}\left(\gamma_{\text{eos}}\right)=\frac{\mathcal{T}^{*}\left(\gamma_{\text{eos}}\right)}{Y_{c}}\label{deltataustargamaeos (B13)}\end{equation}
 (see column $9$ of Table \ref{Table V}), which corresponds to the
effective power law $\Gamma_{e}^{+}\left(\Delta\tau^{*}\right)=\Gamma_{e}^{+}(\gamma_{\text{eos}})\left(\Delta\tau^{*}\right)^{-\gamma_{\text{eos}}}$
.

The residuals $r\%(\Gamma_{e}^{+})=100\left(\frac{\Gamma_{\text{eos}}^{+}}{\Gamma_{e}^{+}(\gamma_{\text{e}\text{os}})}-1\right)$
(see column $10$, Table \ref{Table V}), generally lower than the
typical experimental uncertainty estimated to $10\%$, confirm that
the universal features observed beyond the Ising-like preasymptotic
domain but within the Ising-like extended asymptotic domain, i.e.,
$\Delta\tau^{*}\lesssim\frac{t_{\text{EAD}}^{+}}{\vartheta}=\frac{\mathcal{L}_{\text{EAD}}^{+,\left\{ 1f\right\} }}{Y_{c}}=\mathcal{L}_{\text{EAD}}^{f}$
with $t_{\text{EAD}}^{+}=\Theta^{\left\{ 1f\right\} }\mathcal{L}_{\text{EAD}}^{+,\left\{ 1f\right\} }$
and $\mathcal{L}_{\text{EAD}}^{+,\left\{ 1f\right\} }\simeq0.07-0.1$,
are well-characterized by the two critical scale factors $Y_{c}$
and $Z_{c}$ of each fluid $f$.

\subsection{{}``Universal'' approximation of the logarithmic singularity of
effective amplitudes}

Another practical application of the point to point transformations
given in Fig. \ref{Figure 4} can be obtained focusing our attention
on the logarithmic singularity of any first derivative $\left(\frac{\partial\mathbb{Z}_{P,e}^{+}}{\partial e_{P,e}}\right)_{e_{P,e}\rightarrow e_{P}}$
close to the Ising-like critical point, for any effective amplitude
power law $\mathbb{Z}_{P,e}^{+}\left(t\right)=\frac{F_{P}\left(t\right)}{t^{-e_{P,e}}}$
estimated from any crossover function $F_{P}\left(t\right)$ given
in Ref. \citet{Garrabos2006gb} (with $e_{P,e}\left(t\right)=-\frac{\partial Ln\left[F_{P}\left(t\right)\right]}{\partial Lnt}$)
(see Refs. \citet{Garrabos2006gb,Garrabos2006khiT} for detail). For
the susceptibility case, the logarithmic singularity of $\left(\frac{\partial\mathbb{Z}_{\chi,e}^{+}}{\partial\gamma_{e,th}}\right)_{\gamma_{e,th}\rightarrow\gamma}$
extrapolated beyond the Ising-like preasymptotic domain is illustrated
by the curves labeled $a_{T}$, $a_{M}$, and $a_{P}$ in Fig. \ref{Figure 4}.
For better evaluation within the preasymptotic domain, the related
amplitude singularity in terms of the thermal field dependence is
given in Fig. \ref{Figure 5}b for example by the curve $T_{e}$ of
equation\begin{equation}
\begin{array}{cl}
\mathbb{Z}_{\chi,e}^{+}-\left(\mathbb{Z}_{\chi}^{+}\right)^{-1}= & \left(\mathbb{Z}_{\chi}^{+}\right)^{-1}\mathbb{Z}_{\chi}^{1,+}t^{\Delta\mathbb{Z}_{\chi}^{1,+}t^{\Delta}}\\
 & \times\left[1-log\left(t^{\Delta}\right)\right]t^{\Delta}\end{array}\label{theoretical Zkhi difference (B14)}\end{equation}
 We can approximate Eq. (\ref{theoretical Zkhi difference (B14)})
by the following {}``universal'' power law\begin{equation}
\mathbb{Z}_{\chi,e}^{+}-\left(\mathbb{Z}_{\chi}^{+}\right)^{-1}=\mathbb{Z}_{0}\left(\mathbb{Z}_{\chi}^{+}\right)^{-1}\mathbb{Z}_{\chi}^{1,+}t^{z}\label{practical Zkhi difference (B15)}\end{equation}
 where $\mathbb{Z}_{0}=3.7\pm0.1$ and $z=0.45\pm0.035$ are independent
of the property and the domain. The exponent condition $z<\Delta$,
leading to the $\frac{z}{\Delta}=1-u<1$, is conform to the logarithmic
singularity of the first derivative $\left(\frac{\partial\mathbb{Z}_{\chi,e}^{+}}{\partial\gamma_{e,\text{th}}}\right)_{\gamma_{e,\text{th}}\rightarrow\gamma}$
here approximated by a power law $\left(\frac{\partial\mathbb{Z}_{\chi,e}^{+}}{\partial\gamma_{e,\text{th}}}\right)_{\gamma_{e,\text{th}}\rightarrow\gamma}\propto\left(\gamma-\gamma_{e,\text{th}}\right)^{-u}$.
For practical use, we arbitrarily choose $u=\alpha$, leading to define
$z=\Delta\left(1-\alpha\right)$. The validity of this approximation
is illustrated by the curve $T_{z}$ in Figure \ref{Figure 5}b.

Correspondingly, in Figure \ref{Figure 5}d, the physical asymptotic
representation of $\Gamma_{e}^{+}-\Gamma^{+}$ is now approximated
by the curve $P_{z}$ of asymptotic equation \begin{equation}
\begin{array}{cl}
\Gamma_{e}^{+}-\Gamma^{+} & =\mathbb{Z}_{0}\Gamma^{+}a_{\chi}^{+}\left(\Delta\tau^{*}\right)^{z}\\
 & =\mathbb{X}_{0,\mathcal{L}}\mathbb{Z}_{0}\mathbb{Z}_{\chi}^{1,+}\left(\mathbb{Z}_{\mathcal{X}}^{+}\right)^{-1}\vartheta^{\Delta}\left(\Delta\tau^{*}\right)^{z}\end{array}\label{experimental Gammaplus difference (B16)}\end{equation}
 Using Eqs. (\ref{practical Zkhi difference (B15)}) and (\ref{experimental Gammaplus difference (B16)})
at $t=\Delta\tau^{*}=1$, we obtain\begin{equation}
\Gamma_{e}^{+}\left(1\right)-\Gamma^{+}=\mathbb{X}_{0,\mathcal{L}}^{*}\vartheta^{\Delta}\left[\mathbb{Z}_{\chi,e}^{+}\left(1\right)-\left(\mathbb{Z}_{\chi}^{+}\right)^{-1}\right]\label{Practical Gammaplus difference (B17)}\end{equation}
 The point to point {}``universal'' transformation which approximates
the logarithmic singularity is then illustrated by the two correlated
points (symbol \textbf{x}) at $t=\Delta\tau^{*}=1$, in Figures \ref{Figure 5}b
and \ref{Figure 5}d, respectively. As expected from Figure \ref{Figure 4},
this transformation is given by the product $\mathbb{X}_{0,\mathcal{L}}^{*}\vartheta^{\Delta}$.

Obviously, to close the asymptotic behavior within the Ising-like
preasymtotic domain we can also consider the respective asymptotic
curves labeled $T_{\Delta}$ and $P_{\Delta}$ in Figs. \ref{Figure 5}a
and \ref{Figure 5}c of equations\begin{equation}
\gamma-\gamma_{e,\text{th}}=\Delta\,\mathbb{Z}_{\chi}^{1,+}t^{\Delta}\label{theoretical gamma difference (B18)}\end{equation}
 \begin{equation}
\gamma-\gamma_{e,\text{exp}}=\Delta\, a_{\chi}^{+}\left(\Delta\tau\right)^{\Delta}\label{experimental gamma difference (B19)}\end{equation}
 Here, the point to point transformation at $t=\Delta\tau^{*}=1$
(symbol \textbf{x}), is given by the scale factor universal power
law $\vartheta^{\Delta}$.

The above approximation of the logarithmic singularity has a practical
importance for better analysis of experimental data when the value
$\gamma_{e}$ is found in the range $\gamma_{e}=1.21-1.24$, i.e.
a value which {}``approaches'' the theoretical Ising value. As a
typical example we use the value $\gamma_{e,pVT}=1.211\pm0.025$ obtained
by Levelt-Sengers et al \citet{Levelt1975} from their analysis of
the $pVT$ measurements of Habgood and Schneider \citet{Habgood1954}
in the temperature range $0.2\, K\leq T-T_{c}\leq1.8\, K$ , i. e.
$\Delta\tau_{\text{min}}^{*}=6.9\times10^{-4}$, $\Delta\tau_{\text{max}}^{*}=6.2\times10^{-3}$
and $\left\langle \Delta\tau_{pVT}^{*}\right\rangle =2.07\times10^{-3}$
{[}see line $\#1$, column $4$, Table IV]. This result is then centered
near to the Ising-like borderline of the gray paint domain previously
analyzed. As evidenced by the matching of the corresponding points
labeled $1$ with the curves $P_{\Delta}$ and $P_{z}$ in Figs. \ref{Figure 5}c
and \ref{Figure 5}d, such a result also appears correctly accounted
for using the above approximation. Therefore, using Eqs. (\ref{experimental gamma difference (B19)})
and (\ref{experimental Gammaplus difference (B16)}), we can easily
calculate the two values of the \emph{true} confluent and leading
amplitudes of the two-term Wegner expansion from the following equations\begin{eqnarray}
\left.a_{\chi}^{+}\right|_{pVT} & =\frac{\gamma-\gamma_{e,pVT}}{\Delta\left\langle \Delta\tau_{pVT}^{*}\right\rangle ^{\Delta}}\label{Xe confluent ampltide from pVT (B20)}\\
 & =1.26666\nonumber \\
\left.\Gamma^{+}\right|_{pVT} & =\frac{\Gamma_{e,pVT}^{+}}{1+\frac{\mathbb{Z}_{0}}{\Delta}\left(\gamma-\gamma_{e,pVT}\right)\left\langle \Delta\tau_{pVT}^{*}\right\rangle ^{z-\Delta}}\label{Xe leading amplitude from pVT (B21)}\\
 & =0.057355\nonumber \end{eqnarray}
 These above values are in excellent agreement with the estimated
ones $a_{\chi}^{+}=1.23397$ and $\Gamma^{+}=0.05782$1 \citet{Garrabos2006khiT}
from application of the scale dilatation method. In Eq. (\ref{Xe leading amplitude from pVT (B21)}),
we note the practical importance of the prefactor $\mathbb{Z}_{0}$.

In conclusion, using Figs \ref{Figure 4} and \ref{Figure 5}, we
have explicitely demonstrated that the two dimensionless scale factors
$Y_{c}$ and $Z_{c}$, (or alternatively but equivalently $\vartheta_{\mathcal{L}}\left(\equiv\vartheta\right)$
and $\mathbb{X}_{0,\mathcal{L}}^{*}$), which characterize each one-component
fluid $f$ belonging to $\left\{ 1f\right\} $-subclass, can be used
to calculate the isothermal compressibility over a Ising-like extended
asymptotic domain $\Delta\tau^{*}\lesssim\mathcal{L}_{\text{EAD}}^{f}$.

\end{document}